# First M87 Event Horizon Telescope Results. IV. Imaging the Central Supermassive Black Hole

The Event Horizon Telescope Collaboration
(See the end matter for the full list of authors.)

## Abstract

We present the first Event Horizon Telescope (EHT) images of M87, using observations from April 2017 at 1.3 mm wavelength. These images show a prominent ring with a diameter of ~40 $\mu$as, consistent with the size and shape of the lensed photon orbit encircling the "shadow" of a supermassive black hole. The ring is persistent across four observing nights and shows enhanced brightness in the south. To assess the reliability of these results, we implemented a two-stage imaging procedure. In the first stage, four teams, each blind to the others' work, produced images of M87 using both an established method (CLEAN) and a newer technique (regularized maximum likelihood). This stage allowed us to avoid shared human bias and to assess common features among independent reconstructions. In the second stage, we reconstructed synthetic data from a large survey of imaging parameters and then compared the results with the corresponding ground truth images. This stage allowed us to select parameters objectively to use when reconstructing images of M87. Across all tests in both stages, the ring diameter and asymmetry remained stable, insensitive to the choice of imaging technique. We describe the EHT imaging procedures, the primary image features in M87, and the dependence of these features on imaging assumptions.

## 1. Introduction

Since the discovery of the first astrophysical jet apparently connected to its nucleus (Curtis 1918), the giant elliptical galaxy M87 in the Virgo cluster has been intensively studied with imaging observations. M87's nuclear gas and stellar dynamics, as traced by optical and infrared (IR) spectroscopy, suggest the presence of a nuclear supermassive black hole (SMBH) of mass $M_{\rm BH} \sim (3.3\text{–}6.2) \times 10^9\, M_\odot$ (Macchetto et al. 1997; Gebhardt & Thomas 2009; Gebhardt et al. 2011; Walsh et al. 2013). This high mass, combined with its proximity ($D = 16.8$ Mpc; Blakeslee et al. 2009; Bird et al. 2010; Cantiello et al. 2018; see also EHT Collaboration et al. 2019e, hereafter Paper VI), implies that the nuclear black hole candidate in M87 (hereafter referred to as M87) has an event horizon subtending the second-largest known angular size after Sagittarius A$^*$ (Sgr A$^*$) in the Galactic Center.

Unlike Sgr A$^*$, M87 hosts a powerful kpc-long jet that is bright in the radio, optical, and X-ray bands (e.g., Owen et al. 1989; Sparks et al. 1996; Perlman et al. 1999; Marshall et al. 2002). Weak emission east of the very long baseline interferometry (VLBI) core from the expected counter-jet has also been detected in high-frequency VLBI images (Walker et al. 2018). Material moves down the approaching jet with a maximum apparent speed of $\sim 6c$ (Biretta et al. 1999). On pc and sub-pc scales, VLBI observations show the jet to be edge-brightened and parabolic in shape (Reid et al. 1989; Dodson et al. 2006; Kovalev et al. 2007; Asada & Nakamura 2012; Hada et al. 2013; Nakamura & Asada 2013; Asada et al. 2016) with a characteristic progressive acceleration downstream (Asada et al. 2014; Mertens et al. 2016; Britzen et al. 2017; Hada et al. 2017; Walker et al. 2018; Kim et al. 2018a). High-frequency astrometric VLBI measurements reveal a frequency-dependent shift of the radio core (from optical depth effects), which asymptotically converges to ~40 microarcseconds ($\mu$as) east of the 7 mm core (Hada et al. 2011); this indicates that the jet is launched in the vicinity of the central black hole (e.g., Nakamura et al. 2018) residing within the central ~100 $\mu$as. The high mass and relative proximity of M87 provides an opportunity to image this black hole and jet-launching region on event-horizon scales; however, accessing these scales with ground-based VLBI requires observations with microarcsecond resolution at a wavelength of $\lesssim$1 mm.

To this end, we have developed the Event Horizon Telescope (EHT), a global ad hoc VLBI array operating at 1.3 mm wavelength (EHT Collaboration et al. 2019b, hereafter Paper II). With its longest baselines spanning nearly the diameter of the Earth, the synthesized beam size of the EHT array is approximately 20 $\mu$as. For M87, the EHT beam size corresponds to 3–5 $R_s$, where the Schwarzschild radius $R_s = 2GM_{\rm BH}/c^2$ subtends 3.9–7.3 $\mu$as for the black hole mass range and distance given above. Thus, the EHT can potentially resolve general relativistic effects associated with the SMBH in M87, most notably the "shadow" cast by the black hole on the bright surrounding emission (Bardeen 1973; Luminet 1979; Falcke et al. 2000). This shadow is expected to be encircled by a bright ring at the radius of the lensed photon sphere, with a diameter between approximately 4.8 and 5.2 $R_s$ for a maximally spinning black hole (viewed face-on) and a non-spinning (i.e., Schwarzschild) black hole, respectively (Bardeen 1973; Johannsen & Psaltis 2010). For M87, the expected shadow diameter is 19–38 $\mu$as. Physical models and general relativistic magnetohydrodynamic (GRMHD) simulations show that Doppler-boosted emission from rapidly rotating material near the black hole can result in substantial image brightness asymmetry very near the ring (EHT Collaboration et al. 2019d, hereafter Paper V).

Early EHT observations in 2009 and 2012 detected compact emission with an FWHM size of approximately 40 $\mu$as (Doeleman et al. 2012; Akiyama et al. 2015). However, because of their limited interferometric baseline coverage, these early experiments could not synthesize an image of M87, leaving considerable uncertainty about the nature of the detected emission.

In 2017 April, the EHT conducted an observing campaign using eight stations in six geographic sites.[105] The EHT observed M87 on four days (April 5, 6, 10, and 11), interleaved with observations of other targets. Notably, these observations included 37 telescopes of the Atacama Large Millimeter/submillimeter Array (ALMA) coherently combined to act as a single 73 m diameter telescope (Matthews et al. 2018). The addition of baselines to ALMA significantly increases the sensitivity of the entire EHT array. Additional details of the EHT instrument are given in Paper II; details of the 2017 observations, correlation, and calibration are given in EHT Collaboration et al. (2019c, hereafter Paper III).

We generated images of M87 from the 2017 EHT data in two stages. In the first stage, our aim was to compare the results of four independent imaging teams. Each team used their collective judgment to produce an image, while remaining blind to the others' work. In the second stage, our aim was to minimize reliance on expert judgment, and we instead utilized three imaging pipelines to perform large searches over the imaging parameter spaces. We then selected fiducial imaging parameters for M87 from these searches based on the performance of the parameters when reconstructing images from synthetic data. In both stages, the reconstructed images were consistently dominated by a single feature: a bright ring with a $\sim$40 $\mu$as diameter and azimuthal brightness asymmetry, consistent with the shadow of a $\sim 6.5 \times 10^9 M_\odot$ SMBH in the nucleus of M87.

This Letter presents details of this imaging procedure and analysis; our primary images of M87 are shown in Figures 11 and 15. We begin, in Section 2, by providing a brief review of interferometric measurements and imaging techniques. In Section 3, we describe the EHT observations of M87 in 2017. In Section 4, we estimate image properties for M87 such as total compact flux density and size of the emission region using a non-imaging analysis of the EHT data in combination with results from VLBI imaging of M87 at longer wavelengths. In Section 5, we present the first reconstructed EHT images, generated by four separate imaging teams working independently. In Section 6, we derive fiducial imaging parameters for three different imaging pipelines by performing large parameter surveys and testing on synthetic data. In Section 7, we show the corresponding fiducial images from each pipeline, and we assess their properties and uncertainties in the image and visibility domains. In Section 8, we perform validation tests of the fiducial images via comparisons of gain solutions with the simultaneously observed target 3C 279 and by exploring the dependence of reconstructed images on the participation and calibration of each site. In Section 9, we give a quantitative analysis of the image and ring properties. In Section 10, we summarize our results.

## 2. Background

Every antenna $i$ in an interferometric array records the incoming complex electric field as a function of time, frequency, and polarization: $E_i(t, \nu, P)$. Every pair of sites $(i, j)$ then gives an interferometric visibility, defined as the complex cross-correlation between their recorded electric fields,

$$V_{ij}(t, \nu, P_1, P_2) = \langle E_i(t, \nu, P_1) E_j^*(t, \nu, P_2) \rangle. \tag{1}$$

In practice, radio telescopes record data in dual circular feeds, right circular polarization (RCP) and left circular polarization (LCP), or in an orthogonal linear basis. The set of four possible cross-correlations among the two recorded polarizations at the two sites then provides information about the four Stokes parameters (see, e.g., Roberts et al. 1994). Because this Letter is focused on monochromatic and total intensity imaging, we will suppress the frequency and polarimetric parameters for the remainder of our discussion.

By the van Cittert–Zernike theorem (van Cittert 1934; Zernike 1938), the visibility measured with an ideal interferometer is related to the brightness distribution on the sky $I(x, y)$ via a simple Fourier transform. The interferometer samples a spatial frequency of the image given by the vector baseline $\boldsymbol{b}_{ij}$ joining the sites, projected orthogonal to the line of sight and measured in wavelengths: $\boldsymbol{u}_{ij} = \boldsymbol{b}_{\perp,ij}/\lambda$, where $\lambda = c/\nu$. For a pair of telescopes at fixed locations on the Earth, their projected baseline coordinates $(u, v)$ trace an ellipse as the Earth rotates, sampling a range of spatial frequencies (Thompson et al. 2017). Generalizing to an arbitrary projected baseline $\boldsymbol{u} = (u, v)$, the ideal visibility $\mathcal{V}$ is given by

$$\mathcal{V}(u, v) = \iint e^{-2\pi i (ux+vy)} I(x, y) dx\, dy. \tag{2}$$

In this expression, $x$ and $y$ are angular coordinates on the sky in radians. Equation (2) assumes that the source radiation is spatially incoherent and in the far-field limit. It also assumes that the sky brightness distribution is only non-zero within both a radian and the primary beams of the individual telescopes. All of these criteria are satisfied for EHT observations.[106] In the remainder of this section, we describe the specific interferometric data products used for imaging from EHT data, and we discuss how these data products are affected by practical measurement limitations (Section 2.1). We then summarize the primary methods used to reconstruct images from interferometric data (Section 2.2). For a comprehensive discussion of the EHT data reduction pipeline, see Paper III.

### 2.1. VLBI Data Products and Their Uncertainties

In practice, each measured visibility is contaminated by thermal noise, station-dependent errors, and baseline-dependent errors. However, some combinations of visibilities (so-called "closure" quantities) can be used to construct data products that are independent of station-dependent errors. We now discuss these various data products and define how they will be used to assess consistency between a trial image $I$ and the data.

#### 2.1.1. Interferometric Visibilities

The fundamental interferometric data product is the complex visibility, $V_{ij}$ (see Equation (1)). Each visibility is complex-valued and obeys a conjugation relationship under baseline reversal, $V_{ij} = V_{ji}^*$. Thus, an $N_{\rm el}$-element array samples at most $N_{\rm el}(N_{\rm el} - 1)/2$ different visibilities at each time.

---

[105] M87 (J2000 decl. of $+12°23'28''$) is not visible to one of the EHT stations, the South Pole Telescope (SPT).

[106] EHT baselines are sensitive to image structure subtending $\lesssim$100 $\mu$as $\approx 5 \times 10^{-10}$ radians (Section 4.4).

Visibility measurements have a simple thermal error budget. Neglecting errors in calibration (Section 2.1.2), the measured visibility $V_{ij}$ on a baseline $\boldsymbol{u}_{ij}$ is given by the true visibility $\mathcal{V}_{ij} = \mathcal{V}(\boldsymbol{u}_{ij})$ plus baseline-dependent thermal noise $\epsilon_{ij}$: $V_{ij} = \mathcal{V}_{ij} + \epsilon_{ij}$. For perfect coherent averaging, the thermal noise on measurements obeys a complex normal distribution, with equal variance on the real and imaginary parts. The thermal noise standard deviation, $\sigma_{ij}$, is related to the system equivalent flux densities (SEFDs) of the two sites, the bandwidth $\Delta\nu$, and the coherent integration time $\Delta t$:

$$\sigma_{ij} = \frac{1}{\eta_{\rm Q}}\sqrt{\frac{{\rm SEFD}_i \times {\rm SEFD}_j}{2\,\Delta\nu\,\Delta t}}, \tag{3}$$

where $\eta_{\rm Q}$ is a factor that accounts for sensitivity loss from quantization of the incident electric field. The EHT utilizes standard 2-bit recording, for which $\eta_{\rm Q} = 0.88$ (Thompson et al. 2017).

We quantify agreement between a trial image $I$ and a set of measured visibilities using the mean squared standardized residual. Specifically, for a measured set of $N_{\rm V}$ visibilities, we define

$$\chi^2_{\rm vis}(I) = \frac{1}{2N_{\rm V}}\sum\frac{|\hat{V} - V|^2}{\sigma^2}, \tag{4}$$

where $\hat{V}$ denotes a model visibility corresponding to the trial image, and the sum ranges over all measured visibilities (which may include different baselines, times, and frequencies). The quantity $\chi^2_{\rm vis}$ does not formally correspond to a reduced $\chi^2$, and its interpretation requires caution. For example, we do not include a correction for the effective number of image degrees of freedom, and the results can be sensitive to data processing procedures such as self-calibration. Nevertheless, we expect $\chi^2_{\rm vis} \approx 1$ for final imaging results.

We also use a mean squared standardized residual to quantify agreement between a trial image and a set of measured visibility amplitudes:

$$\chi^2_{\rm amp}(I) = \frac{1}{N_{\rm V}}\sum\frac{|\hat{A} - A|^2}{\sigma^2}, \tag{5}$$

where $\hat{A} = |\hat{V}|$ denotes a model visibility amplitude, and

$$A = \sqrt{(|V|^2 - \sigma^2)\Theta(|V| - \sigma)} \tag{6}$$

is a measured visibility amplitude after correcting for an upward bias from thermal noise. Here, the Heaviside $\Theta$-function ensures that the debiased amplitudes are always real and positive.

### 2.1.2. Closure Quantities

In addition to thermal noise, measured visibilities have systematic errors on both their amplitudes and phases. These errors can often be factorized into multiplicative station-based "gains," which are complex-valued and may vary in both time and frequency. In general, these gains act as linear transformations of the incident polarized electric fields at each site; here, we focus on a simpler model using scalar electric fields. These complex systematic errors modify the relationship between a measurement $V_{ij}$ and the corresponding ideal visibility $\mathcal{V}_{ij}$ as

$$V_{ij} \approx g_i g_j^* \mathcal{V}_{ij} + \epsilon_{ij}, \tag{7}$$

where $\epsilon_{ij}$ is the thermal noise. Equation (7) is an approximation, as it neglects cross-talk between orthogonal receivers, the non-factorizability of the station-dependent bandpass after frequency averaging, the non-factorizability of station-dependent temporal fluctuations after coherent averaging, and other imperfections. We discuss these imperfections and their implications for the EHT error budget in Section 4.2 (for a comprehensive discussion, see Paper III).[107]

The site dependence of the gains allows for the construction of data products that are insensitive to these systematic errors. Such quantities are called "closure" quantities. For instance, the sum of three visibility phases $\psi = \arg(V)$ around a directed triangle $i$–$j$–$k$ of sites is insensitive to the phase of the complex station-based gains $g_i$. Namely, neglecting thermal noise,

$$\psi_{{\rm C},ijk} = \psi_{ij} + \psi_{jk} + \psi_{ki} = \Psi_{ij} + \Psi_{jk} + \Psi_{ki}, \tag{8}$$

where $\Psi = \arg(\mathcal{V})$ is the phase of the ideal visibility. This phase $\psi_{{\rm C},ijk}$ is called the closure phase of the $i$–$j$–$k$ triangle (Jennison 1958). Because phase stability is challenging for millimeter VLBI, closure phase is a useful measurement of intrinsic source phase. For an $N_{\rm el}$-element interferometer, there are $N_{\rm el}(N_{\rm el}-1)(N_{\rm el}-2)/6$ closure phase measurements, but only $(N_{\rm el}-1)(N_{\rm el}-2)/2$ of these are non-redundant.

When the signal-to-noise ratio (S/N) of each component visibility is at least $\sim$3, closure phase errors are well approximated as Gaussian and depend only on the three component S/N values $S$ (e.g., Rogers et al. 1995):

$$\sigma_{\psi_{\rm C},ijk} \approx \sqrt{S_{ij}^{-2} + S_{jk}^{-2} + S_{ki}^{-2}}. \tag{9}$$

In this expression, $\sigma_{\psi_{\rm C}}$ is the closure phase standard deviation in radians.

We compute $\chi^2_{\rm CP}$ statistics to quantify image-data agreement analogously to Equation (4):

$$\chi^2_{\rm CP}(I) = \frac{1}{N_{\psi_{\rm C}}}\sum\frac{|e^{i\hat{\psi}_{\rm C}} - e^{i\psi_{\rm C}}|^2}{\sigma^2_{\psi_{\rm C}}}, \tag{10}$$

where $\hat{\psi}_{\rm C}$ denotes a model closure phase from the trial image, $N_{\psi_{\rm C}}$ is the number of closure phases in the sum, and the sum ranges over all measurements (which may include different triangles, times, and frequencies).

Similar to closure phase, certain combinations of four visibility amplitudes on a quadrangle $ijk\ell$ are insensitive to gain amplitudes $|g_i|$. For example, neglecting thermal noise,

$$A_{{\rm C},ijk\ell} = \frac{A_{ij}A_{k\ell}}{A_{ik}A_{j\ell}} = \frac{\mathcal{A}_{ij}\mathcal{A}_{k\ell}}{\mathcal{A}_{ik}\mathcal{A}_{j\ell}}, \tag{11}$$

where $\mathcal{A} = |\mathcal{V}|$. Such combinations $A_{{\rm C},ijk\ell}$ are called closure amplitudes (Twiss et al. 1960). While $N_{\rm el}$ sites give at most $N_{\rm el}(N_{\rm el}-1)(N_{\rm el}-2)(N_{\rm el}-3)/8$ closure amplitudes, only $N_{\rm el}(N_{\rm el}-3)/2$ of these are non-redundant. Also, while the noise statistics of closure amplitudes are approximately Gaussian if each of the four amplitudes has high S/N, the distribution is highly non-Gaussian when any of the measurements is low S/N, especially those in the denominator (L. Blackburn et al. 2019, in preparation; A. E. Broderick et al. 2019, in preparation). To mitigate this behavior and to symmetrize the properties of numerator and denominator terms, we use the logarithm of the closure amplitudes, $\ln A_{\rm C}$, for analysis. We compute $\chi^2_{\log {\rm CA}}$ statistics as

$$\chi^2_{\log {\rm CA}}(I) = \frac{1}{N_{\ln A_{\rm C}}} \sum \frac{|\ln \hat{A}_{\rm C} - \ln A_{\rm C}|^2}{\sigma^2_{\ln A_{\rm C}}}, \quad (12)$$

where the uncertainty on the log closure amplitude is, in the high-S/N limit,

$$\sigma_{\ln A_{\rm C}, ijk\ell} \approx \sqrt{S_{ij}^{-2} + S_{k\ell}^{-2} + S_{ik}^{-2} + S_{j\ell}^{-2}}. \quad (13)$$

[107] Note that the simulated data used later in this Letter do not use the simplified Equation (7) but instead employ a full Jones matrix formulation (Appendix C).

Note that closure quantities are generally not independent, and their uncertainties are only approximately Gaussian.

### 2.2. Interferometric Imaging Methods

Because an interferometer incompletely samples the visibility domain, the inverse problem of determining an image from a measured set of visibilities is ill-posed. Consequently, reconstructed images are not unique—they always require information, assumptions, or constraints beyond the interferometer measurements. One strong imaging constraint is image positivity. A second strong constraint is a restricted field of view (FOV) for the reconstructed image (i.e., the gridded FOV).[108] The choice of FOV must be made with care, as incorrect restrictions can result in false image structure. Imaging algorithms can additionally impose or favor other physically motivated properties related to the image (e.g., image smoothness) or to the instrument (e.g., a maximal resolution of reconstructed features).

Imaging algorithms can be broadly categorized into two methodologies: inverse modeling and forward modeling. The former includes deconvolution methods such as CLEAN (Högbom 1974; Clark 1980), while the latter includes regularized maximum likelihood (RML) methods such as the classical maximum entropy method (MEM; e.g., Narayan & Nityananda 1986). We now discuss each of these in more depth, with an emphasis on aspects and algorithms that are most relevant for EHT imaging.

#### 2.2.1. Imaging via Inverse Modeling

Conventional inverse modeling approaches to imaging typically begin with an inverse Fourier transform of the sampled visibilities (giving the so-called "dirty image"). They then deconvolve the effects associated with the limited baseline coverage (giving the so-called "dirty beam"). For both VLBI and connected-element radio interferometry, the standard reconstruction algorithm in this category is CLEAN (e.g., Högbom 1974; Schwarz 1978; Clark 1980; Schwab 1984; Cornwell et al. 1999; Cornwell 2008).

The classical CLEAN algorithm (Högbom 1974; Clark 1980) models the image as a collection of point sources and determines the locations and flux densities of these point sources iteratively. After reaching a specified stopping criterion, CLEAN typically convolves the many-point-source image model with a "clean beam." This beam is obtained by matching a Gaussian to the central component of the dirty beam, and it approximates the point-spread function of the interferometric data. However, it is the *unconvolved* point-source model that is compared with the data to assess goodness of fit; the beam-convolved image displayed will not fit the measured visibility amplitudes.

Lack of absolute phase information and a priori calibration uncertainties in EHT measurements require multiple consecutive iterations of CLEAN followed by self-calibration, a routine that solves for station gains to maximize consistency with visibilities of a specified trial image (Wilkinson et al. 1977; Readhead et al. 1980; Cornwell & Wilkinson 1981; Pearson & Readhead 1984; Cornwell & Fomalont 1999). For the CLEAN imaging in this Letter, we used the `DIFMAP` software (Shepherd 1997, 2011).

#### 2.2.2. Imaging via Forward Modeling

Forward modeling approaches to imaging usually represent the image as an array of pixels and only require a Fourier transform of this array to evaluate consistency between the image and data. These methods can easily integrate nonlinear image-data consistency measures and constraints on the reconstructed image properties, such as sparsity or smoothness. While these types of imaging algorithms have been developed for decades (e.g., Frieden 1972; Gull & Daniell 1978; Cornwell & Evans 1985; Narayan & Nityananda 1986; Briggs 1995; Wiaux et al. 2009a, 2009b) and are commonly utilized in optical interferometry (e.g., Buscher 1994; Baron et al. 2010; Thiébaut 2013; Thiébaut & Young 2017), they are used less frequently than CLEAN for radio interferometry. However, forward modeling methods have been intensively developed for the EHT (e.g., Honma et al. 2014; Bouman et al. 2016, 2018; Chael et al. 2016, 2018; Ikeda et al. 2016; Akiyama et al. 2017a, 2017b; Johnson et al. 2017; Kuramochi et al. 2018).

These methods are often derived using probabilistic arguments, although they do not typically produce results that have formal probabilistic interpretations. The general approach in this type of imaging is to find the image $I$ that minimizes a specified objective function,

$$J(I) = \sum_{\text{data terms}} \alpha_D \chi^2_D(I) - \sum_{\text{regularizers}} \beta_R S_R(I). \quad (14)$$

In this expression, each $\chi^2_D$ is a goodness-of-fit function corresponding to the data term $D$ (Section 2.1), and each $S_R$ is a regularization term corresponding to the regularizer $R$ (Appendix A). The "hyperparameters" $\alpha_D$ and $\beta_R$ control the relative weighting of the data and regularization terms. Placing too much weight on the regularization terms will result in final reconstructed images that are not acceptably compatible with the data (i.e., the image will not achieve $\chi^2_D \sim 1$ for all data terms). Because this function is analogous to the log-likelihood of a posterior probability function, we refer to this category of imaging methods as RML methods.

Regularizers explored in VLBI include image entropy (e.g., Narayan & Nityananda 1986), smoothness (e.g., Bouman et al. 2016; Chael et al. 2016; Kuramochi et al. 2018), and sparsity in the image or its gradient domain (e.g., Wiaux et al. 2009a, 2009b; Honma et al. 2014; Akiyama et al. 2017b). Regularizing functions can be combined; e.g., simultaneously favoring both sparsity and smoothness through regularization can mitigate limitations of using only one or the other (e.g., Akiyama et al. 2017a).

[108] The $\mathrm{FOV_{arr}}$ of an interferometric array depends on the primary beams of the antennas and the averaging time and bandwidth (Thompson et al. 2017). For VLBI, $\mathrm{FOV_{arr}}$ is usually substantially larger than the FOV of reconstructed images. Throughout this Letter, we will exclusively use FOV to refer to the angular extent of reconstructed images.

Note that while regularization is an important choice in RML methods, other decisions such as data preparation, the optimization methodology, and the FOV of the reconstructed image can be equally important in influencing the final image. Unlike CLEAN, no final restoring beam is required when using RML, and these methods often achieve a modest degree of "super resolution" (i.e., finer than the nominal diffraction limit, $\theta \sim \lambda/|\boldsymbol{u}|_{\max}$).

For the EHT, RML methods mitigate some difficulties of CLEAN reconstructions through increased flexibility in the data products used for imaging. For instance, RML methods can directly fit to robust data products such as closure quantities (e.g., Buscher 1994; Baron et al. 2010; Bouman et al. 2016; Chael et al. 2016, 2018; Akiyama et al. 2017a, 2017b). However, RML methods often introduce other difficulties, such as requiring the solution to a complex nonlinear optimization problem (minimizing Equation (14)). As with CLEAN, RML methods sometimes employ an iterative imaging loop, alternately imaging and then self-calibrating the data. For the RML reconstructions in this Letter, we used two open-source software libraries that have been developed specifically for the EHT: `eht-imaging`[109] (Chael et al. 2016, 2018, 2019) and `SMILI`[110] (Akiyama et al. 2017a, 2017b, 2019). Appendix A gives definitions of the regularizers used throughout this Letter, as implemented in these libraries.

## 3. Observations and Data Processing

### 3.1. EHT Observations and Data

The EHT observed M87 with seven stations at five geographic sites on 2017 April 5, 6, 10, and 11. The participating facilities were the phased Atacama Large Millimeter/submillimeter Array (ALMA) and Atacama Pathfinder Experiment telescope (APEX) in the Atacama Desert in Chile, the James Clerk Maxwell Telescope (JCMT) and the phased Submillimeter Array (SMA) on Maunakea in Hawai'i, the Arizona Radio Observatory Sub-Millimeter Telescope (SMT) on Mt. Graham in Arizona, the IRAM 30 m (PV) telescope on Pico Veleta in Spain, and the Large Millimeter Telescope Alfonso Serrano (LMT) on Sierra Negra in Mexico. These observations of M87 were interleaved with other targets (e.g., the quasar 3C 279), some of which were visible to an eighth EHT station, the South Pole Telescope (SPT).

Data were recorded in two polarizations and two frequency bands. All sites except ALMA and the JCMT recorded dual circular polarization (RCP and LCP). ALMA recorded dual linear polarization that was later converted to a circular basis via `PolConvert` (Martí-Vidal et al. 2016; Matthews et al. 2018; Goddi et al. 2019), and the JCMT recorded a single circular polarization (the recorded polarization varied from day to day).[111] All sites recorded two 2 GHz bands centered on 227.1 and 229.1 GHz (henceforth, low and high band, respectively). Paper II provides details on the setup, equipment, and station upgrades leading up to the 2017 observations.

Paper III outlines the correlation, calibration, and validation of these data. In particular, the data reduction utilized the sensitive baselines to ALMA to estimate and correct for stable instrumental phase offsets, RCP–LCP delays, and stochastic phase variations within scans. After these corrections, the data have sufficient phase stability to coherently average over scans. The data were also amplitude calibrated using station-specific measurements; stations with an intra-site partner (i.e., ALMA, APEX, SMA, and JCMT) were then "network calibrated" to further improve the amplitude calibration accuracy and stability via constraints among redundant baselines. The final network-calibrated data sets were frequency averaged per band and coherently averaged in 10 s intervals before being used for our imaging analysis. All data presented and analyzed in this work are Stokes $I$ (or pseudo $I$) visibilities processed via the EHT–Haystack Observatory Processing System (HOPS) pipeline (Paper III; Blackburn et al. 2019). Information about accessing SR1 data and the software used for analysis can be found on the Event Horizon Telescope website's data portal.[112]

### 3.2. Data Properties

Figure 1 shows the baseline $(u, v)$ coverage for EHT observations of M87. The shortest baselines in the EHT are intra-site (i.e., the SMA and JCMT are separated by 0.16 km; ALMA and APEX are separated by 2.6 km). These intra-site baselines are sensitive to arcsecond-scale structure (see Section 4.3). In contrast, our longest baselines (joining the SMA or JCMT to PV) are sensitive to microarcsecond-scale structure. Baseline coverage on individual days (bottom panels of Figure 1) is comparable for April 5, 6, and 11 (18, 25, and 22 scans, respectively). However, April 10 had significantly less coverage, with only 7 scans.

Figure 2 (left panel) shows the S/N as a function of baseline length for M87 on April 11, after coherently averaging scans. The split in S/N distributions at various baseline lengths is due to the sharp difference in sensitivity for the co-located Atacama sites ALMA and APEX. The right panel of Figure 2 shows the visibility amplitude (correlated flux density) for M87 on April 11 after amplitude and network calibration.

There is a prominent secondary peak in the network-calibrated visibility amplitudes between two deep minima ("nulls"), the first at $\sim$3.4 G$\lambda$ and the second at $\sim$8.3 G$\lambda$. The amplitudes along the secondary peak are weakly dependent on baseline position angle, suggesting some degree of source symmetry, and their overall trends are consistent for all days (see Paper III, Figure 13). However, evidence for source anisotropy can be seen at the location of the first null, where the east–west oriented Hawai'i–LMT baseline gives significantly lower amplitudes than the north–south oriented ALMA–LMT baseline at the same projected baseline length (see also Paper VI). This anisotropy is further supported by multiple measurements of non-zero closure phase (Figure 3). The majority of notable low-amplitude outliers across days are due to reduced performance of the JCMT or the LMT on a select number of scans. Despite the amplitudes of these data being low, the derived closure quantities remain stable (Paper III).

Similarly, most closure quantities for M87 are broadly consistent across all days, although day-to-day variations are significant for some sensitive closure combinations involving long baselines to PV or to the Hawai'i stations. Figure 3 shows examples of closure phases for various triangles and levels of variability: a "trivial" triangle including co-located sites (ALMA–APEX–SMT, left panel) that is expected to be consistent with zero, a non-trivial and mostly non-variable triangle (ALMA–LMT–SMT, center panel) with largely persistent structure across all days, and a non-trivial triangle (LMT–SMA–SMT, right panel) showing intrinsic source structure evolution in M87 between the two sets of observations on April 5, 6 and 10, 11.

[109] https://github.com/achael/eht-imaging

[110] https://github.com/astrosmili/smili

[111] Because the JCMT recorded a single circular polarization, baselines to JCMT use Stokes "pseudo $I$." Namely, we use parallel-hand visibilities to approximate Stokes $I$ under the assumption that the source is weakly circularly polarized.

[112] https://eventhorizontelescope.org/for-astronomers/data

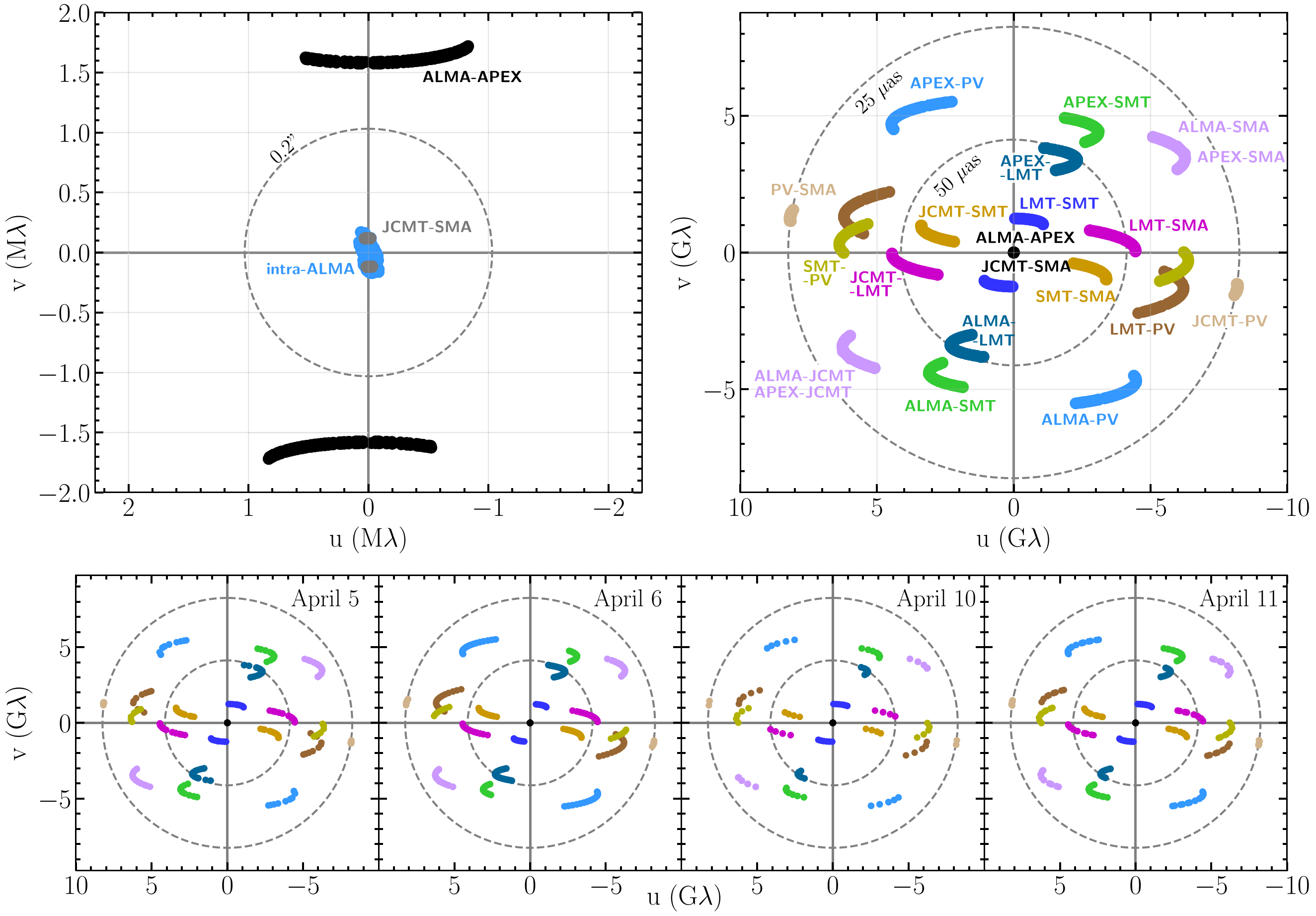

**Figure 1.** Top panels: aggregate baseline coverage for EHT observations of M87, combining observations on all four days. The left panel shows short-baseline coverage, comprised of ALMA interferometer baselines and intra-site EHT baselines (SMA–JCMT and ALMA–APEX). These short baselines probe angular scales larger than 0.1". The right panel shows long-baseline coverage, comprised of all inter-site EHT baselines. These long baselines span angular scales from 25 to 170 $\mu$as. Each point denotes a single scan, which range in duration from 4 to 7 minutes. Bottom panels: the full baseline coverage on M87 for each observation. In all panels, the dashed circles show baseline lengths corresponding to the indicated fringe spacings (0.2" for the upper-left panel; 25 and 50 $\mu$as for the remaining panels).

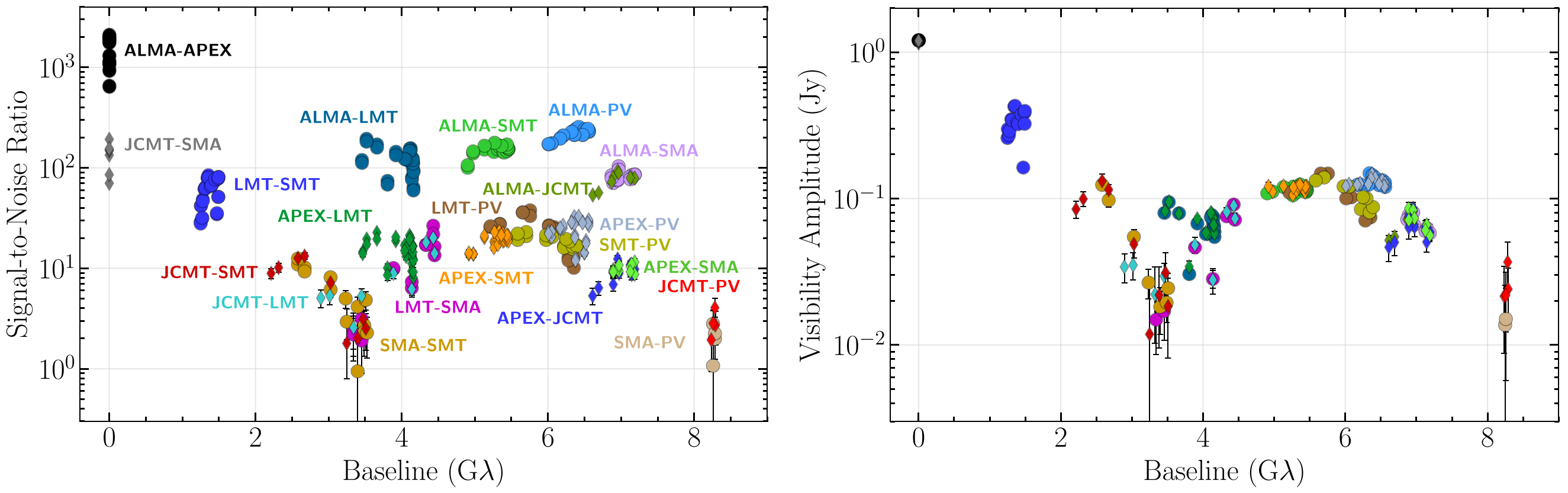

**Figure 2.** Left panel: S/N as a function of projected baseline length for EHT observations of M87 on April 11. Each point denotes a visibility amplitude coherently averaged over a full scan (4–7 minutes). Points are colored by baseline. Right panel: visibility amplitudes (correlated flux density) as a function of projected baseline length after a priori and network calibration. The amplitudes are corrected for upward bias from thermal noise (Equation (6)). Error bars denote $\pm 1\sigma$ uncertainty from thermal noise and do not include expected uncertainties in the a priori calibration (see Paper III and Section 4.1).

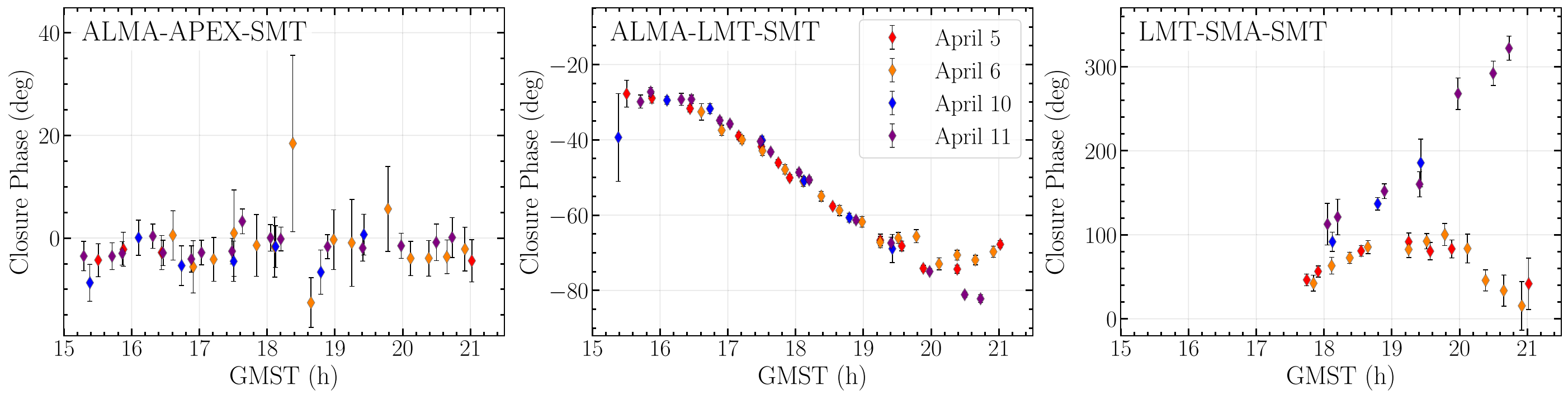

**Figure 3.** Selected closure phases from coherently averaged visibilities on three triangles as a function of Greenwich Mean Sidereal Time (GMST) using data from all four days. Error bars denote $\pm 1\sigma$ uncertainties from thermal noise. The trivial ALMA–APEX–SMT triangle (left panel) has closure phases near zero on all days, as expected because this triangle includes an intra-site baseline. Deviations from zero arise from a combination of thermal and systematic errors (Paper III). The ALMA–LMT–SMT triangle (middle panel) shows persistent structure across all days, while the large LMT–SMA–SMT triangle (right panel) shows source evolution between the first two days and last two days.

## 4. Pre-imaging Considerations

In this section, we analyze the measured M87 visibilities directly to assess what conclusions are supported by the data irrespective of choices made in the imaging procedures. This non-imaging analysis provides estimates of the quality of the a priori calibration, the level of non-closing systematic errors, and the degree of temporal variability in the source. It also guides the generation of realistic synthetic data sets (Section 6.1) and motivates the choice of imaging parameters (e.g., the FOV of reconstructed images) used to define parameter ranges in the imaging parameter surveys (Section 6.2).

### *4.1. Expected Amplitude Calibration Limitations*

The amplitude calibration error budget is determined from uncertainties on individual measurements of station performance, as described in Paper III. The error budget is only representative of global station performance and is not specified for individual measurements. While this procedure is adequate for stations with stable performance and weather during the observing run, the error budget may be underestimated for stations with variable performance. The SMT, PV, SMA, JCMT, APEX, and ALMA stations are well characterized either through years of studies or via network calibration. More recent additions to the EHT (the LMT and the SPT) may have more variable behavior, as their observing systems are not yet well exercised and because they do not have sufficiently close sites to permit network calibration.

Specifically for M87, the LMT is the most undercharacterized station. The LMT began observing M87 in the evening, when the dish is still affected by thermal gradients in the back-up structure or panel distortions from daytime heating, both of which are significant for open-air telescopes. These effects are common for sensitive millimeter-wave dishes and cause surface instability. In addition, evening conditions are inadequate for accurate pointing and focusing of the telescope (particularly on weaker sources), leading to performance that can vary substantially across scans and from day to day. A defocused dish can measure persistently low amplitudes on baselines to that station between focus attempts (typically every one to two hours). Changes in telescope pointing can cause amplitudes to fluctuate significantly from scan to scan (from the telescope moving to and from the source) and between pointing attempts (typically every half-hour). Issues in telescope focus can also lead to uncertainties in the a priori calibration for other sources observed during the same time period, such as 3C 279. However, pointing errors for 3C 279 are expected to be less severe, as it is bright enough for the LMT to point directly on-source prior to VLBI scans. Thus, the corrections needed for the LMT are expected to better match the a priori amplitude error budget during observations of 3C 279 (mostly correcting for focus errors) than during observations of M87 (correcting for both focus and pointing errors).

In Section 8.1 and Appendix F, we compare estimated residual gains for the SMT and the LMT from imaging M87 and 3C 279. In Paper VI, we compare these results with the estimated residual gains when fitting parametric models to the interferometric data.

### *4.2. Estimates of Non-closing Systematic Errors*

Apart from the thermal errors, which are well understood and can be evaluated from first principles, there are several sources of additional systematic error present in the EHT data. These include polarimetric leakage (proportional to the magnitude of station-dependent leakage terms and the source fractional polarization), residual polarimetric-gain calibration offsets, loss of amplitude with long coherent averaging, the effects of biased S/N estimators, and the limitations of a linear error propagation model for closure quantities.

These systematic errors can affect both visibilities and closure quantities. While these effects are small, in a very high S/N regime systematic uncertainties may dominate over thermal uncertainties. In Paper III, we quantified the magnitude of systematic uncertainties in the EHT 2017 data both with a series of statistical tests on the distributions of trivial closure quantities and with cross-validation tests of data products across different frequency bands and polarizations. For M87, the characteristic magnitude of systematic errors was found to be only a fraction of the thermal errors, even for heavily averaged data. In the case of 3C 279, which has significant intrinsic polarization, systematic errors can dominate over the thermal errors. In both cases the characteristic magnitudes are small: less than 2° for closure phases, and less than 4% for closure amplitudes.

### *4.3. Constraints on the Total Compact Flux Density*

EHT baselines sample angular scales in M87 that span nearly five orders of magnitude. The largest gap in coverage occurs between the arcsecond scales sampled by intra-site baselines (ALMA–APEX and SMA–JCMT) and the microarcsecond scales sampled by inter-site baselines. The intra-site baselines are insensitive to detailed structure on microarcsecond scales; their measurements are consistent with an unresolved point source having $\sim$1.2 Jy of flux density (i.e., the "core" of M87, corresponding to the bright, upstream end of the jet) in addition to faint compact features in the large-scale jet of M87, such as "Knot A" (Owen et al. 1989). Long inter-site baselines resolve out much of this emission, and are sensitive to only the compact central source. The shortest inter-site EHT baseline is SMT–LMT, which has a fringe spacing of 139–166 $\mu$as for M87. Thus, there are two relevant estimates of the total flux density: the total flux density seen in the core on arcsecond scales, $F_{\rm tot} \approx 1.2$ Jy, and the total compact flux density, $F_{\rm cpct} \leqslant F_{\rm tot}$, which is the relevant quantity for the integrated flux density of a reconstructed image with an FOV limited to only a few hundred microarcseconds.

In Appendix B, we derive a series of constraints on $F_{\rm cpct}$ using the 2017 EHT data and quasi-simultaneous observations at longer wavelengths. These constraints depend on the effective size of the compact emission region (expressed as an equivalent Gaussian FWHM). We also compare these constraints with the estimates of source size and compact flux density reported in Doeleman et al. (2012) and Akiyama et al. (2015) from observations of M87 in 2009 and 2012, respectively. These previous results used a three-site array, which included the Combined Array for Research in Millimeter-wave Astronomy (CARMA; in California) in addition to the SMT, JCMT, and SMA. The addition of CARMA provided a short inter-site baseline (SMT–CARMA, with $|\boldsymbol{u}| \sim 0.6$ G$\lambda$), while the two remaining baselines, SMT–JCMT/SMA and CARMA–JCMT/SMA, sampled $2.2 < |\boldsymbol{u}| < 3.5$ G$\lambda$.

Taken together, the EHT constraints give $F_{\rm cpct} = 0.64^{+0.39}_{-0.08}$ Jy; combined with the multi-wavelength constraints, they give $F_{\rm cpct} = 0.66^{+0.16}_{-0.10}$ Jy. For both these estimates, the central value gives the maximum a posteriori estimate and the uncertainties give the 95% confidence interval after marginalizing over the source size. Because all of the EHT constraints use the SMT–LMT baseline, they correspond to the source size projected along that direction. However, because the visibility amplitudes are highly symmetric, we expect these estimates to apply at all position angles (see Section 3). We infer a value of $F_{\rm cpct}$ that is significantly lower than what was found in observations in 2009 and 2012.[113]

As there is a range of possible values for the total compact flux density, we do not enforce a common value for all reconstructed images of M87. Instead, before evaluating image-data consistency through $\chi^2$ metrics, we add a large Gaussian component (FWHM $> 500$ $\mu$as) to each reconstruction, which only affects visibilities on the intra-site baselines. The integrated flux density of this Gaussian component is set so that the total flux density of the modified image matches the intra-site visibility amplitudes.

We emphasize that the constraints on the size and total flux density of the compact emission were derived by making rigid assumptions about the station gains, by combining constraints among all four observing days (over which M87 may have varied in $F_{\rm cpct}$ and size), and by imposing short-baseline approximations for the visibility function. Thus, while we will use these constraints to guide our imaging and analysis, we will not strictly enforce them.

[113] In addition to the EHT measurements in 2009, quasi-simultaneous Global mm-VLBI Array (GMVA) 86 GHz observations of M87 in 2009 May found an elevated core flux density of $\sim$1 Jy (Kim et al. 2018a), while in more recent epochs the total flux density decreased close to the value of $F_{\rm cpct}$ estimated for 2017 (see Hada et al. 2016; Kim et al. 2018a).

**Table 1**
Metrics of EHT Angular Resolution for the 2017 Observations of M87

| | FWHM$_{\rm maj}$ ($\mu$as) | FWHM$_{\rm min}$ ($\mu$as) | P.A. (°) |
|---|---|---|---|
| Minimum Fringe Spacing $1/|\boldsymbol{u}|_{\rm max}$ (All Baselines) | | | |
| April 5–11 | 24.8 | ⋯ | ⋯ |
| Minimum Fringe Spacing (ALMA Baselines) | | | |
| April 5–11 | 28.6 | ⋯ | ⋯ |
| CLEAN Beam (Uniform Weighting) | | | |
| April 5 | 25.8 | 17.9 | 10.1 |
| April 6 | 24.9 | 17.5 | 2.3 |
| April 10 | 25.3 | 17.2 | 6.7 |
| April 11 | 25.4 | 17.4 | 6.0 |
| CLEAN Restoring Beam (Used in this Letter) | | | |
| April 5–11 | 20 | 20 | ⋯ |

**Note.** Because the EHT coverage is fairly symmetric, to avoid asymmetries introduced by restoring beams and to homogenize the images among epochs, we adopt a circular Gaussian restoring beam with 20 $\mu$as FWHM for all CLEAN reconstructions.

### *4.4. Image Resolution and Degrees of Freedom*

The diffraction-limited resolution of an interferometer depends upon its finest fringe spacing: $\theta_{\rm min} = f_\theta/|\boldsymbol{u}|_{\rm max}$, where $f_\theta$ is a scaling coefficient. The resolution of reconstructed images also depends upon the baseline coverage and sensitivity and is commonly quantified using procedures to fit the central peak of the point-spread function (or "dirty beam"). Table 1 gives several common representations of the EHT resolution for observations of M87.

The maximum number of independent degrees of freedom in a reconstructed image is then $N_{\rm dof} \lesssim (\theta_{\rm fov}/\theta_{\rm min})^2$ , where $\theta_{\rm fov}$ is the FOV of the image. Of course, the image may only be non-zero in a small fraction of the full FOV, reducing $N_{\rm dof}$ correspondingly. We emphasize that the number of degrees of freedom in a reconstructed image is not determined by the number of pixels in the image. Likewise, for a constrained FOV, it is not possible to arbitrarily fit (or overfit) interferometric data by making the pixel resolution increasingly fine because the baseline correlation length for measurements in the visibility domain is determined by the FOV rather than by the pixel resolution. Consequently, a reconstructed image and its effective number of degrees of freedom is sensitive to the FOV of the image, but not to the chosen pixel size.

### *4.5. Image Conventions*

Throughout this Letter, we present images using their equivalent brightness temperatures defined by the Rayleigh–Jeans law: $T_{\rm b} = \frac{c^2}{2\nu^2 k} I_\nu$, where $I_\nu$ is the specific intensity, $c$ is the speed of light, $\nu$ is the observing frequency, and $k$ is the Boltzmann constant (e.g., Rybicki & Lightman 1979). We use brightness temperature rather than the standard radio convention of flux density per beam (e.g., Jy/beam) because our images are spatially resolved and because RML methods do not have a natural associated beam. However, we emphasize that brightness temperature does not necessarily correspond to any physical temperature of the radio-emitting plasma. The radio spectrum of M87 is not a blackbody, and its 230 GHz emission is from synchrotron radiation (Paper V). Finally, for visualization of our images, we use perceptually uniform colormaps from the ehtplot[114] library.

Throughout this Letter, we restore CLEAN images using a circular Gaussian beam with 20 $\mu$as FWHM, which is comparable to the geometric mean of the principal axes of the CLEAN beam (see Table 1). Unless otherwise stated, any image that has been restored will include the restoring beam in the lower right. Also, for consistency with RML methods but in contrast with standard practice, our presented CLEAN images do not include the residual image, corresponding to the inverse Fourier transform of gridded residual visibilities. In Section 7.2, we describe characteristics of the residual images.

## 5. First M87 Images from Blind Imaging

VLBI images are sensitive to choices made in the imaging and self-calibration process. Choices required in using any imaging method include deciding which data are used (e.g., low and/or high band, flagging), specifying the self-calibration procedures, and fixing the reconstructed image FOV. In addition, imaging methods also require choices that are particular to their assumptions and methodology. For CLEAN, these choices include choosing a set of CLEAN windows and a data-weighting scheme. For RML methods, choices include the selection of which data and regularizer terms and weights to use in the objective function (Equation (14)). With this abundance of user input, it can be difficult to assess what image properties are reliable from a given imaging method.

The dangers of false confidence and collective confirmation bias are magnified for the EHT because the array has fewer sites than typical VLBI arrays, there are no previous VLBI images of any source at 1.3 mm wavelength, and there are no comparable black hole images on event-horizon scales at any wavelength. To minimize the risk of collective bias influencing our final images, in our first stage of analysis we reconstructed images of M87 in four independent imaging teams.

### *5.1. Imaging Procedure and Team Structure*

We subdivided our first M87 imaging efforts into four separate imaging teams. The teams were blind to each others' work, prohibited from discussing their imaging results and even from discussing aspects of the data that might influence imaging (e.g., which stations or data might be of poor quality). No restrictions were imposed on the data pre-processing or imaging procedures used by each team. Teams 1 and 2 focused on RML methods, while Teams 3 and 4 primarily used CLEAN. In addition to independently imaging M87, teams also independently imaged other sources observed by the EHT in 2017 to test the blind imaging procedure.

Blind imaging procedures have long been used to reduce the risk of group bias. Prior to the 2017 observations, we organized a series of "imaging challenges" that used synthetic data to assess how conventional and newly developed imaging algorithms would perform for the EHT (Bouman 2017).[115] Reconstructing images independently in these challenges helped us identify which image features were likely intrinsic, and which were likely to be spurious. To compare EHT 2017 results among teams while keeping submissions blind, we built a website that allowed users to independently upload images and automatically compare them to the ground truth images and submissions from other users (Bouman 2017).

### *5.2. First M87 Imaging Results*

The first M87 imaging analysis used an early-release engineering data set (ER4; Paper III). These data had a priori and network calibration applied but did not have calibrated relative RCP–LCP gains. Consequently, each team imaged the data using only parallel-hand products (i.e., RCP–RCP or LCP–LCP) to approximate total intensity. The April 11 data set was selected for the first comparison, as it had the best coverage for the M87/3C 279 pair and the most stable a priori amplitude calibration (especially for the LMT).

The imaging teams worked on the data independently, without communication, for seven weeks, after which teams submitted images to the image comparison website using LCP data (because the JCMT recorded LCP on April 11). After ensuring image consistency through a variety of blind metrics (including normalized cross-correlation, Equation (15)), we compared the independently reconstructed images from the four teams.

Figure 4 shows these first four images of M87. All four images show an asymmetric ring structure. For both RML teams and both CLEAN teams, the ring has a diameter of approximately 40 $\mu$as, with brighter emission in the south. In contrast, the ring azimuthal profile, thickness, and brightness varies substantially among the images. Some of these differences are attributable to different assumptions about the total compact flux density and systematic uncertainties (see Table 2).

The initial blind imaging stage indicated that the image of M87 is dominated by a $\sim$40 $\mu$as ring. The ring persists across the imaging methods. Next, we moved to a second, non-blind imaging stage that focused on exploring the space of acceptable images for each method. The independent team structure was only used for the first stage of imaging; the remainder of this Letter will categorize results by imaging methodology.

## 6. Determining Imaging Parameters via Surveys on Synthetic Data

To explore the dependence of the reconstructed images on imaging assumptions and impartially determine a combination of fiducial imaging parameters, we introduced a second stage of image production and analysis: performing scripted parameter surveys for three imaging pipelines. To objectively evaluate the fidelity of the images reconstructed by our surveys—i.e., to select imaging parameters that were independent of expert judgment—we performed these surveys on synthetic data from a suite of model images as well as on the M87 data. The synthetic data sets were designed to have properties that are similar to the EHT M87 visibility amplitudes (e.g., prominent amplitude nulls). This suite of synthetic data allowed us to test the scripted reconstructions with knowledge of the corresponding ground truth images and, thereby, select fiducial imaging parameters for each method. These fiducial parameters were selected to perform well across a variety of source

[114] https://github.com/chanchikwan/ehtplot

[115] Similar blind procedures have also been used in the optical interferometry community to evaluate and compare imaging methods (e.g., Lawson et al. 2004).

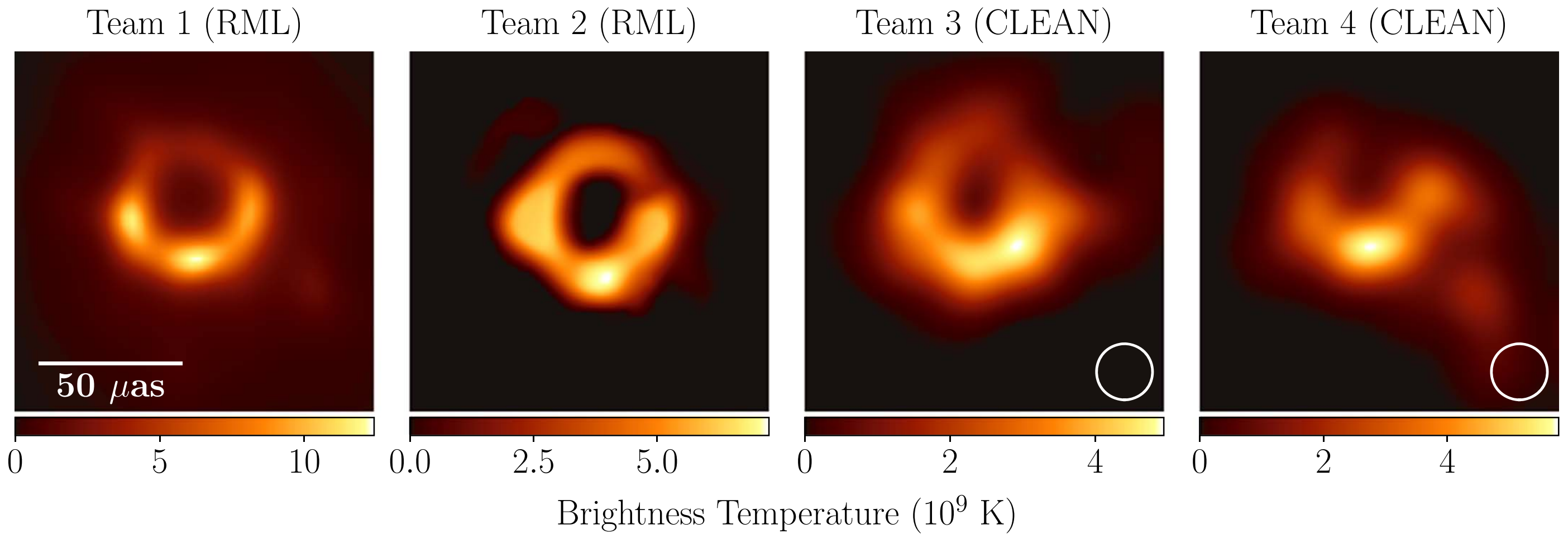

**Figure 4.** The first EHT images of M87, blindly reconstructed by four independent imaging teams using an early, engineering release of data from the April 11 observations. These images all used a single polarization (LCP) rather than Stokes $I$, which is used in the remainder of this Letter. Images from Teams 1 and 2 used RML methods (no restoring beam); images from Teams 3 and 4 used CLEAN (restored with a circular 20 $\mu$as beam, shown in the lower right). The images all show similar morphology, although the reconstructions show significant differences in brightness temperature because of different assumptions regarding the total compact flux density (see Table 2) and because restoring beams are applied only to CLEAN images.

**Table 2**
Image Properties and Data Consistency Metrics for the First M87 Images (See Figure 4)

| | Team 1 | Team 2 | Team 3 | Team 4 |
|---|---|---|---|---|
| Image Properties | | | | |
| Method | RML | RML | CLEAN | CLEAN |
| $F_{\rm cpct}$ (Jy) | 0.94 | 0.43 | 0.42 | 0.42 |
| Engineering Data (10 s avg., LCP, 0% sys. error) | | | | |
| $\chi^2_{\rm CP}$ | 2.06 | 2.48 | 2.44 | 2.33 |
| $\chi^2_{\rm log\,CA}$ | 1.20 | 2.16 | 2.15 | 1.43 |
| Science Release (scan-avg., Stokes $I$, 0% sys. error) | | | | |
| $\chi^2_{\rm CP}$ | 1.13 | 5.40 | 2.28 | 1.89 |
| $\chi^2_{\rm log\,CA}$ | 2.12 | 5.41 | 3.90 | 5.32 |
| Science Release (scan-avg., Stokes $I$, 1% sys. error) | | | | |
| $\chi^2_{\rm CP}$ | 1.00 | 3.85 | 2.04 | 1.55 |
| $\chi^2_{\rm log\,CA}$ | 1.96 | 5.07 | 3.64 | 4.8 |
| Science Release (scan-avg., Stokes $I$, 10% sys. error) | | | | |
| $\chi^2_{\rm CP}$ | 0.49 | 0.95 | 1.11 | 0.48 |
| $\chi^2_{\rm log\,CA}$ | 0.46 | 1.36 | 0.98 | 0.79 |

**Note.** Data metrics are shown as originally computed on April 11 data (using 10 s averaged engineering data with LCP) and using the data from the first EHT science release (scan-averaged, Stokes $I$) when 0%, 1% and 10% systematic error has been included. Teams 2–4 chose to exclude the intra-site baselines in their imaging. However, for consistency with our later $\chi^2$ values computed from science release data, we include these baselines when computing $\chi^2$ after adding an extended component to these images containing the missing flux density.

structures, including sources without the prominent ring observed in our images of M87.

We emphasize that the ensemble of results from these parameter surveys do not correspond to a posterior distribution of reconstructed images. Our surveys are coarse-grained and do not completely explore the choices in the imaging process. Nonetheless, they identify regions of imaging parameter space that consistently produce faithful image reconstructions on synthetic data, and they help us identify which features of our reconstructions are consistent and which features vary with specific parameter choices.

### *6.1. Synthetic Source Models and Data*

To create a testing suite of synthetic data, we selected simple geometric models that have corresponding visibility amplitudes that are similar to those observed in M87 (Figure 2). The primary data properties that we used to define similarity are (1) a large decrease in flux density on baselines between 0 and 1 G$\lambda$, indicating extended structure, (2) visibility nulls at $\sim$3.4 and $\sim$8.3 G$\lambda$, and (3) a high secondary peak between the nulls at $\sim$6 G$\lambda$, which recovers $\sim$15% of the total compact flux density.

We selected four models with distinct compact morphologies that each reproduce these features of the M87 data. The four models are (1) a tapered ring with 44 $\mu$as ring diameter, (2) a tapered crescent of the same diameter with its brightest point oriented directly south, (3) a tapered disk with 70 $\mu$as diameter, and (4) two different circular Gaussian components separated by 32.3 $\mu$as at a position angle of 292°. To ensure rough consistency and compatibility with the M87 parameters estimated in Section 4, we adopted a total compact flux density of 0.6 Jy for all these simple geometric models. Note that none of the synthetic EHT data sets generated from these simple models reproduces all features seen in the M87 data. For example, the ring and disk models both have point symmetry, so all their closure phases are either 0° or 180°.

To simulate the effects of a large-scale jet on our data (which only significantly affects intra-site visibilities), we added a three-component Gaussian model that approximates the inner M87 jet at 3 mm (e.g., Kim et al. 2018a). The jet also has 0.6 Jy of total flux density, giving a total image flux density in each case (compact+jet) of 1.2 Jy. To produce non-closing systematic errors from polarimetric leakage, we also included linear polarization in each model. For additional details on these simulated models and data, see Appendix C.1. Figure 5 shows these model images.

We generated synthetic data from each image using the `eht-imaging` software library. The synthetic data were produced with the baseline coverage and sensitivity of the EHT on all four days of the 2017 observations. Station-based errors were added in a Jones matrix formalism (Thompson et al. 2017; see Appendix C.2). To simulate a lack of absolute phase calibration, a random station phase is adopted for each scan. Independent station-based gains were applied to each visiblity, with two components of each gain factor drawn from normal distributions: one term that was constant over the observation, and one that varied randomly among scans. We captured the increased uncertainty and variability in station gains at the LMT (Section 4.1) by giving these gain terms larger variances. Time-independent polarimetric leakage terms were drawn from a complex Gaussian distribution with 5% standard deviation, motivated by previous estimates of the polarization leakage at EHT sites (Johnson et al. 2015). Identical gain, phase, and leakage contamination was applied to the high- and low-band data. Figure 5 shows four examples of visibility amplitudes generated using this procedure. Appendix C.2 provides full details about the synthetic data generation procedure.

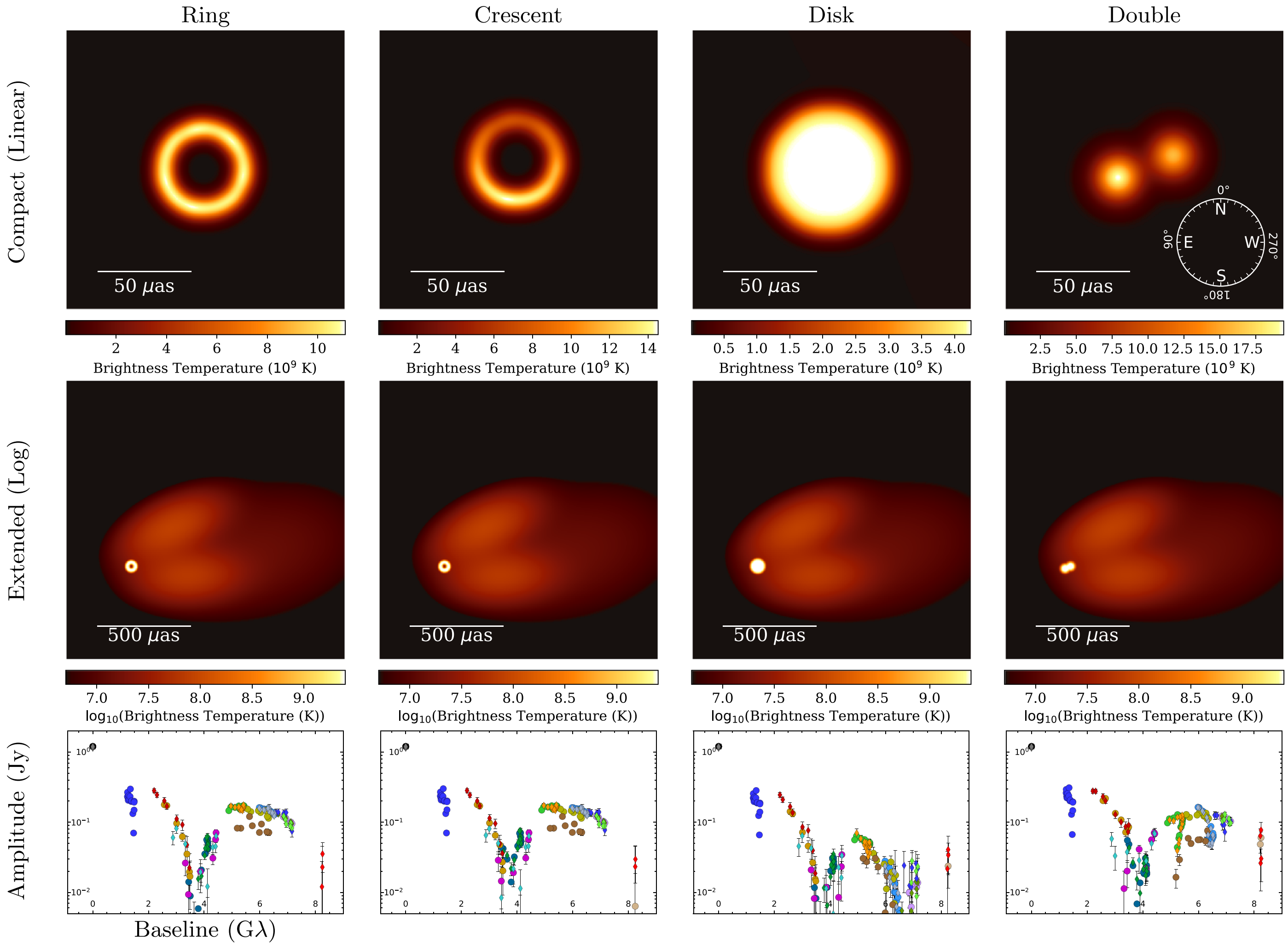

**Figure 5.** The four simple geometric models and synthetic data sets used in the parameter surveys (see Appendix C for details). Top: linear scale images, highlighting the compact structure of the models. Middle: logarithmic scale images, highlighting the larger-scale jet added to each model image. Bottom: one realization of simulated visibility amplitudes corresponding to the April 11 observations of M87. We indicate the conventions for cardinal direction and position angle used throughout this Letter on the upper-right panel. Note that east is oriented to the left, and position angles are defined east of north.

Furthermore, to test whether or not our results are sensitive to specific choices made in the synthetic data generation, we compared our generated synthetic data and reconstructed images to results from another VLBI data simulator, `MeqSilhouette+rPICARD` (Blecher et al. 2017; Janssen et al. 2019). `MeqSilhouette+rPICARD` takes a more physical approach to mm-VLBI synthetic data generation (with added corruption based on, e.g., measured weather parameters and antenna pointing offsets) and it uses the full `CASA`-based EHT reduction pipeline for calibration. Despite the differences in approach, both synthetic data pipelines yield comparable results (see Appendix C.3 and Figure 31). We use `eht-imaging` for all further synthetic data in this Letter.

### 6.2. Imaging Pipelines and Parameter Survey Space

To survey the space of imaging parameters relevant to each imaging software package and to test their effects on reconstructions of simulated data, we designed scripted imaging pipelines in three software packages: `DIFMAP`, `eht-imaging`, and `SMILI`. Each pipeline has some choices that are fixed (e.g., the convergence criterion, the pixel size, etc.) but takes additional parameters (e.g., the regularizer weights, the total compact flux density) as arguments. We then reconstructed images from all M87 and synthetic data sets using all possible parameter combinations on a coarse grid in the space of these parameters. We chose large ranges for each parameter, deliberately including values that we expected to produce poor reconstructions.

The parameter choices and the values surveyed vary among the three pipelines, and the pipelines also differ in which frequency bands were used. However, a consistent feature of the three pipelines is that none of them do any data flagging, even of intra-baselines.[116] We now describe each pipeline in detail.

### 6.2.1. DIFMAP

While CLEAN was implemented by hand in our first stage of analysis (Section 4), a scripted version of CLEAN was developed in DIFMAP to carry out the parameter search in this second stage. This script has five parameters taken as arguments: the total assumed compact component flux density, the stop condition, the relative weight correction factor for ALMA in the self-calibration process, the diameter of the cleaning region, and a parameter controlling how the $(u, v)$ grid weights are scaled by the visibility errors. CLEAN imaging with DIFMAP was performed separately for the low- and high-band data. The DIFMAP images we present in this Letter are derived from the low-band data averaged to 10 s in time.

The DIFMAP script begins by gridding the visibilities in the $(u, v)$ plane, after which the direct Fourier transform, or "dirty image," and the point-spread function, or "dirty beam," are computed. A uniform $(u, v)$ density weighting function is used for gridding the visibilities. This choice gives equal weight to each gridded spatial frequency containing a measurement, irrespective of sample density. The visibility weights in $(u, v)$ gridding can be further multiplied by $\sigma^{\kappa}$, where $\sigma$ is the thermal noise, with our parameter survey exploring $\kappa = -2, -1$, and 0.

In addition to the $(u, v)$ weighting, one or more antennas can be downweighted in the self-calibration process. When fitting for complex antenna gains, the self-calibration algorithm of DIFMAP uses weights that are proportional to $\sigma^{-2}$. In the case of EHT data, it is particularly helpful to downweight data on baselines to phased ALMA—which has a much higher S/N than other stations—so that these ALMA baselines do not dominate the corrections to phases and amplitudes. In the parameter survey, we have scaled the ALMA weights by a factor ranging from 0.01 to 1.0. All other station weights were kept at 1.0 (i.e., at their original values).

A disk-shaped set of CLEAN windows, or imaging mask, aligned with the previously found geometrical center of the underlying emission structure (e.g., the ring for M87) and with a specified diameter is then supplied; these windows define the area on the image where the CLEAN algorithm searches for point sources. We limited the cleaning windows to image only the compact structure ($\leqslant$100 $\mu$as) in order to prevent CLEAN from adding larger-scale emission features that are poorly constrained by the lack of short EHT baselines. The emission structure common to all early trials (Section 5) was used as a guide for setting the disk-shaped window for scripted reconstructions, but different diameters from 40 to 100 $\mu$as were tested in the parameter search.

Two main stopping criteria are implemented in the DIFMAP script to control the amount of cleaning; imaging may stop (1) when the desired total compact flux density is reached, or (2) when the relative decrease in the rms of the residual image over the image noise estimated from the visibilities dips below a given threshold. In the parameter survey, two cases were tested: (a) criterion 1 alone, and (b) a combination of criteria 1 and 2, such that cleaning will stop whenever one of the two criteria is reached first. In both cases, the assumed total compact flux density was varied from 0.5 to 0.8 Jy. The threshold for the relative decrease in the residual image rms in the criterion 2 was always fixed to $10^{-4}$.

After the script finishes a round of cleaning, it introduces a large ($\gtrsim$2 mas) Gaussian component that recovers the extended missing flux density and does not affect the compact emission structure. By fitting the most extended structure, the Gaussian component also acts as an "anchor" for the intra-site baselines, minimizing the variations in the antenna gains during the amplitude self-calibration loops.

The initial image, produced by iterative cleaning and phase self-calibration, is used to amplitude self-calibrate the LMT (with a solution interval equal to the scan length), which is known to have poor a priori amplitude calibration (see Section 4.1). In this step, the gains of the other antennas in the array are fixed. The initial model is then removed and further iterations of cleaning loops are performed until a new model that recovers all the compact flux density is obtained. This round is followed by self-calibration at all sites in both phase and amplitude. The loop of cleaning and self-calibration is then repeated with progressively shorter solution intervals in the amplitude self-calibration step, from an initial solution equal to the whole duration of the observations down to a minimum solution interval of 10 s. For consistency with the outputs of the eht-imaging and SMILI pipelines, the final CLEAN image omits the large Gaussian component.

Figure 6 shows a 2D slice of the DIFMAP parameter survey results from varying the weights on ALMA data in the self-calibration step and the diameter of the circular mask of CLEAN windows on synthetic data. We show the results for synthetic data of a crescent model and for M87 data from April 11. It is important to note that although DIFMAP images are displayed using a restoring beam of 20 $\mu$as (FWHM shown on figures), as is standard in CLEAN imaging, the underlying $\delta$-function CLEAN components are used for computing $\chi^2$ so as to avoid modifying the reconstructed amplitudes through Gaussian convolution.

### 6.2.2. eht-imaging

The eht-imaging parameter survey used a template imaging script that takes seven parameters as arguments. These are the total assumed compact component flux density, the fractional systematic error on the measured visibilities, the FWHM of the circular Gaussian initial image (also used as a prior for maximum entropy regularization), and the relative weights of four regularization terms: MEM, the $\ell_1$ norm ($\ell_1$), total variation (TV), and total squared variation (TSV). For details on each of these regularization terms, see Appendix A.

The eht-imaging script begins by preparing the data for imaging. After scan-averaging the data and combining the high- and low-band data, it rescales the amplitude of zero baseline measurements to account for the unresolved extended components. It then performs an initial self-calibration of the LMT by using only the LMT–SMT baseline and adjusting only the LMT amplitude gains to match a Gaussian image with a total flux density of 0.6 Jy and a size of 60 $\mu$as, based on the constraints determined in Section 4.

The script also incorporates additional systematic uncertainty into the error budget during imaging. Namely, for each visibility, a specified fraction of the visibility amplitude is added in quadrature to the thermal noise; this inflated thermal noise is then used to estimate uncertainties on all data products, thereby accounting for small, non-closing systematic errors from, e.g., polarimetric leakage. We surveyed values of 0%, 1%, 2%, and 5% in computing this additional systematic uncertainty. Note that these levels of systematic uncertainty do not necessarily correspond with the estimated level of non-closing errors presented in Paper III; however, additional systematic uncertainty beyond the a priori estimate can help in avoiding overfitting to a few sensitive baselines (e.g., those to ALMA) during image optimization.

[116] In the first imaging stage (Section 5), teams explored keeping or flagging the intra-site baselines and found that this choice does not significantly affect the recovered images. Although the inclusion of the intra-site baselines does not improve the $(u, v)$ coverage, these baselines add non-trivial closure amplitudes that can improve imaging (Chael et al. 2018).

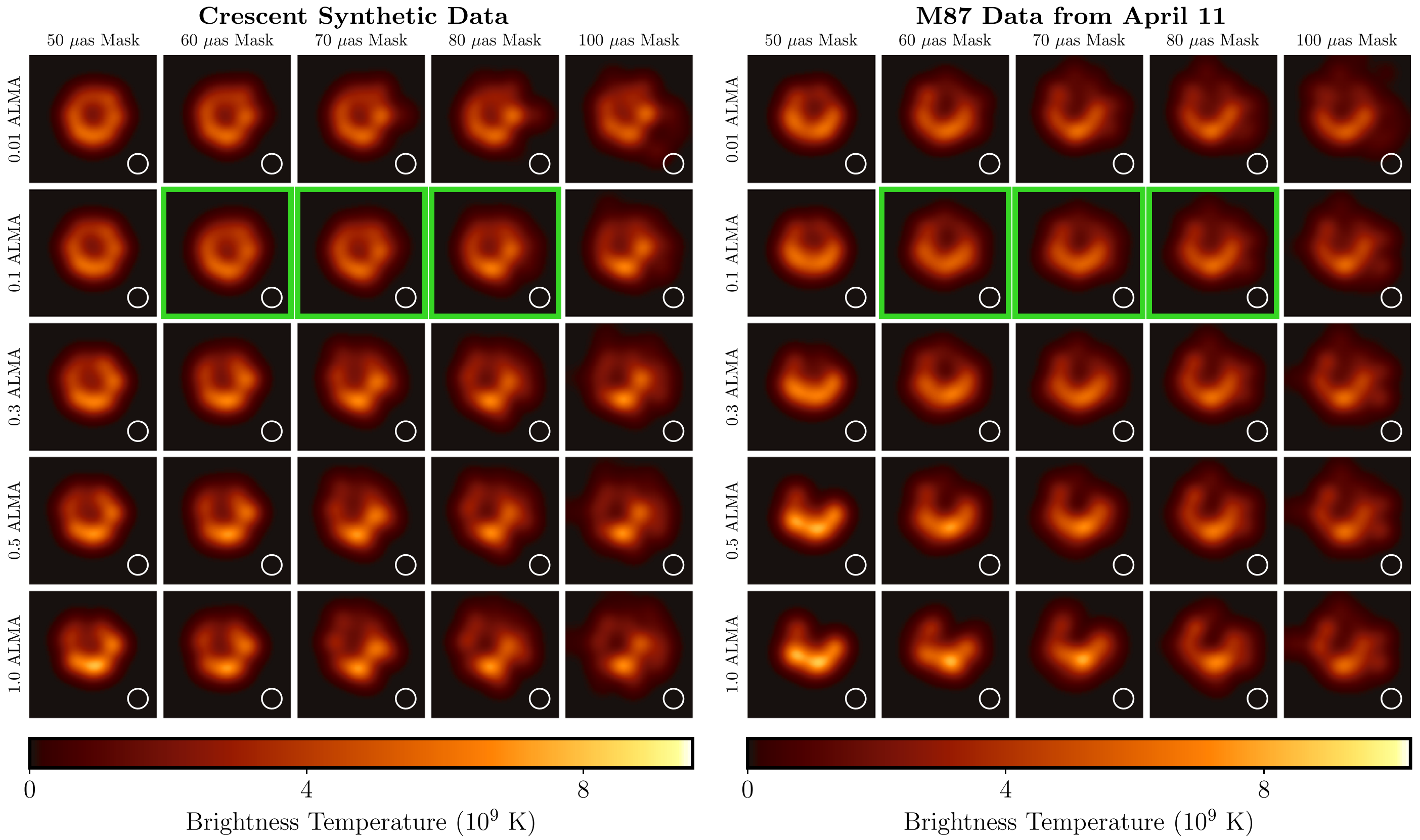

**Figure 6.** Selection of the DIFMAP (CLEAN) parameter survey results on real and synthetic data with April 11 EHT baseline coverage. A 2D slice of the 5D parameter space is displayed, corresponding to different diameters of the circular mask and the data weight on ALMA in self-calibration. All other parameters are kept constant (Compact Flux = 0.5 Jy, $\kappa = -1$, Stop Condition = Flux Reached). The left panel shows results of the parameter search on the Crescent synthetic data, while the right panel shows reconstructions for the same parameters on M87 data. Images that meet the threshold for the Top Set are outlined in green (see Section 6.3.1).

The final pre-imaging step is to *inverse* taper the visibility amplitudes. Specifically, we divide the visibility amplitudes and their estimated error by the Fourier transform of a 5 $\mu$as FWHM Gaussian, thereby raising the amplitudes on long baselines while preserving the S/N. At the end of the script, after producing an image from the inverse-tapered data, we convolve the reconstructed image with a 5 $\mu$as FWHM Gaussian to ensure that it will fit the original, untapered data. This procedure enforces a limiting angular resolution of 5 $\mu$as in the final reconstructed images.

After the data have been prepared, the eht-imaging script proceeds to imaging, using a $64 \times 64$ pixel grid with a 128 $\mu$as FOV. The script alternates between solving for the optimal image under the current data constraints and self-calibrating the complex station gains (Equation (7)). The first imaging iteration uses only visibility amplitudes and closure quantities. Because the amplitudes are uncalibrated in this iteration, systematic error tolerance is added on top of the (potentially inflated) thermal uncertainty for the visibility amplitudes, using the systematic error budget provided in Paper III for all sites except LMT, which has its budget increased by 15%. After phase self-calibration, complex visibilities are used for optimization in combination with closure quantities.

Between imaging iterations, the image is convolved with a circular Gaussian with a FWHM corresponding to the nominal array resolution, $\lambda/|\boldsymbol{u}|_{\max} \approx 25$ $\mu$as. This procedure aids convergence to a global minimum by moving intermediate images away from local minima. In each iteration, we increase the weight on data terms relative to the regularizers and reduce the systematic noise tolerance for amplitude calibration uncertainties. The final reconstructed image is convolved with the same 5 $\mu$as Gaussian that was used for the inverse taper, ensuring consistency with the original data while imposing a constraint that no features in the reconstructed image can be finer than 5 $\mu$as.

Figure 7 shows a 2D slice of the eht-imaging parameter survey results from varying the weights on the MEM and the TV regularizers. In this figure, we show the results for both synthetic data from a crescent model and for EHT M87 data from April 11.

### 6.2.3. SMILI

In the SMILI imaging pipeline, we reconstructed images using low-band EHT data and utilized weighted-$\ell_1$ ($\ell_1^{\rm w}$), TV, and TSV regularizers (see Appendix A). $\ell_1^{\rm w}$ favors sparsity in the image domain, using a circular Gaussian prior image as a "soft mask," which increasingly penalizes pixel intensities farther from the image center. In contrast, TV favors sparsity in the gradient domain (leading to piecewise-smooth images), and TSV favors overall image smoothness. The SMILI search adopts an FOV of 128 $\mu$as (matching the `eht-imaging` pipeline). The survey included five values for the FWHM of the prior image for $\ell_1^{\rm w}$, ranging from 40 to 80 $\mu$as. The pipeline also surveyed five values for the total compact flux density, between 0.4 and 0.8 Jy.

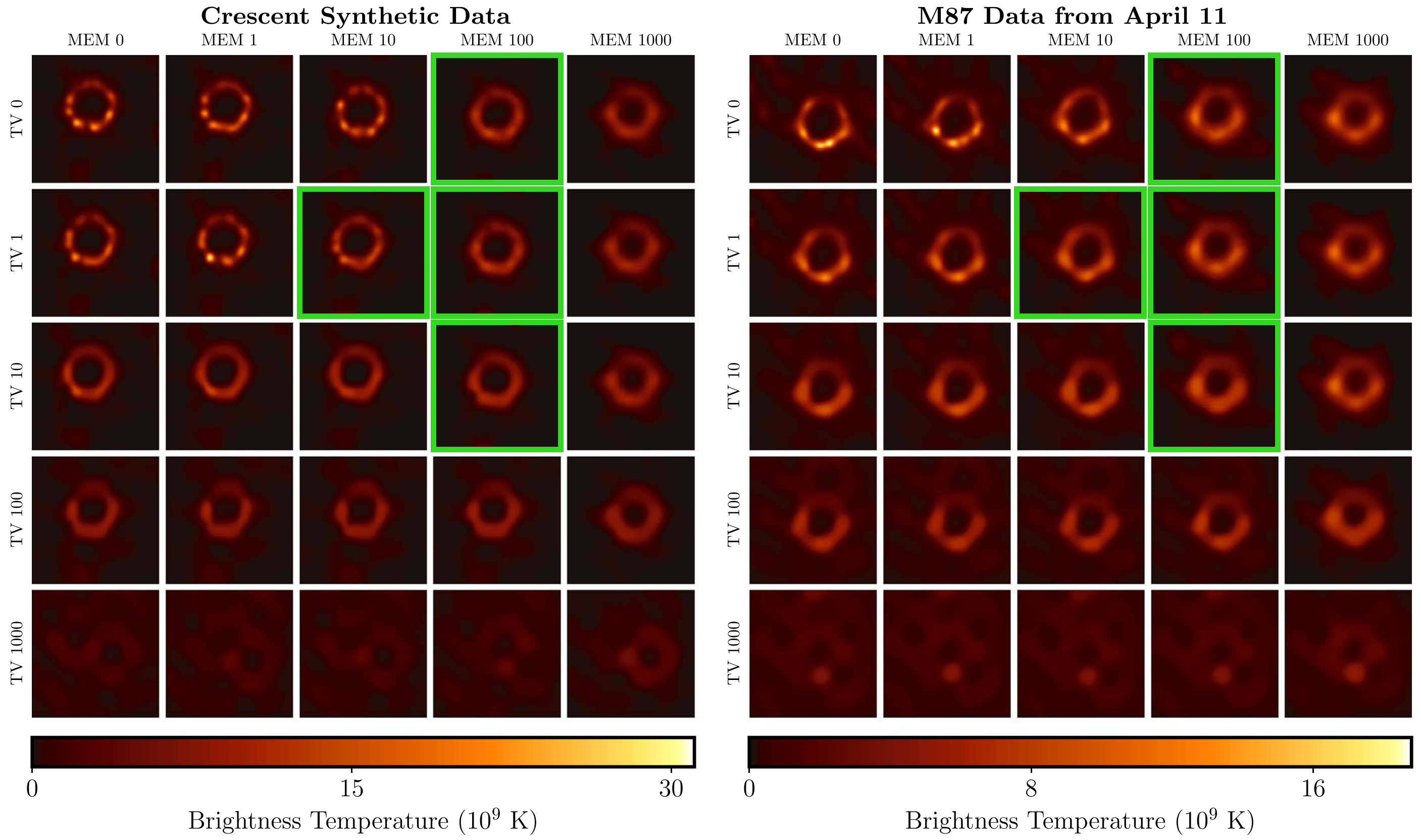

**Figure 7.** Selection of the `eht-imaging` (RML) parameter survey results on real and synthetic data with April 11 EHT baseline coverage. A 2D slice of the 7D parameter space is displayed, corresponding to different weights on the MEM and TV regularizers. All other parameters are kept constant (Compact Flux = 0.6 Jy, Initial/MEM FWHM = 40 $\mu$as, Systematic Error = 1%, TSV = 0, and $\ell_1 = 0$). The left panel shows results of the parameter search on the Crescent synthetic data, while the right panel shows reconstructions for the same parameters on M87 data. Images that meet the threshold for the Top Set are outlined in green. Note that the upper-left reconstruction has *no* regularization; it is produced by enforcing only image positivity and a constrained FOV.

Prior to imaging, the script coherently averages the visibilities over scans and then pre-calibrates the LMT gains using an assumed Gaussian model, as in the procedure described above for the `eht-imaging` script. The script then adds systematic uncertainties in quadrature to the thermal noise of complex visibilities to account for small, non-closing errors. The survey included options for 0% (no systematic error) and 1% systematic error, which correspond to the fraction of each visibility amplitude that is added in quadrature to the reported thermal noise $\sigma$ values.

Similar to the `eht-imaging` and `DIFMAP` scripts, the SMILI imaging procedure is iterative, with four stages of imaging and self-calibration. Reconstructions at each stage begin with a circular Gaussian with an FWHM of 20 $\mu$as as the initial image. After the first imaging attempt in each stage, subsequent initializations use the previously obtained image convolved with the initial Gaussian. We perform three imaging cycles for each self-calibration stage, self-calibrating only to the final reconstructed image in each of the three cycles.

In the first two self-calibration stages, the imaging step uses closure data (closure amplitudes and phases) and visibility amplitudes. To account for errors in the amplitude calibration, during these stages the visibility amplitudes have additional 5% and 30% uncertainties added in quadrature to the thermal noise (on top of the systematic error tolerance noted above) for non-LMT and LMT baselines, respectively. For the final two self-calibration stages, the imaging uses closure data and complex visibilities.

Figure 8 shows a 2D slice of the `SMILI` parameter survey results from varying the weights on the TSV regularizer and the prior radius. We show the results for synthetic data from a crescent model and for M87 data from April 11. Although the `SMILI` and `eht-imaging` pipelines have a number of similarities, these pipelines use different regularizers, calibrate and constrain data differently, and were implemented using independent software libraries.

### 6.3. Selecting Imaging Parameters

We derived two principal outputs from each parameter survey: a "Top Set" of parameter combinations that each produce acceptable images on the synthetic data, and a single combination of best-performing, "fiducial" parameters for each pipeline. In this section, we describe our ranking procedure to find acceptable parameter combinations, the possible limitations of this procedure, and the resulting acceptable parameter combinations for each imaging pipeline.

#### 6.3.1. Ranking Imaging Parameters

To determine the Top Sets, we designed a two-step selection procedure. The first step requires that all reconstructions of M87

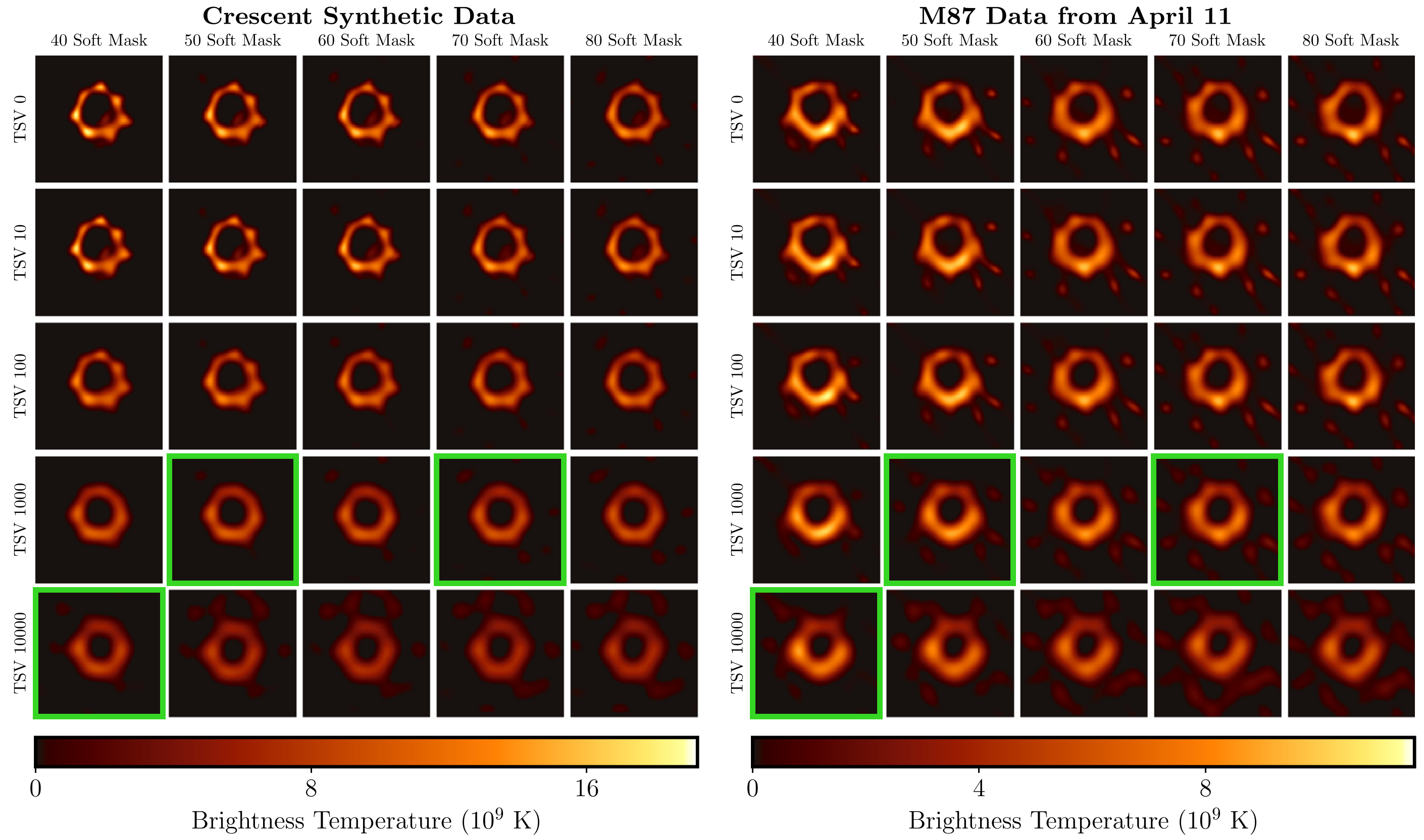

**Figure 8.** Selection of the SMILI (RML) parameter survey results on real and synthetic data with April 11 EHT baseline coverage. A 2D slice of the 6D parameter space is displayed, corresponding to varying the diameter of the soft mask region and the weight on the TSV regularizer. All other parameters are kept constant (Compact Flux = 0.5 Jy, Systematic Error = 0%, TV = 0, $\ell_1^{\mathrm{W}} = 1$). The left panel shows results of the parameter search on the Crescent synthetic data, while the right panel shows reconstructions for the same parameters on M87 data. Images that meet the threshold for the Top Set are outlined in green.

with a given combination of parameters are consistent with their underlying data; the second step requires that the reconstructions of simulated data are sufficiently similar to their corresponding ground truth images. Parameter combinations that meet both criteria define the Top Set, and the parameter combination in this set that gives the best performance on the simulated data is used to define the fiducial images of M87 for each day.

Each parameter survey spans a large range of parameter space, including combinations that are poorly adapted to the M87 data (e.g., very large regularizer weights in the RML methods). To filter parameter combinations that lead to final images with poor fits to the data, we calculate $\chi^2$ values on closure phase and log closure amplitudes using scan-averaged low-band data. For the $\chi^2$ computation, we only construct closure quantities from averaged visibilities that have synchronized start times. Any combinations with $\chi^2 \geqslant 2$ on any day of EHT observations are excluded from the Top Set. However, the DIFMAP imaging pipeline uses 10 s averages (while we estimate $\chi^2$ values using scan averages), and the pipeline downweights the extremely high S/N baselines to ALMA during the iterative self-calibration process. Consequently, DIFMAP reconstructions tend to fail this strict $\chi^2$ requirement unless a systematic uncertainty is included in the error budget. As a result, before performing this initial $\chi^2 < 2$ thresholding on DIFMAP reconstructions, we include 10% systematic uncertainty in quadrature to the thermal uncertainty of every visibility.

Next, to determine the fidelity of the different reconstructions on synthetic data, we compute the normalized cross-correlation $\rho_{\mathrm{NX}}$ between each surviving image and the corresponding ground truth image. For a given reconstruction $X$ and its ground truth image $Y$, the normalized cross-correlation is

$$\rho_{\mathrm{NX}}(X, Y) = \frac{1}{N}\sum_i \frac{(X_i - \langle X\rangle)(Y_i - \langle Y\rangle)}{\sigma_X\, \sigma_Y}, \qquad (15)$$

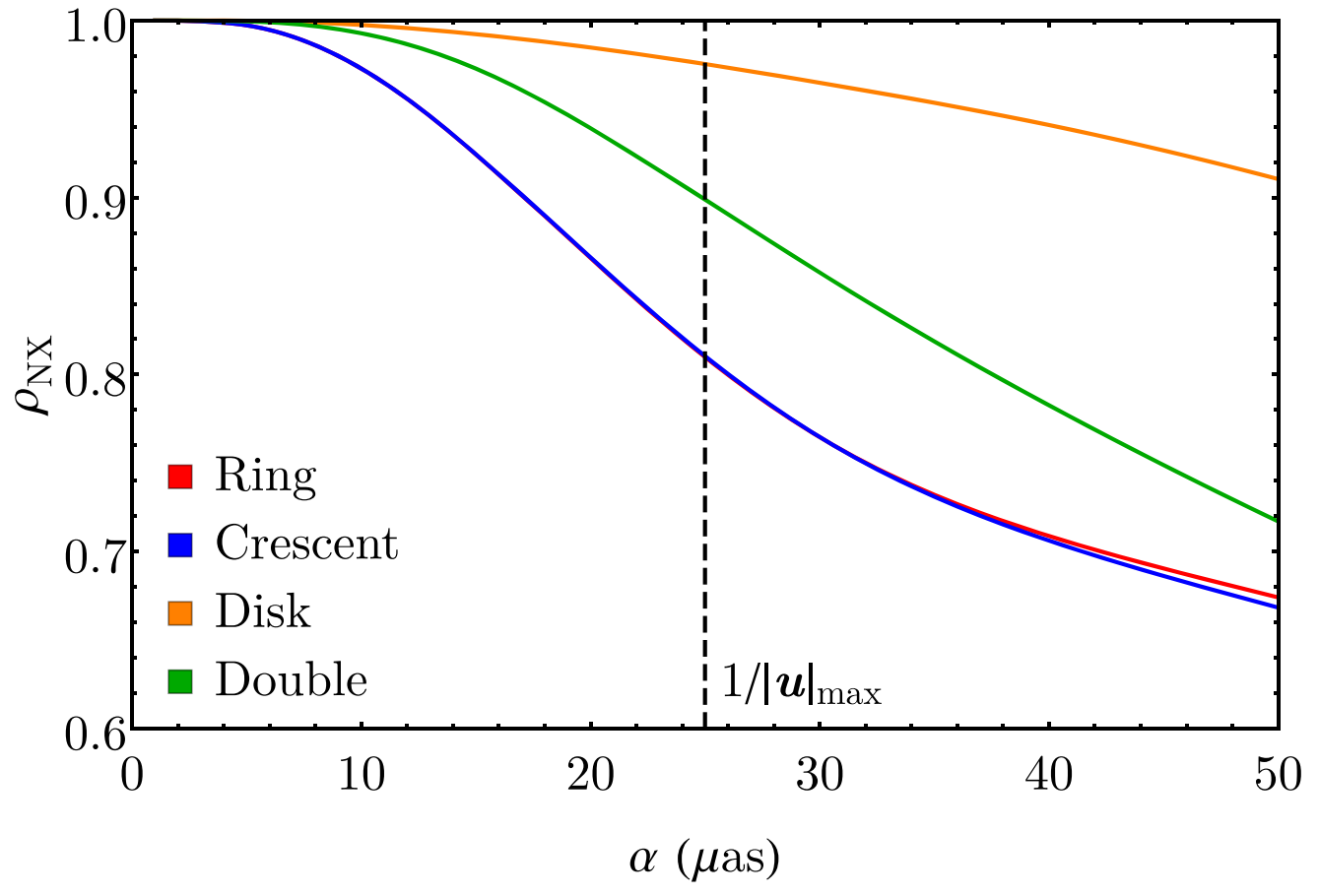

**Figure 9.** Normalized cross-correlation, $\rho_{\mathrm{NX}}$, of the four simulated images compared before and after convolution with a circular Gaussian kernel with FWHM $\alpha$. The vertical dashed line shows the nominal diffraction-limited resolution of the EHT (see Table 1). For each source, we invert these curves to obtain from a given $\rho_{\mathrm{NX}}$ an equivalent blurring kernel size $\alpha$. The Top Set reconstructions are defined as those with this $\alpha$ smaller than the nominal EHT resolution.

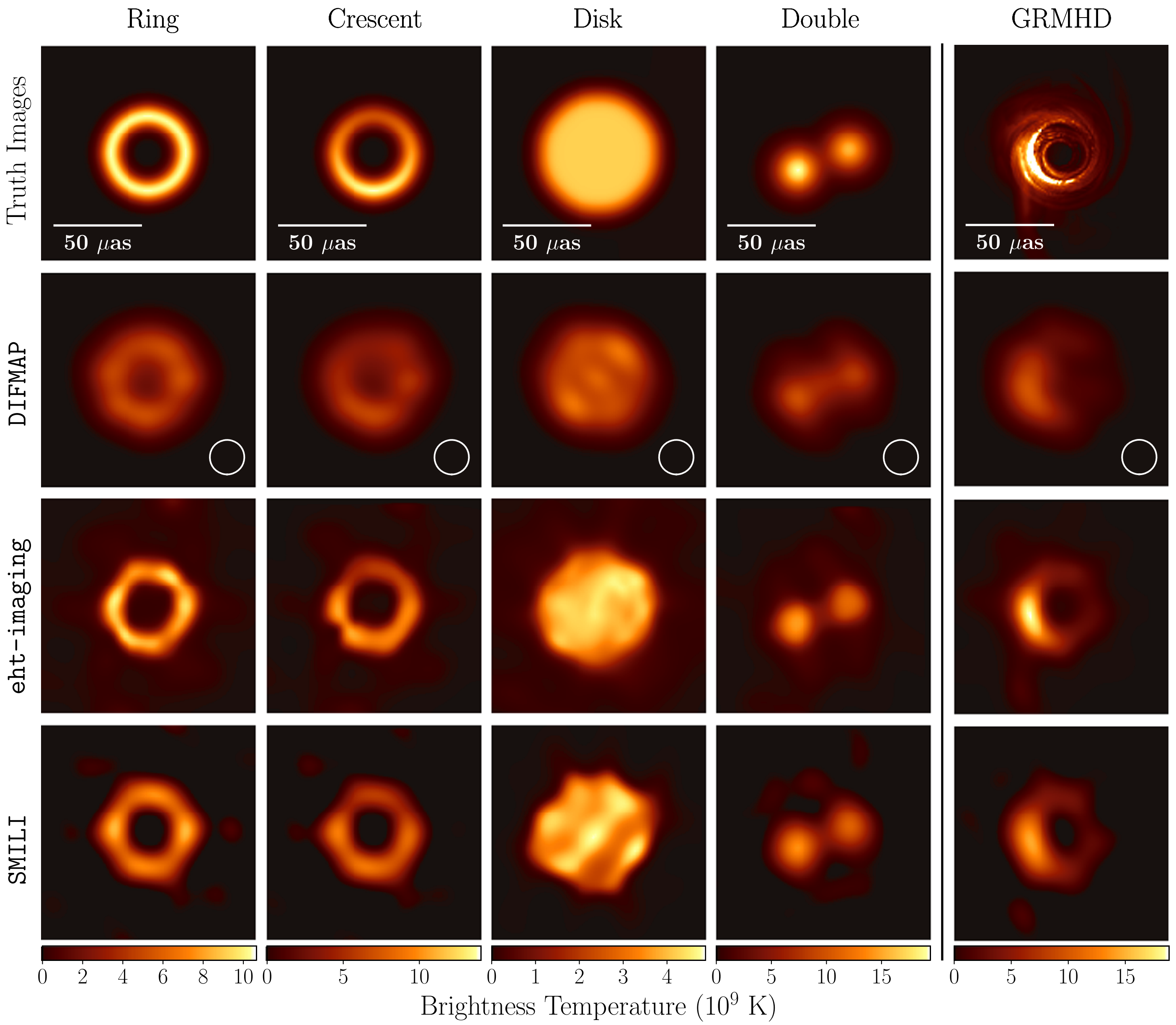

**Figure 10.** Cross-validation of the imaging parameter selection procedure. In each of the left four columns, we show reconstructed images for the simple geometric source models. These reconstructions do not use the fiducial imaging parameters identified by the full training set; instead, we selected the imaging parameters for each geometric source model after excluding that particular model from the parameter selection process. For example, in the disk reconstructions, the parameters were selected by assessing reconstructions of only the ring, crescent, and double source models. Thus, the selected parameters vary among these four columns, but we can verify that the training sets do not overly constrain the outcomes. In the fifth column, we show reconstructions of a GRMHD snapshot (Paper V) using the fiducial M87 parameters selected from all four geometric models. That is, the script and parameters used to produce these GRMHD image reconstructions are identical to those used to produce our fiducial M87 images (shown in Figure 11). Because the GRMHD snapshot has a substantially higher peak brightness than the reconstructions, its column has been scaled to the peak brightness of the `eht-imaging` reconstruction.

where the sum is over all $N$ pixels in the two images, $\langle X \rangle$ and $\langle Y \rangle$ are the mean pixel values in the images, and $\sigma_X$ and $\sigma_Y$ are the standard deviations of the pixel values in the two images. The normalized cross-correlation quantifies image similarity, where a value of 1 implies perfect correlation between images, 0 implies no correlation, and $-1$ implies perfect anti-correlation. Because the absolute visibility phase is unknown and, thus, the image centroid for reconstructions is not fixed, we compute $\rho_{\rm NX}$ for each possible discrete shift between the images and select the largest value. Before computing $\rho_{\rm NX}$ for `DIFMAP` images, the CLEAN components are convolved with a restoring beam with FWHM of 20 $\mu$as.

We choose the median value of $\rho_{\rm NX}$ across all days as the representative value for a given combination of imaging parameters on a particular model. Taking the median allows us to incorporate days with different baseline coverage in the scoring procedure while preventing the poorest reconstructions (e.g., those on April 10) from dominating the outcome.

While $\rho_{\rm NX}$ is a useful measure of relative agreement among images, it does not provide an absolute score of reconstruction quality. For instance, Figure 9 shows that convolving the four geometric source models with the same circular Gaussian of FWHM $\alpha$ produces $\rho_{\rm NX}$ values (computed against the original image) that are different depending on the underlying model. Thus, to better standardize the interpretation of $\rho_{\rm NX}$ for reconstructions of different underlying images, we invert the curves in Figure 9 to compute, for each value of $\rho_{\rm NX}$, an effective kernel size $\alpha$. That is, we quantify the fidelity of a reconstruction in terms of an effective blurring kernel FWHM $\alpha$ that would give the same $\rho_{\rm NX}$ as we compute from the reconstruction.

We compute a characteristic value $\alpha_{\rm char}$ for each combination of imaging parameters by taking the mean $\alpha$ across the four sources, and we then rank the parameters according to this characteristic value. The top-ranked combination of parameters is then denoted as the "fiducial parameters" of that method. We also select a Top Set of parameter combinations that obtain an $\alpha_{\rm char}$ smaller than the nominal EHT resolution. That is, the Top Set includes all imaging parameter combinations with $\alpha_{\rm char} < 1/|\boldsymbol{u}|_{\rm max}$. A selection of the Top Set reconstructions is highlighted in green for the `DIFMAP`, `eht-imaging`, and `SMILI` parameter searches in Figures 6–8, respectively.

### *6.3.2. Parameter Selection Considerations*

In Figure 10, we test whether or not our parameter selection procedure leads to reconstructed images that are unduly similar to the training set. In this test, we repeat the parameter selection process after withholding a specified geometric source model from the training set. For example, in Figure 10, the parameter combinations used to produce the images of the disk model were determined by selecting the best-performing parameters on only the ring, crescent, and double. Despite excluding the disk from the training set, all three pipelines produce an image with a disk morphology that does not resemble any of the other images in the training set.

A common feature of all four geometric models in the training set is that they restrict the compact emission to a region spanning $\lesssim$70 $\mu$as. One possible concern is that the resultant Top Set parameter combinations may then fail to reconstruct images of sources that have compact components extending over larger regions. However, such bright compact components would introduce variations over time in the visibilities that could not be fit with an overly restricted FOV. Thus, our requirement that Top Set reconstructions of M87 have $\chi^2 < 2$ ensures that they are not missing bright extended components.

Other common features in all our training set models are the lack of small-scale structure (the finest features in each image is $\geqslant$10 $\mu$as) and a fixed total compact flux density of 0.6. We test that our fiducial scripts produce accurate results for an image with small-scale structure using using a GRMHD snapshot image (see Figure 10), and we explore systematic uncertainties related to the total compact flux density in Appendix H.

### *6.3.3. Results of Parameter Selection*

In Table 3, we summarize the specifications and outcomes of the three parameter surveys. In this table, for each parameter value, we list the fraction of parameter combinations in the Top Set that include that value. We also provide the total number of parameter combinations surveyed, the number of these combinations in the Top Set, and the fiducial imaging parameters for each pipeline. Table 4 quantifies the fidelity of the fiducial and Top Set images reconstructed from simulated data by reporting the equivalent blurring kernel $\alpha$ obtained with each method.

The distributions in Table 3 show that the parameter survey eliminates regions of parameter space from consideration based on our selection criteria. For instance, in the `eht-imaging` results, reconstructions using large hyperparameters (>10) on the TV and TSV regularizers or those using a large initial Gaussian (FWHM > 50 $\mu$as) perform poorly regardless of the other parameter values used. These parameters are not represented in the Top Set.

In addition, differences in the procedures for each pipeline lead to different fiducial parameters. In some cases, these are

**Table 3**
Parameters and Their Surveyed Values for the `DIFMAP` (Top), `eht-imaging` (Middle), and `SMILI` (Bottom) Pipelines

| | | | | | | | |
|---|---|---|---|---|---|---|---|
| `DIFMAP` (1008 Param. Combinations; 30 in Top Set) | | | | | | | |
| Compact | **0.5** | **0.6** | **0.7** | **0.8** | | | |
| Flux (Jy) | 27% | 27% | 30% | 17% | | | |
| Stop | | Flux Reached | | | $\Delta$rms $\leqslant 10^{-4}$ | | |
| Condition | | 70% | | | 30% | | |
| ALMA | **0.01** | **0.1** | **0.3** | **0.5** | **0.7** | **1.0** | |
| Weight Factor | 17% | 60% | 20% | 3% | 0% | 0% | |
| Mask | **40** | **50** | **60** | **70** | **80** | **90** | **100** |
| Diam. ($\mu$as) | 0% | 0% | 47% | 27% | 23% | 3% | 0% |
| UV Weight | **0** | **−1** | **−2** | | | | |
| Exponent $\kappa$ | 10% | 60% | 30% | | | | |
| `eht-imaging` (37500 Param. Combinations; 1572 in Top Set) | | | | | | | |
| Compact | **0.4** | **0.5** | **0.6** | **0.7** | **0.8** | | |
| Flux (Jy) | 12% | 19% | 24% | 23% | 22% | | |
| Init./MEM | **40** | **50** | **60** | | | | |
| FWHM ($\mu$as) | 58% | 42% | 0% | | | | |
| Systematic | **0%** | **1%** | **2%** | **5%** | | | |
| Error | 26% | 27% | 26% | 20% | | | |
| Regularizer: | **0** | **1** | **10** | **$10^2$** | **$10^3$** | | |
| MEM | 0% | 0% | 8% | 92% | 0% | | |
| TV | 31% | 35% | 33% | 0% | 0% | | |
| TSV | 31% | 34% | 32% | 3% | 0% | | |
| $\ell_1$ | 23% | 24% | 24% | 22% | 7% | | |
| `SMILI` (10800 Param. Combinations; 529 in Top Set) | | | | | | | |
| Compact | **0.4** | **0.5** | **0.6** | **0.7** | **0.8** | | |
| Flux (Jy) | 22% | 31% | 25% | 14% | 8% | | |
| $\ell_1^{\rm w}$ Soft Mask | **40** | **50** | **60** | **70** | **80** | | |
| FWHM. ($\mu$as) | 29% | 19% | 21% | 21% | 15% | | |
| Systematic | **0%** | **1%** | | | | | |
| Error | 50% | 50% | | | | | |
| Regularizer: | **0** | **10** | **$10^2$** | **$10^3$** | **$10^4$** | **$10^5$** | |
| TV | 9% | 9% | 11% | 38% | 32% | 0% | |
| TSV | 13% | 14% | 13% | 24% | 36% | 0% | |
| | **0** | **$10^{-2}$** | **$10^{-1}$** | **1** | **10** | **$10^2$** | |
| $\ell_1^{\rm w}$ | 0% | 0% | 9% | 47% | 44% | 0% | |

**Note.** For instance, the `eht-imaging` survey produced $5 \times 3 \times 4 \times 5^4 =$ 37,500 parameter combinations, and 1572 of these passed the criteria for inclusion in the Top Set. Below each parameter value we specify the fraction of the Top Set parameter combinations that include that value (see Section 6.3.1). Boxed parameters are those corresponding to the fiducial images. Note that the fiducial parameters are determined by identifying the parameter combination that jointly performs best on all the synthetic data sets; the fiducial parameters do not necessarily correspond with the parameters that have the largest share in the Top Set.

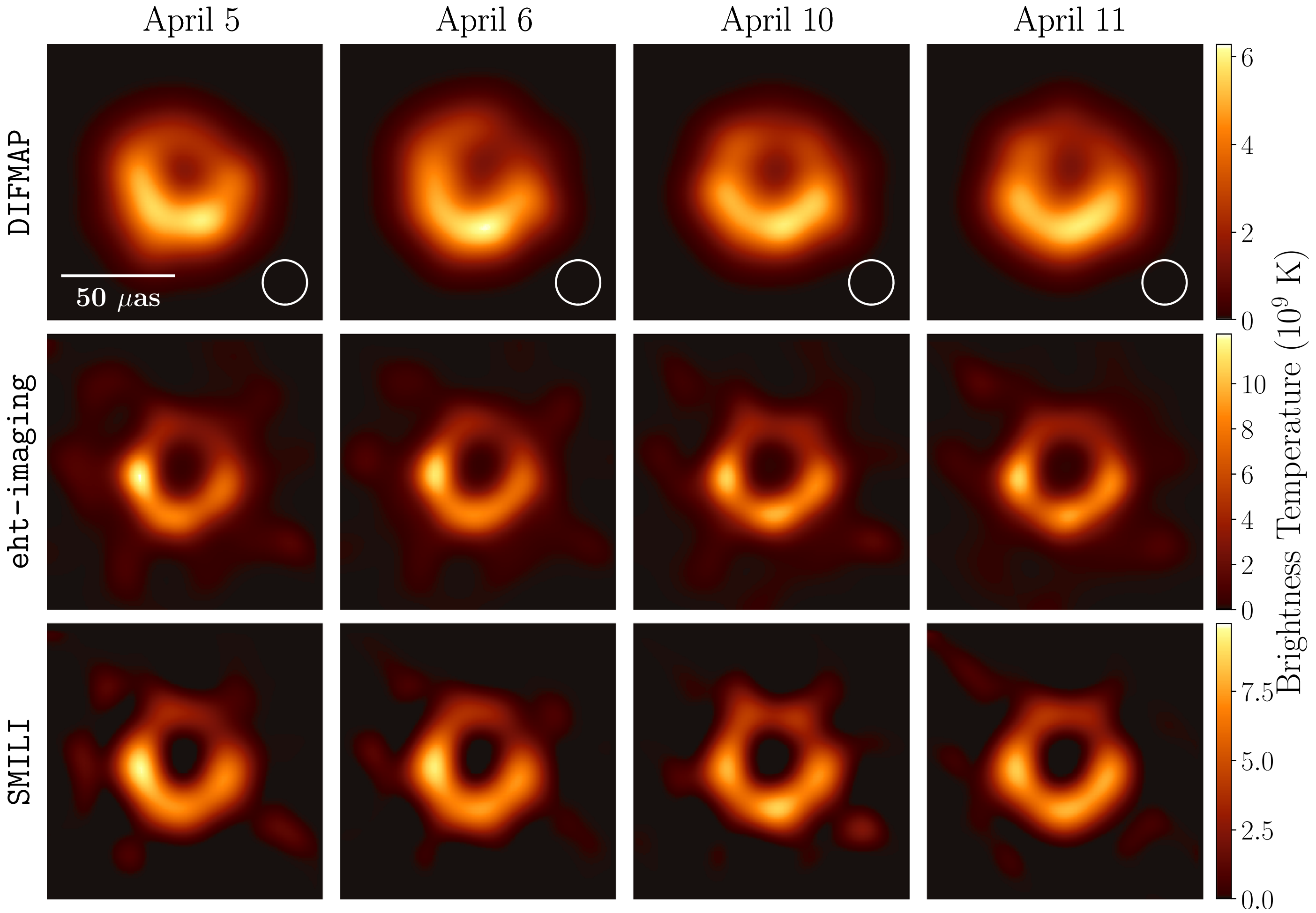

**Figure 11.** Fiducial images of M87 on all four observed days from each of the three imaging pipelines. CLEAN images (from `DIFMAP`) are shown after convolution with a 20 $\mu$as beam; `eht-imaging` and `SMILI` results have no restoring beam applied. Different selected fiducial imaging parameters (e.g., compact flux) result in different peak brightness temperatures for each method, as indicated by the unique color bars for each row.

**Table 4**
Equivalent Blurring Values $\alpha$ from the Fiducial and Top Set Images from Each Pipeline on Each of the Four Simple Geometric Models in the Training Set

| | $\alpha$ ($\mu$as), Ring | | | $\alpha$ ($\mu$as), Crescent | | | $\alpha$ ($\mu$as), Disk | | | $\alpha$ ($\mu$as), Double | | |
|---|---|---|---|---|---|---|---|---|---|---|---|---|
| | Fiducial | Mean | Std. Dev. | Fiducial | Mean | Std. Dev. | Fiducial | Mean | Std. Dev. | Fiducial | Mean | Std. Dev. |
| `DIFMAP` | 19.4 | 21.2 | 1.5 | 19.0 | 21.6 | 1.5 | 18.5 | 21.2 | 2.1 | 19.8 | 21.6 | 1.5 |
| `eht-imaging` | 9.4 | 13.4 | 3.3 | 9.3 | 12.2 | 1.8 | 18.0 | 20.5 | 2.0 | 11.3 | 15.8 | 3.4 |
| `SMILI` | 10.6 | 12.2 | 2.8 | 10.6 | 12.2 | 2.3 | 12.3 | 18.7 | 3.1 | 11.3 | 14.8 | 3.4 |

**Note.** For each pipeline, we give $\alpha$ for the fiducial image, as well as the mean and standard deviation of $\alpha$ across the Top Set.

expected from differences in the underlying methods. For instance, the prior and mask sizes are treated differently in different scripts: a hard mask was used for `DIFMAP`, a softer mask for `SMILI`, and only weak constraints on compactness were used for `eht-imaging`. However, other choices are unexpected. For example, the top-ranked parameter combinations (i.e., the fiducial parameters) for the `DIFMAP` and `SMILI` pipelines use a total compact flux density of 0.5 Jy, favoring this value over the true total compact flux density of 0.6 Jy. However, these preferences are mild, and the Top Set from each pipeline includes images with all values of total compact flux density surveyed, from 0.4 to 0.8 Jy (see Table 3). We explore the dependence of the assumed compact flux density in Appendix H.

## 7. Fiducial M87 Images: Properties and Uncertainties

Having determined fiducial imaging parameters via tests on synthetic data (Section 6.3.1), we now show the corresponding fiducial M87 images from each imaging method for each day of the 2017 EHT observations. We also assess the consistency of the fiducial images with the data, discuss other visibility-domain properties of these images, and estimate image uncertainties resulting from imaging parameter choices.

### 7.1. Fiducial M87 Images

Figure 11 shows the fiducial images from each day of EHT observations and each imaging method. These fiducial images are broadly consistent; all twelve images have a prominent,

**Table 5**
Closure Quantity $\chi^2$ Values for the Fiducial M87 Images and $\chi^2$ Statistics (Mean and Standard Deviation) for Top Set Images

| | | DIFMAP | | | eht-imaging | | | SMILI | | |
|---|---|---|---|---|---|---|---|---|---|---|
| | | 0% | 1% | 10% | 0% | 1% | 10% | 0% | 1% | 10% |
| April 5 | | | | | | | | | | |
| Fiducial: | $\chi^2_{\rm CP}$ | 6.03 | 4.55 | 0.85 | 0.98 | 0.94 | 0.32 | 0.98 | 0.94 | 0.32 |
| | $\chi^2_{\log {\rm CA}}$ | 4.07 | 3.87 | 1.04 | 1.02 | 0.87 | 0.30 | 1.10 | 1.01 | 0.39 |
| Top Set: | $\chi^2_{\rm CP}$ | 9.40 ± 3.35 | 6.98 ± 2.33 | 1.20 ± 0.33 | 1.00 ± 0.13 | 0.96 ± 0.11 | 0.32 ± 0.02 | 1.09 ± 0.21 | 1.04 ± 0.20 | 0.34 ± 0.05 |
| | $\chi^2_{\log {\rm CA}}$ | 4.99 ± 1.17 | 4.74 ± 1.11 | 1.28 ± 0.29 | 0.97 ± 0.27 | 0.84 ± 0.17 | 0.29 ± 0.02 | 1.11 ± 0.28 | 1.02 ± 0.25 | 0.39 ± 0.07 |
| April 6 | | | | | | | | | | |
| Fiducial: | $\chi^2_{\rm CP}$ | 2.30 | 2.07 | 0.66 | 1.51 | 1.46 | 0.54 | 1.44 | 1.39 | 0.52 |
| | $\chi^2_{\log {\rm CA}}$ | 1.90 | 1.85 | 0.70 | 1.04 | 0.96 | 0.39 | 1.24 | 1.16 | 0.49 |
| Top Set: | $\chi^2_{\rm CP}$ | 4.40 ± 1.45 | 3.96 ± 1.34 | 1.18 ± 0.45 | 1.59 ± 0.16 | 1.53 ± 0.15 | 0.55 ± 0.04 | 1.49 ± 0.23 | 1.44 ± 0.22 | 0.53 ± 0.07 |
| | $\chi^2_{\log {\rm CA}}$ | 3.49 ± 1.51 | 3.38 ± 1.45 | 1.13 ± 0.36 | 1.00 ± 0.14 | 0.93 ± 0.11 | 0.38 ± 0.02 | 1.22 ± 0.25 | 1.14 ± 0.23 | 0.48 ± 0.07 |
| April 10 | | | | | | | | | | |
| Fiducial: | $\chi^2_{\rm CP}$ | 2.22 | 1.95 | 0.56 | 0.77 | 0.75 | 0.37 | 0.68 | 0.66 | 0.35 |
| | $\chi^2_{\log {\rm CA}}$ | 1.07 | 1.05 | 0.43 | 1.24 | 1.16 | 0.56 | 1.14 | 1.07 | 0.58 |
| Top Set: | $\chi^2_{\rm CP}$ | 3.93 ± 2.10 | 3.40 ± 1.50 | 0.96 ± 0.26 | 0.83 ± 0.10 | 0.80 ± 0.09 | 0.39 ± 0.03 | 0.75 ± 0.14 | 0.73 ± 0.14 | 0.37 ± 0.08 |
| | $\chi^2_{\log {\rm CA}}$ | 1.57 ± 0.64 | 1.52 ± 0.62 | 0.60 ± 0.21 | 1.30 ± 0.17 | 1.21 ± 0.14 | 0.60 ± 0.07 | 1.11 ± 0.32 | 1.05 ± 0.30 | 0.57 ± 0.17 |
| April 11 | | | | | | | | | | |
| Fiducial: | $\chi^2_{\rm CP}$ | 1.90 | 1.53 | 0.49 | 0.96 | 0.90 | 0.27 | 1.07 | 1.01 | 0.38 |
| | $\chi^2_{\log {\rm CA}}$ | 2.48 | 2.35 | 0.67 | 0.97 | 0.84 | 0.36 | 1.08 | 0.97 | 0.46 |
| Top Set: | $\chi^2_{\rm CP}$ | 4.01 ± 1.67 | 2.87 ± 1.05 | 0.61 ± 0.11 | 0.97 ± 0.10 | 0.91 ± 0.09 | 0.27 ± 0.02 | 1.23 ± 0.28 | 1.16 ± 0.27 | 0.40 ± 0.06 |
| | $\chi^2_{\log {\rm CA}}$ | 3.33 ± 1.01 | 3.16 ± 0.95 | 0.92 ± 0.24 | 0.99 ± 0.13 | 0.86 ± 0.09 | 0.36 ± 0.02 | 1.14 ± 0.13 | 1.02 ± 0.12 | 0.49 ± 0.07 |

**Note.** For each, values are shown after including systematic uncertainties of 0%, 1%, and 10%. Inclusion in the top set requires $\chi^2 < 2$ for both closure phases and amplitudes with 0% systematic uncertainty for eht-imaging and SMILI, and with 10% systematic uncertainty for DIFMAP. See Section 2.1 for definitions and limitations of these metrics.

asymmetric ring feature, approximately 40 $\mu$as in diameter with enhanced brightness toward the south. Each image has a prominent depression in the center of the ring and a peak brightness temperature along the ring of $\sim 10^{10}$ K. Because they are restored with a 20 $\mu$as beam, the DIFMAP images have broader structure, leading to a weaker central depression and lower peak brightness temperatures. For all pipelines, the finite resolution of the reconstructions results in brightness temperatures that are only lower limits on the maximum intrinsic brightness temperature.

As expected from tests on synthetic data (Section 6), the details of the reconstructions differ between the imaging methods. Although the RML images (eht-imaging and SMILI) are generally more similar to each other than to the DIFMAP images, SMILI reconstructions are fainter inside and outside the ring than the eht-imaging reconstructions from the strong sparsity regularization. Based on the results of the parameter survey, M87 data was reconstructed with a total compact flux density of 0.5 Jy for DIFMAP and SMILI, and with 0.6 Jy for eht-imaging. While 0.5 Jy is slightly below the range of acceptable values for $F_{\rm cpct}$ identified in Section 4, the constraints presented in Section 4 made strict assumptions about the gains that are not rigidly enforced in imaging, and our parameter surveys using synthetic data demonstrate that some imaging methods produce higher fidelity (i.e., higher $\rho_{\rm NX}$) images when reconstructing an image with less compact flux density than the true value. We explore the effects of the assumed compact flux density further in Appendix H.

In Table 5, we provide $\chi^2$ values for the 12 fiducial images as well as $\chi^2$ statistics across the Top Sets. We compute these values after including varying amounts of systematic uncertainty (0%, 1%, and 10% of visibility amplitudes) added in quadrature to the thermal uncertainty. This additional uncertainty accounts for possible systematic errors (e.g., from polarimetric leakage) and also ensures that the statistics quantify overall image-data consistency without being overwhelmed by the extremely sensitive ALMA baselines. For the largest value, 10%, the systematic uncertainty allows modest tolerance for source structure that is absent from the reconstructed images (e.g., from extended structure in the M87 jet outside the masked region or the FOV of the reconstructed image).

Figures 12 and 13 compare self-calibrated visibility amplitudes and closure phases on three selected triangles with the corresponding model visibility amplitudes and closure phases computed from each April 11 fiducial image. As indicated in the $\chi^2$ values listed in Table 5, the two RML methods produce fiducial images that are consistent with the data to within roughly their thermal uncertainties, and the $\chi^2$ values have little scatter across the Top Sets. The DIFMAP fiducial image also shows agreement with the data, but requires modest systematic tolerance for comparable image-data consistency (because of the different time averaging used in DIFMAP imaging the ALMA downweighting), and the $\chi^2$ values show more scatter across the Top Set. All three fiducial images fit the nulls in visibility amplitudes at 3.4 and 8.3 G$\lambda$.

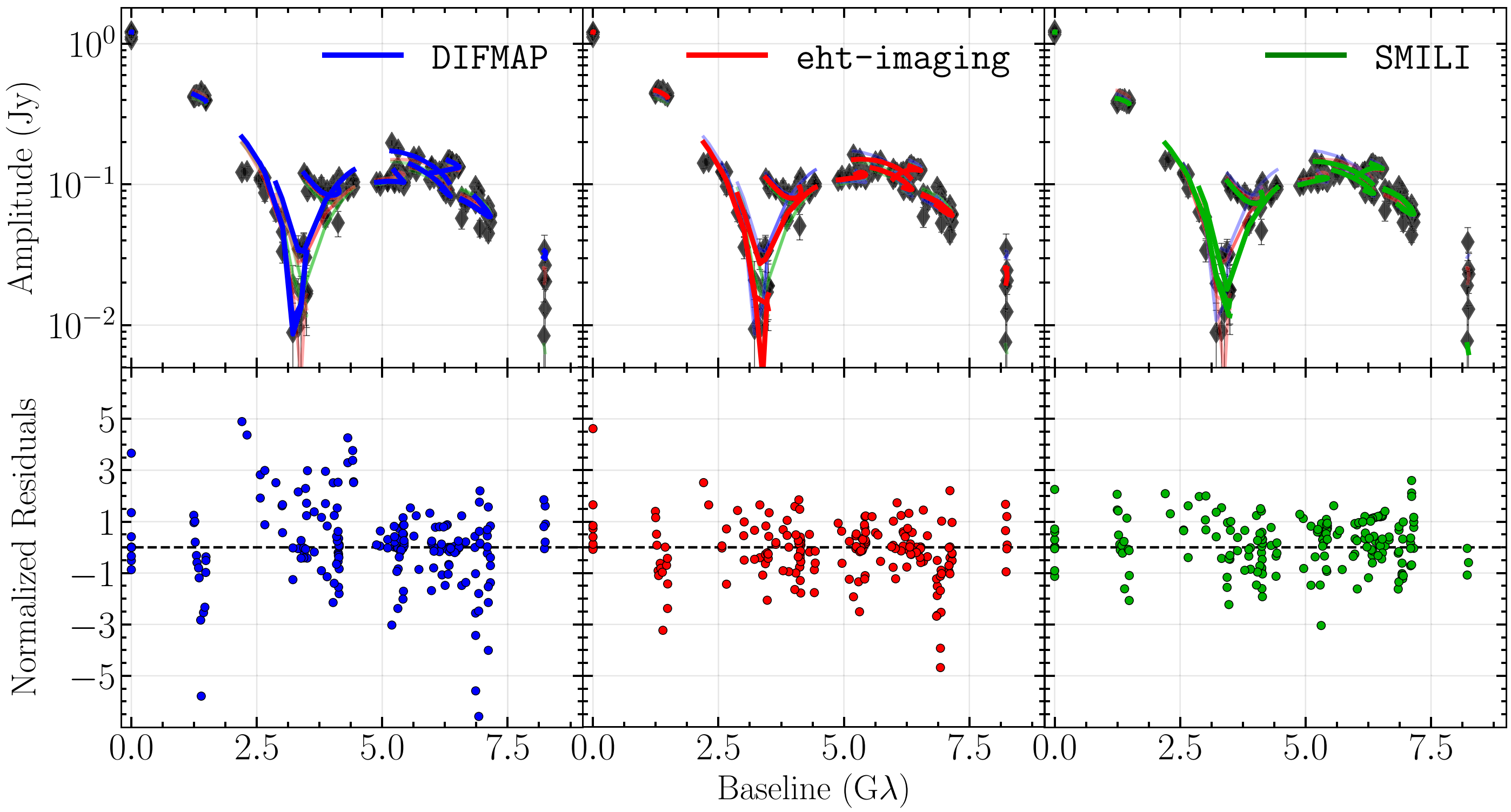

**Figure 12.** Top: visibility amplitudes of self-calibrated data on April 11 as a function of baseline length compared with corresponding values for the three fiducial images. In each panel, visibilities are plotted after self-calibrating the data to the indicated fiducial image; thus, the visibility amplitudes differ across the three panels. For each image, we add an additional large-scale Gaussian component before self-calibration to account for the excess flux density seen on intra-site baselines (see Section 4.3). Bottom: visibility amplitude residuals for each fiducial image, after normalizing by the thermal noise with 1% systematic uncertainty added in quadrature.

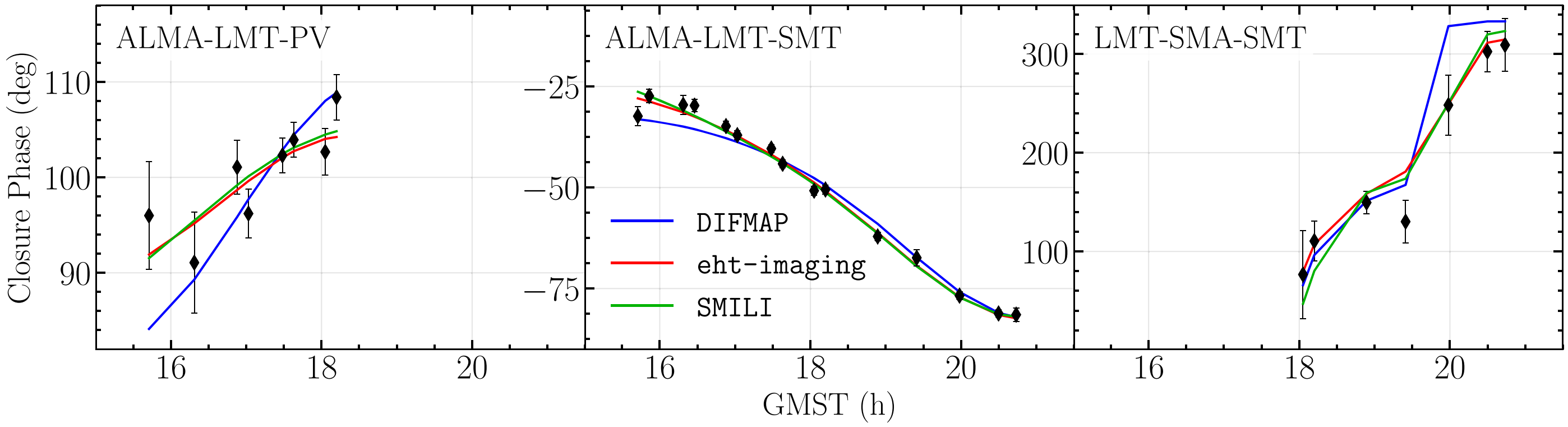

**Figure 13.** Closure phases plotted as function of GMST on three selected triangles from the April 11 observations. The solid lines indicate the corresponding closure phase curves from the fiducial images produced by the three pipelines.

Our images of M87 do not show extended emission outside the observed ring, although a jet is prominent in VLBI images of M87 at longer wavelengths. Likewise, while the geometric models in the training set (Section 6.1) include emission from a simple extended jet model, none of the Top Set reconstructions from simulated EHT data recover this extended component nor are their recovered compact structure affected by the missing jet components. Furthermore, our fiducial images in all three pipelines do not recover the more realistic faint jet emission in the GRMHD simulation reconstructions in Figure 10. All of these results reinforce the expectation (see Section 4.3) that the current EHT array lacks the short-baseline coverage and dynamic range necessary to recover a faint, extended jet.

In Figure 14, we show the fiducial images from the three pipelines on April 11 after blurring each of them separately to achieve a common, conservative resolution. Specifically, we restore the `DIFMAP` image with the same 20 $\mu$as beam used throughout this Letter. For `eht-imaging` and `SMILI`, we use the circular Gaussian convolution kernels that maximize the normalized cross-correlation between the restored RML image and the restored `DIFMAP` image (17.1 $\mu$as for `eht-imaging`; 18.6 $\mu$as for `SMILI`). The three blurred images have a consistent ring diameter ($\approx$40 $\mu$as) and the same overall asymmetry (oriented south). Figure 15 shows the simple average of the three equivalently blurred fiducial images for each day. We adopt the images in

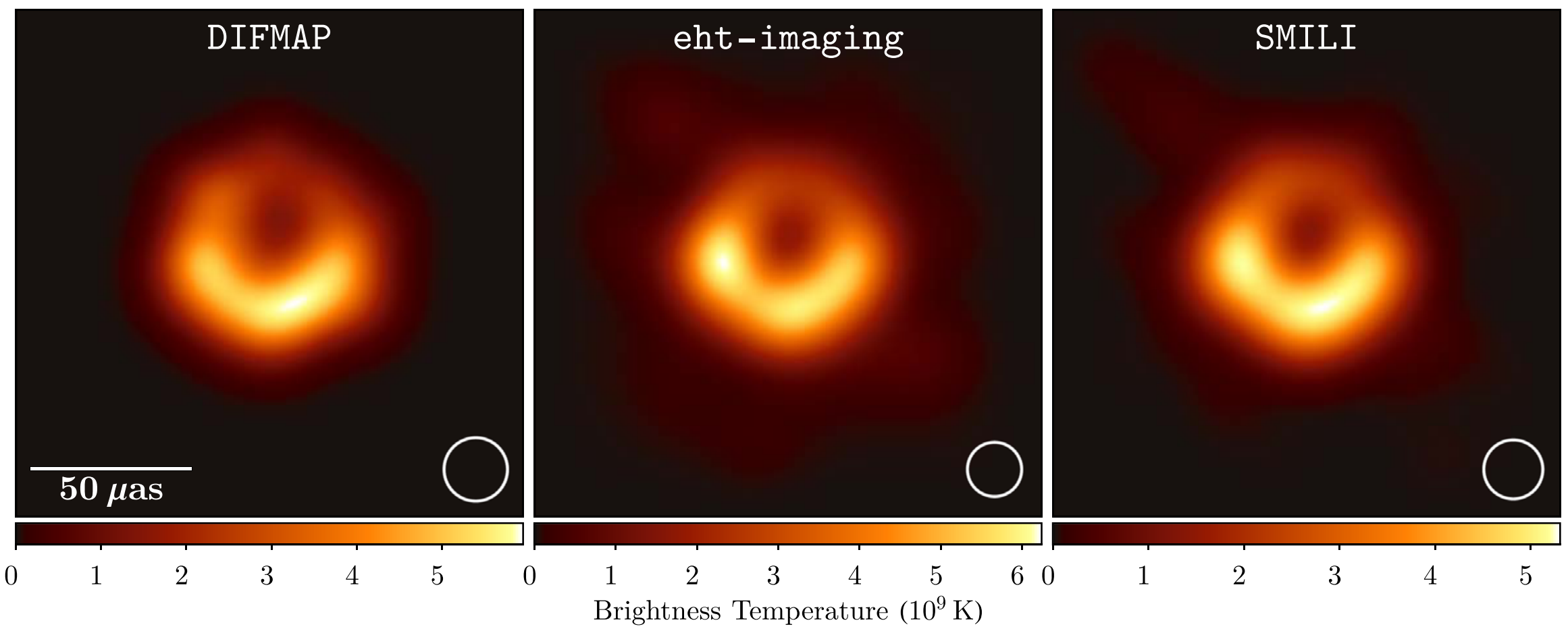

**Figure 14.** Fiducial images of M87 on April 11 from our three separate imaging pipelines after restoring each to an equivalent resolution. The `eht-imaging` and `SMILI` images have been restored with 17.1 and 18.6 $\mu$as FWHM Gaussian beams, respectively, to match the resolution of the `DIFMAP` reconstruction restored with a 20 $\mu$as beam.

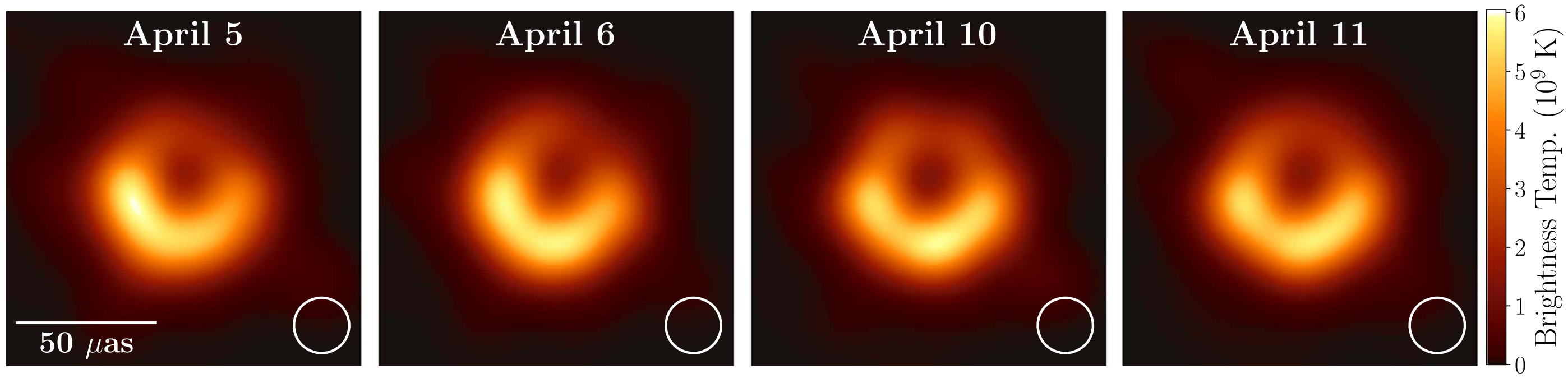

**Figure 15.** Averages of the three fiducial images of M87 for each of the four observed days after restoring each to an equivalent resolution, as in Figure 14. The indicated beam is 20 $\mu$as (i.e., that of `DIFMAP`, which is always the largest of the three individual beams).

Figure 15 as a conservative representation of our final M87 imaging results.

The fiducial images from each pipeline (Figure 11), as well as the conservative, blurred, pipeline-averaged images (Figure 15) provide some evidence for evolution in the ring structure between April 5, 6 and 10, 11. We discuss this evolution in more detail in Appendix E (Figure 33). Some change in the image structure between April 5, 6 and 10, 11 is necessitated by the variations seen in the underlying closure phases (Paper III). We find more variation in the image pairs that are separated by larger intervals, suggesting that these variations are intrinsic. However, we cannot unambiguously associate the observed variability with coherent evolution of a specific image feature.

Figure 16 shows the visibility amplitude and phase for each of the three April 11 fiducial images as a function of vector baseline. Note that no restoring beam is required for CLEAN in this visibility-domain analysis. Each image produces nulls in the visibility amplitude near the SMA–SMT baseline, consistent with the observed amplitudes (see Figure 2). The visibility phase shows rapid swings at these nulls. The visibilities of the images from the different pipelines are most similar near the EHT measurements, as expected. On longer baselines than those sampled by the EHT, the `DIFMAP` image produces notably higher visibility amplitudes than those of the `eht-imaging` and `SMILI` images, as expected from the fact that the `DIFMAP` image is fundamentally a collection of point sources.

### 7.2. Image Uncertainties

Measuring the variation in images produced in a parameter survey Top Set allows us to evaluate image uncertainties due to the explored imaging choices. Figure 17 shows uncertainties related to imaging assumptions from the largest Top Set (that of the `eht-imaging` parameter survey) on April 11 data.

Reconstructed image uncertainties are concentrated in the regions with enhanced brightness, notably in the three "knots" in the lower half of the ring (Figure 17; top panel). These are also the regions that show the most variation among different imaging methods (Appendix I compares their azimuthal profiles). Visibility-domain modeling provides another method to assess image structure. In Paper VI, we explore fitting simple crescent models to the data. For instance, a crescent with a brightness gradient and blurring reproduces the north–south asymmetry in images without additional azimuthal structure (the "blurred and slashed with LSG" crescent of Paper VI). However, this model gives $\chi^2_{\rm CP}$ between 3.2 and 11.5 and $\chi^2_{\log {\rm CA}}$ between 2.2 and 6.6 for different days and bands when assuming 0% systematic error (compare with Table 5). Adding additional degrees of freedom in the form of three elliptical Gaussian components to the crescent yields better fits, similar to those found with imaging. These Gaussian components sometimes reside near or on the crescent, affecting the azimuthal structure (see Appendix B.2.2 in Paper VI). Thus, our current image reconstructions and visibility-domain analyses are not able to confirm or reject the reality of the "knots" seen in images, and these features should be interpreted with caution.

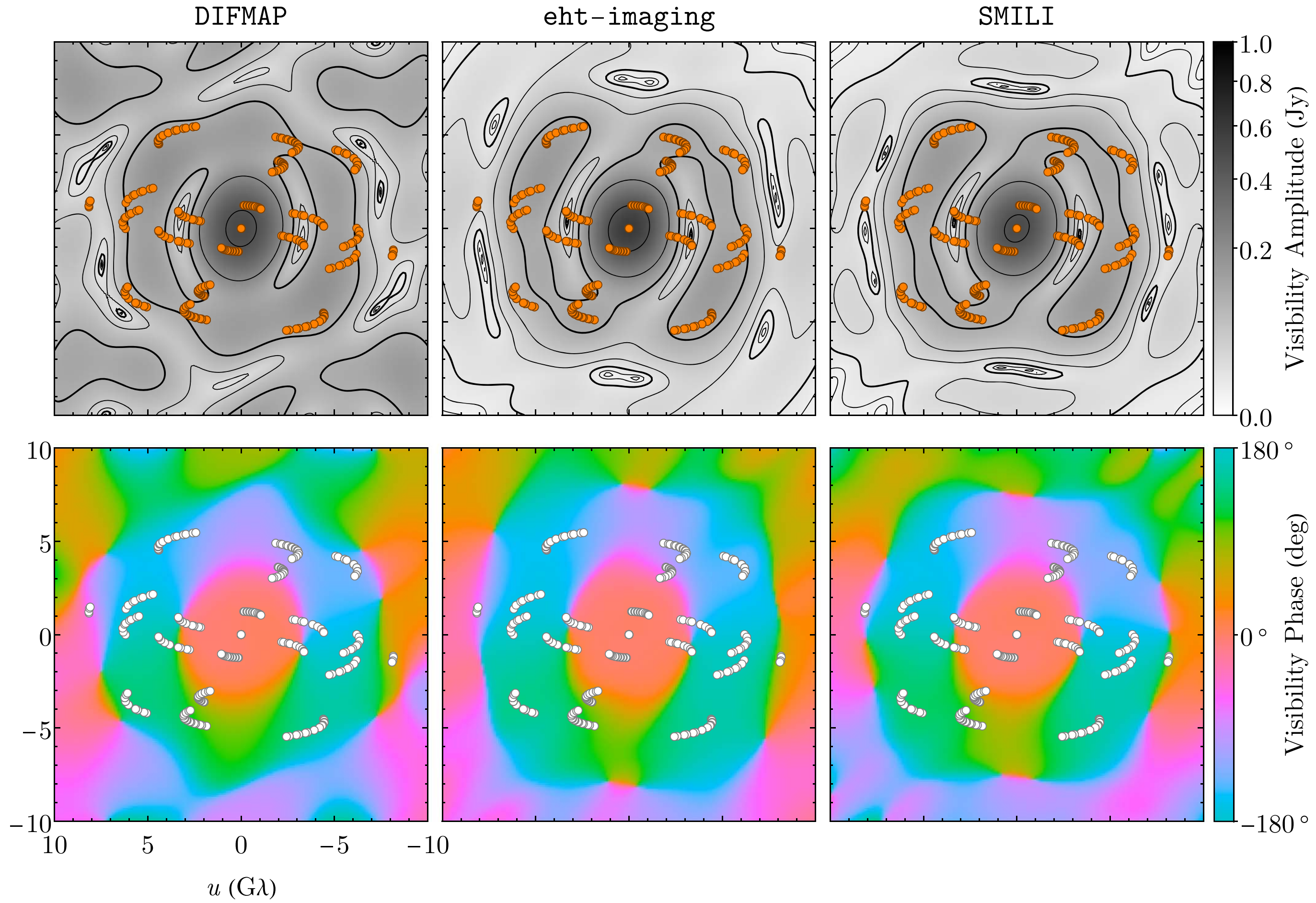

**Figure 16.** Visibility-domain representation of the three fiducial pipeline images for April 11. The top panels show visibility amplitude as a function of vector baseline; the bottom panels show visibility phase as a function of vector baseline. Amplitude contours are spaced logarithmically, with three contours per decade. Thicker contours correspond to 10 and 100 mJy; the lowest contour is at $10^{1/3} \approx 2.2$ mJy, although the images produce nulls in the visibility domain. Plotted points show the April 11 EHT coverage.

In the visibility domain, the uncertainties are concentrated at the origin, reflecting the spread of total compact flux density choices present in the Top Set (Figure 17; bottom panel). After normalizing the total flux densities of the Top Set images, the visibility domain uncertainties at the origin are zero; the uncertainties peak on baselines at $\sim$2 G$\lambda$, then fall gradually with increasing baseline length. The fractional uncertainties are the most straightforward to interpret, as they are low ($\lesssim$10%) in regions near EHT coverage and grow to values near or exceeding unity for baselines that are separated by more than a few G$\lambda$ from an EHT measurement.

The strong correlation between the visibility-domain fractional uncertainties and gaps in EHT coverage suggests that imaging uncertainties are dominated by imaging parameter choices rather than by thermal or systematic noise. That is, the fiducial image fidelity is primarily limited by the sparse EHT baseline coverage, rather than by sensitivity or calibration accuracy. For example, the standard deviation of visibility amplitudes across the Top Set for M87 is $\sim$50 mJy on baselines at $\sim$2 G$\lambda$, falling to $\sim$10 mJy at $\sim$8 G$\lambda$. In contrast, the median thermal noise of non-ALMA baselines is 7 mJy and of ALMA baselines is 0.7 mJy.

In the image domain, the median rms variation is $1.6 \times 10^8$ K for the Top Set images of M87, and $1.3 \times 10^8$ K for the normalized Top Set images. Because the peak image brightness is $\sim 10^{10}$ K, these variations indicate that our images achieve a dynamic range of $\sim$50. Across the image, the minimum fractional uncertainty from imaging parameter choices is 8.7% across the Top Set images without flux density normalization, and 7.9% with normalization. For comparison, the rms of the DIFMAP residual images is significantly smaller: $\sim 3 \times 10^7$ K across all days for the fiducial parameters. Thus, adding in the residual images would have negligible effect on our final reconstructions, especially relative to image differences between images from different imaging methods and from different parameter combinations using a single method.

## 8. Image Validation

Having determined fiducial images of M87 from each imaging method on each observing day, we now perform additional validation tests to assess their reliability. First, in Section 8.1, we compare the residual telescope gains estimated for M87 with those for the calibrator source 3C 279 to determine whether or not the significant variation seen in the inferred LMT residual gains are consistent between the two sources. Next, in Section 8.2, we perform data resampling tests, using identical scripted reconstructions on data sets that exclude individual sites, to assess how sensitive the image morphology is to the loss of different combinations of data. Finally, in Section 8.3, we test whether or not the time variability seen in the fiducial images is related to differences in the sampled ($u$, $v$) coverage across the days.

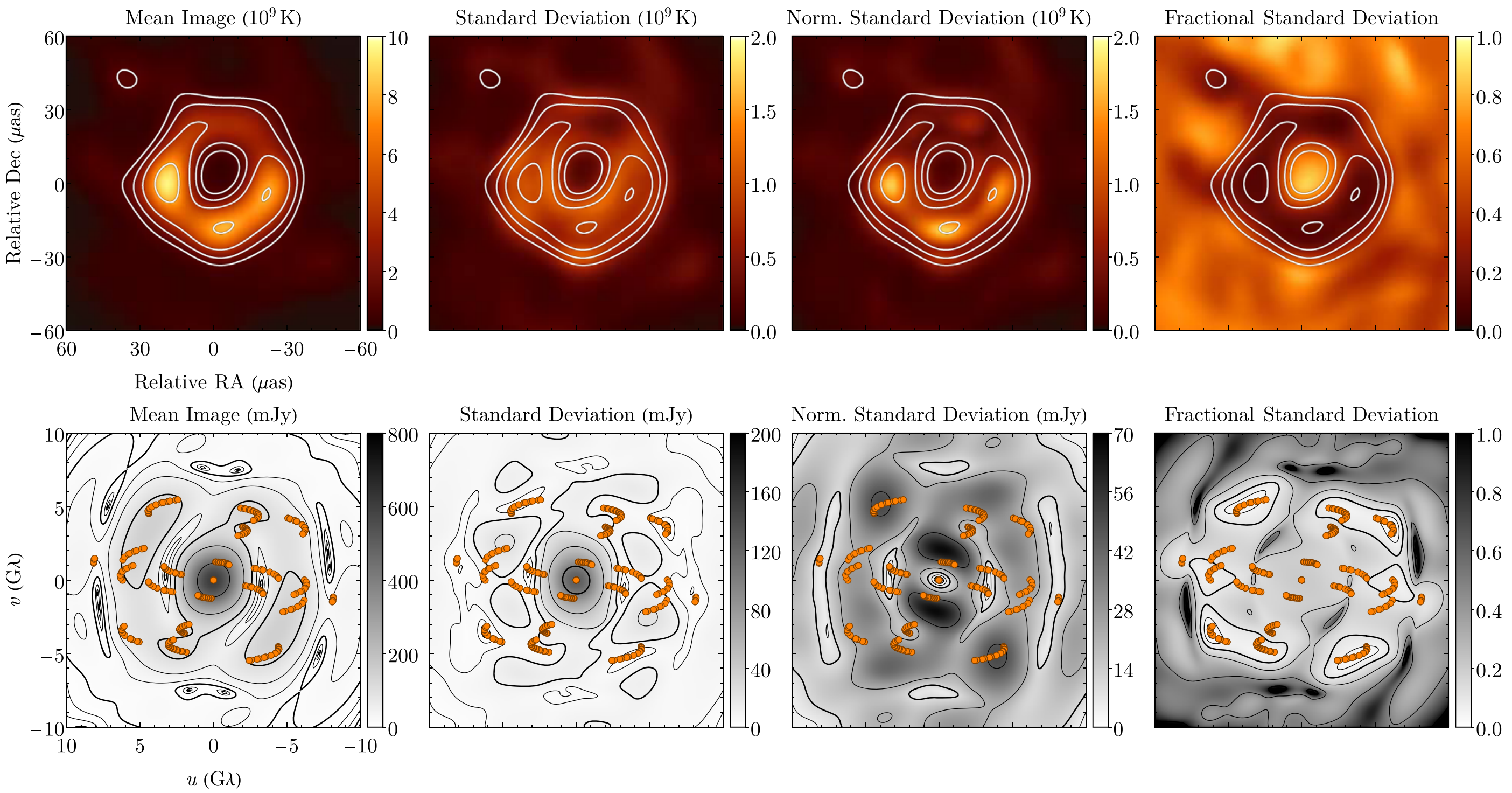

**Figure 17.** Uncertainties related to imaging assumptions from the `eht-imaging` parameter search on April 11 data. The top row shows the mean image and associated uncertainties in the image domain, while the bottom row shows visibility amplitudes of the mean image and their uncertainties in the visibility domain. From left to right, panels correspond to the mean Top Set image, the standard deviation of the Top Set reconstructions, the standard deviation after normalizing each reconstructed image to the total flux density of the mean image (0.62 Jy), and the standard deviation divided by the mean. In the image domain panels, contours are those of the mean image, at 10%, 20%, 40%, and 80% of the peak. In the visibility domain panels, contours are equally spaced logarithmically with three contours per decade; the thicker contours are 10 and 100 mJy (left three) and 10% (right).

### *8.1. Gain Validation with the Calibrator 3C 279*

The 2017 EHT observations of M87 were interleaved with those of the active galactic nucleus 3C 279. We have reconstructed images of 3C 279 using the three software packages used for M87 imaging in order to assess the consistency of these 3C 279 images with the M87 reconstructions in terms of the inferred time-variable residual gains. Because the sources are nearby on the sky (separation = 19°), the inferred time-variable residual gains at each site for the same day should be similar. We do *not* use the observations of 3C 279 to derive gain corrections for M87; instead, we compare the derived gains on both sources after independent imaging as a post hoc consistency test.

Figure 18 shows the aggregate baseline coverage for the interleaved EHT observations of 3C 279 in 2017 April. While the SPT could not observe M87, it participated in the observations of 3C 279, viewing it at an elevation of ~6° (i.e., a relative air mass of ~10). The addition of the SPT significantly improves the north–south resolution of the array. Figure 19 shows the April 11 visibility amplitudes from 3C 279 after a priori and network calibration. From ALMA interferometric measurements, the total flux density for 3C 279 (8–10 Jy) is nearly an order of magnitude higher than that of M87 (Goddi et al. 2019). As expected for a source with bright, linear jet features, the 3C 279 amplitudes vary strongly with baseline position angle and have a more complex structure on long baselines than M87 (Figure 2).

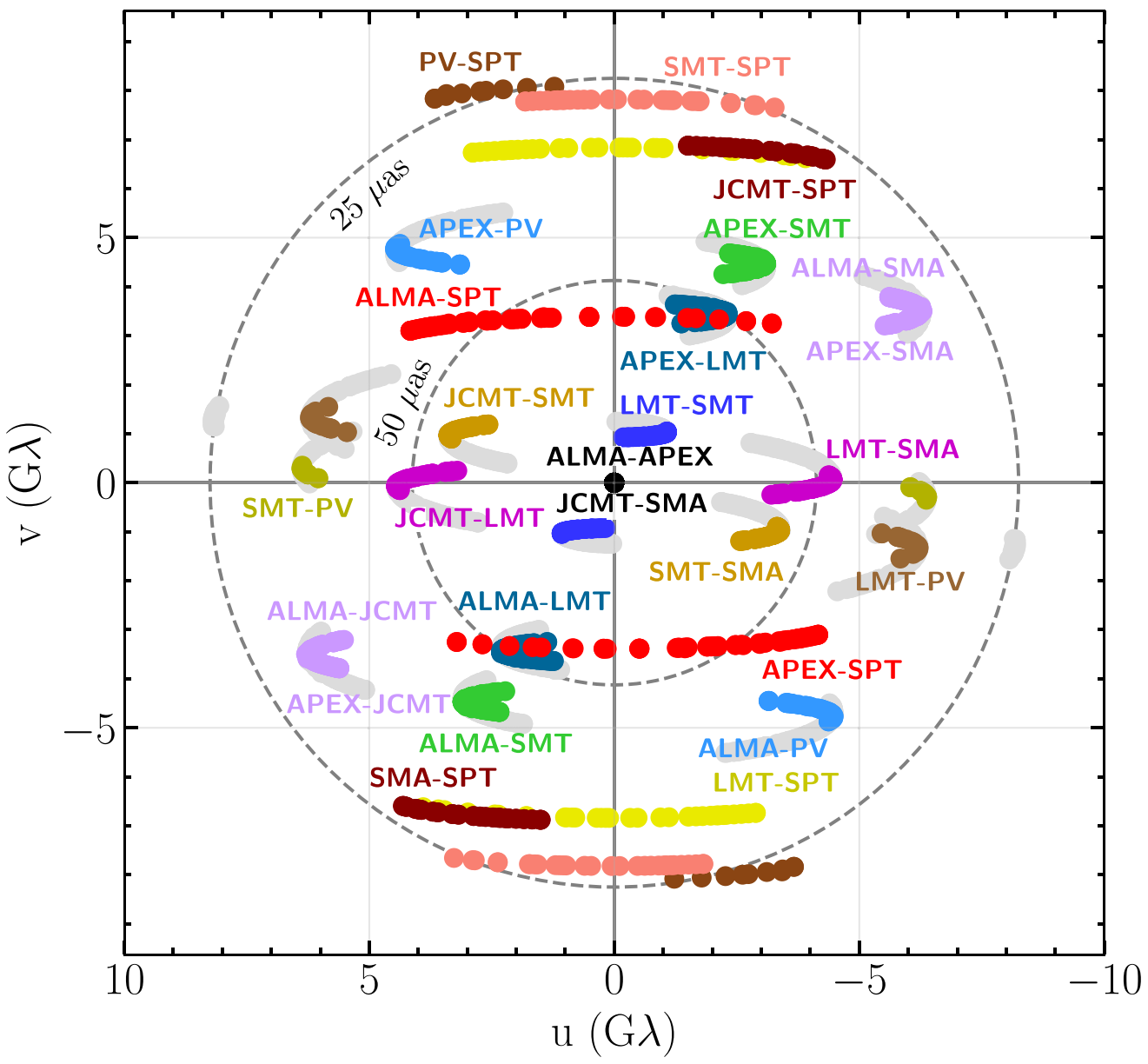

**Figure 18.** Aggregate baseline coverage for EHT observations of 3C 279 in 2017 April. The coverage of M87 is shown in light gray for comparison.

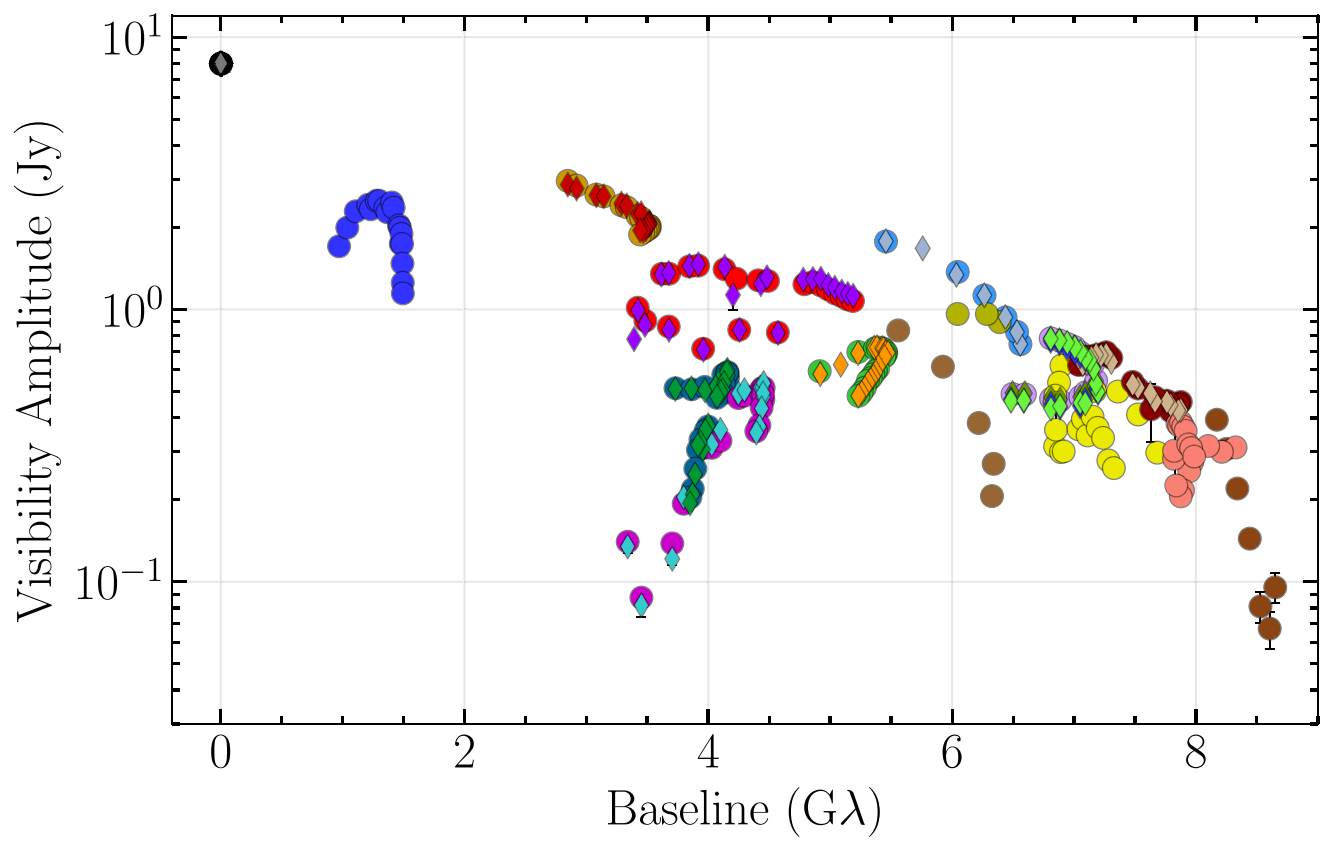

**Figure 19.** Visibility amplitudes of 3C 279 on April 11 as a function of projected baseline length after a priori and network calibration. The amplitudes are corrected for upward bias from thermal noise (Equation (6)). Points are colored by baseline as in Figure 18.

We imaged 3C 279 using both traditional CLEAN (DIFMAP) and RML (eht-imaging and SMILI) methods. Because of pronounced differences between the sources in different characteristics (total flux density, FOV, compact structure morphology), we do not use the same fiducial scripts derived for M87 on the 3C 279 data. Furthermore, as a consequence of having very few short baselines in the EHT array, we have found imaging 3C 279 to be more difficult than M87 because of its more extended structure. A full description of 3C 279 imaging procedures and results will be presented separately.

Figure 20 shows reconstructed images of 3C 279 from all three methods using data from April 11. The source exhibits two compact and bright features with a separation of $\sim$100 $\mu$as: a primary component extended in the northwest to southeast direction, and a secondary component perpendicular to the first. This secondary component extends from the core in the direction of the 3C 279 jet observed at lower frequencies (see, e.g., Lister et al. 2016; Jorstad et al. 2017).

In Figure 21, we compare the interleaved multiplicative station gains for M87 and 3C 279 on April 5. The station gains were derived via self-calibration to the fiducial images of M87 from the three imaging pipelines (Section 7) and on the three images of 3C 279 (Figure 20). In Table 6, we present the median gains for the two sources and compare them to the expected a priori visibility amplitude error budget, assuming nominal pointing and focus (Paper III).

The gains derived from images produced with the three imaging methods are consistent. For all stations except the LMT, the derived gains are time-variable, but close to unity. Issues with the performance of the LMT (Section 4, and Paper III) result in a larger median gain correction at that station, in line with the gain corrections derived directly from the visibility amplitudes via the use of crossing tracks on M87 described in Appendix B. Furthermore, the interleaved LMT gain curve has large variations with time on both sources. In particular, the M87 gains at the LMT have large, single-scan excursions not seen in the 3C 279 gain trends. These excursions are likely due to poor pointing on M87, which is nearly an order of magnitude fainter than 3C 279.

In Table 14 of Appendix F, we present the median gains for the two sources, compared to the expected a priori visibility amplitude error budget, across all days. Figure 34 in Appendix F compares the gain variations recovered from imaging M87 and 3C 279 across all days for the SMT and LMT stations. In some cases, the absolute gain is not identical across the different sources and imaging pipelines. This variation can be partly ascribed to the large uncertainty in the total flux density of 3C 279. However, in all cases the relative gain trends with time are broadly consistent, except for the occasional large excursions seen in the LMT M87 gains.

The consistency of the derived gain variations between the two sources and the different imaging methods for all stations on all days, particularly for the large corrections at LMT, provides confidence that these corrections are not the result of imaging artifacts or missing structure in M87 reconstructions. In Paper VI we show that derived gains from fitting parametric models are similar to those derived from imaging.

### *8.2. Image Dependence on Sites*

Our comparisons of reconstructed images on different days show that the primary image features in M87 are not sensitive to small changes in baseline coverage (Figure 11). We now explore how more severe changes in baseline coverage affect imaging. In particular, we assess whether or not the image structure is closely tied to the data from any one site.

The top row of Figure 22 shows reconstructed images of M87 using the fiducial eht-imaging parameters after excluding all data on baselines to one of the five EHT sites that observed the source. As shown in Figure 1, removing some sites entirely from the array does not lead to a significant decrease in the length of the longest baseline in either the north–south or east–west direction. The notable exceptions are Chile (ALMA and APEX) and Pico Veleta (PV) in Spain. The Chilean sites are part of all baselines with north–south lengths $|v| > 2.5$ G$\lambda$, while PV provides the only baselines in the northeast direction with lengths longer than 5 G$\lambda$. Thus, the loss of either of these sites significantly elongates the synthesized beam and impedes the reconstruction of the ring feature.

The presence of a ring in our reconstructed images of M87 is resilient to several site losses, even though the fiducial imaging parameters were not chosen using a reduced ($u$, $v$) coverage. The ring is successfully reconstructed when we entirely omit both Hawai'i stations (SMA and JCMT), even though these stations are required to sample the two visibility nulls (at $\sim$3.4 and $\sim$8.3 G$\lambda$; see Section 2.1). The fiducial scripts also reconstruct a ring when LMT data are entirely omitted, but the ring orientation is reversed. Unsurprisingly, the site losses that result in the poorest reconstructions of the ring are those that most degrade the synthesized beam, namely Chile and PV.

We also tested whether or not severe calibration errors could have affected imaging. To do so, we applied our fiducial scripts after omitting calibrated visibility amplitudes to a specified site while still retaining closure phases and closure amplitudes. The lower panels of Figure 22 show this test for M87 reconstructions using the eht-imaging pipeline (for similar results with SMILI, see Appendix D).[117] Comparing the top and bottom row of Figure 22, it is apparent that even sites with poor amplitude calibration (e.g., the LMT) can provide essential constraints through closure quantities. For this reason, we do not flag data with poor amplitude calibration before imaging. We also show an image that uses only closure quantities for the reconstruction, therefore using no calibration information for

[117] The fiducial script was minimally adapted to allow this test.

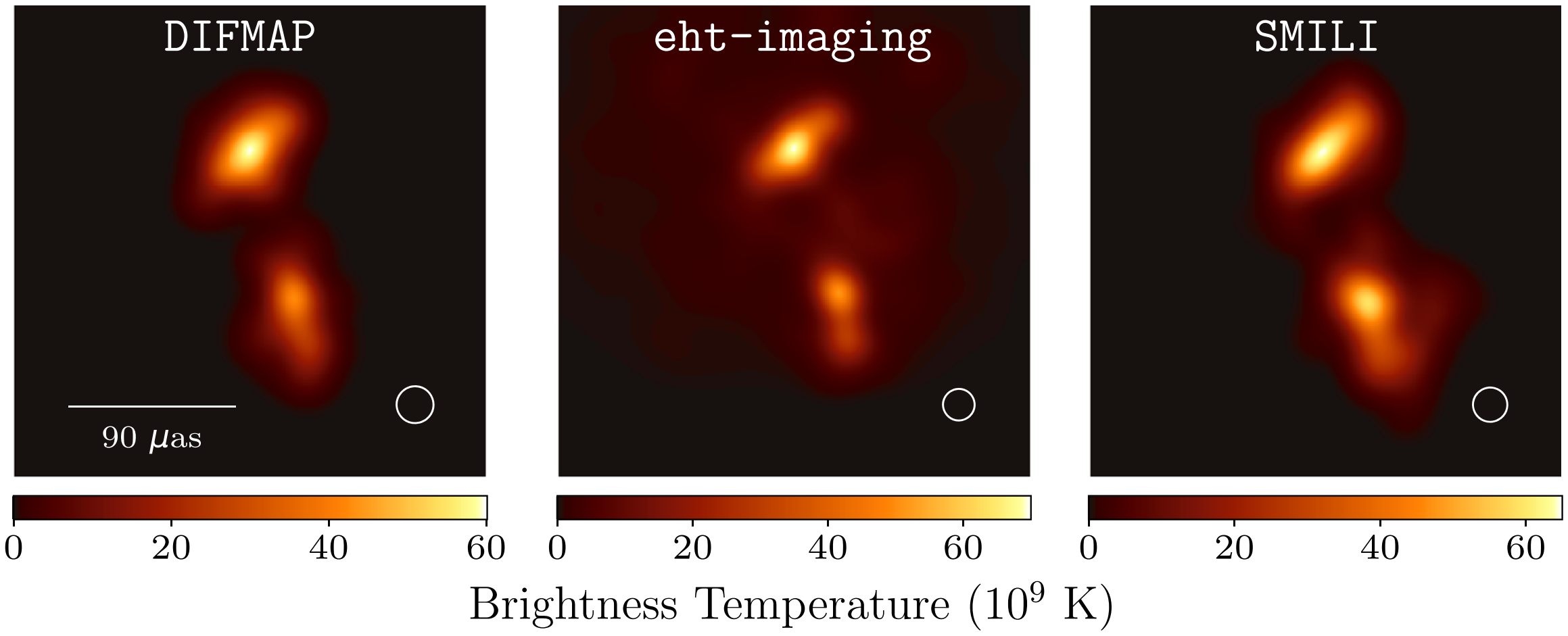

**Figure 20.** Representative images of 3C 279 from the April 11 EHT observations produced using `DIFMAP`, `eht-imaging`, and `SMILI`. To simplify visual comparisons and display the images at similar resolutions, the images are restored with circular Gaussian beams of 20, 17.1, and 18.6 $\mu$as FWHM, respectively.

**Table 6**
Median, 25th, and 75th Percentile Residual Gain Corrections for M87 and 3C 279 on April 5

| Station | Fiducial M87 Median Gain | | | 3C 279 Median Gain | | | A Priori Budget (%) |
|---|---|---|---|---|---|---|---|
| | `DIFMAP` | `eht-imaging` | `SMILI` | `DIFMAP` | `eht-imaging` | `SMILI` | |
| ALMA | $0.97^{+0.02}_{-0.03}$ | $0.97^{+0.02}_{-0.01}$ | $0.98^{+0.01}_{-0.01}$ | $0.99^{+0.02}_{-0.09}$ | $1.11^{+0.04}_{-0.02}$ | $0.97^{+0.03}_{-0.01}$ | 5.0 |
| APEX | $1.05^{+0.04}_{-0.02}$ | $1.02^{+0.02}_{-0.01}$ | $1.01^{+0.01}_{-0.01}$ | $0.99^{+0.01}_{-0.00}$ | $0.90^{+0.01}_{-0.00}$ | $1.05^{+0.01}_{-0.02}$ | 5.5 |
| SMT | $1.13^{+0.01}_{-0.06}$ | $1.02^{+0.04}_{-0.01}$ | $0.99^{+0.02}_{-0.01}$ | $1.06^{+0.01}_{-0.02}$ | $0.97^{+0.01}_{-0.02}$ | $1.04^{+0.02}_{-0.01}$ | 3.5 |
| JCMT | $1.00^{+0.02}_{-0.00}$ | $1.00^{+0.00}_{-0.00}$ | $1.00^{+0.00}_{-0.00}$ | $1.00^{+0.01}_{-0.00}$ | $1.02^{+0.02}_{-0.01}$ | $1.00^{+0.00}_{-0.01}$ | 7.0 |
| LMT | $1.46^{+0.76}_{-0.21}$ | $1.55^{+0.93}_{-0.19}$ | $1.47^{+0.91}_{-0.19}$ | $1.08^{+0.16}_{-0.04}$ | $1.21^{+0.13}_{-0.11}$ | $1.35^{+0.27}_{-0.05}$ | 11.0 |
| SMA | $1.00^{+0.02}_{-0.01}$ | $1.00^{+0.01}_{-0.00}$ | $1.00^{+0.00}_{-0.00}$ | $1.01^{+0.01}_{-0.01}$ | $1.04^{+0.03}_{-0.00}$ | $0.99^{+0.00}_{-0.01}$ | 7.5 |
| PV | $1.14^{+0.04}_{-0.04}$ | $0.96^{+0.02}_{-0.07}$ | $0.98^{+0.02}_{-0.07}$ | $1.33^{+0.12}_{-0.04}$ | $1.14^{+0.04}_{-0.05}$ | $0.94^{+0.05}_{-0.02}$ | 5.0 |

**Note.** These gains were derived via self-calibration (with no systematic error included). The error budget on a priori calibration is derived in Paper III. Note that the median gain corrections for ALMA, APEX, SMA, and JCMT can reasonably be much smaller than this error budget because network calibration has already been applied. The variation in the recovered gains among pipelines is partly due to the large uncertainty in the total flux density (between 8 and 10 Jy) and total compact flux density of 3C 279.

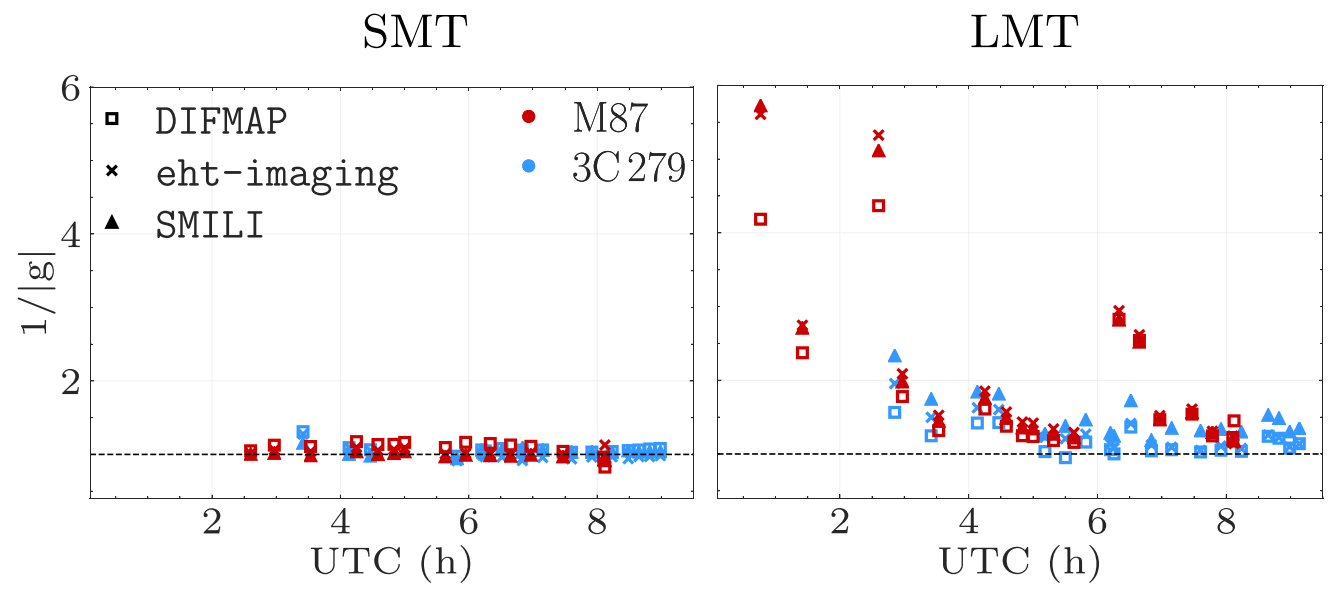

**Figure 21.** Multiplicative residual station gains for the SMT (left) and LMT (right) derived from the 3C 279 images (Figure 20) and fiducial M87 images (Figure 14) from the three imaging pipelines on April 5. Gains for the fiducial images of M87 are shown in red; those for 3C 279 are shown in blue. The particularly large excursions on the LMT M87 gains are likely due to poor pointing. Note that LMT could not observe 3C 279 before $2^{h}30$ UTC.

any site. This image demonstrates that closure quantities are sufficient to recover the diameter and asymmetry of the ring in the fiducial images.

### *8.3. Tests of the Observed Time Variability*

Our fiducial images of M87 (Figure 11) over the four observing days show an evolution in the ring structure between the two observations on April 5, 6 and the two observations on April 10, 11 (Figure 33). However, the data used to reconstruct these images do not identically sample the ($u$, $v$) plane (Figure 1). To verify whether or not the observed evolution in the fiducial M87 images is intrinsic, it is necessary to confirm that the evolution is not an artifact imprinted on the reconstructions by the changing ($u$, $v$) coverage sampled by the EHT on the different observing days.

To this end, we constructed reduced data sets from the April 6 and 11 data with "overlapping" baseline coverage. Specifically, we flag all ($u$, $v$) points along a particular baseline track that are not common to the April 6 and 11 observations (i.e., we flag points on April 6 with no measurement on April 11 within 0.01 G$\lambda$, and vice versa). The number of (10 s averaged) visibilities is reduced from ~18,000 to ~5,000 in the overlapping data set. We then used the `eht-imaging` script with the identified fiducial parameters (Section 6.2.2) to reconstruct images on both days with these subsampled data sets.

Figure 23 compares the reconstructions made from these overlapping ($u$, $v$) data to our original images from the full data sets. The results show no significant variation in the morphology due to the changing ($u$, $v$) coverage. In particular, the enhanced southwest brightness in the April 11 image relative to the April 6 image is still prominent in the overlapping coverage reconstructions. This test shows that the reconstructed evolution is not due, at least solely, to the changes in ($u$, $v$) coverage between days, although each image is still affected by the limited ($u$, $v$) coverage. The bottom panels of Figure 23 show selected closure phases on April 6 and 11 that are significantly different; reconstructed images with both the full and overlapping data sets recover this significant evolution of closure phases. We discuss the observed time variability further in Appendix E.

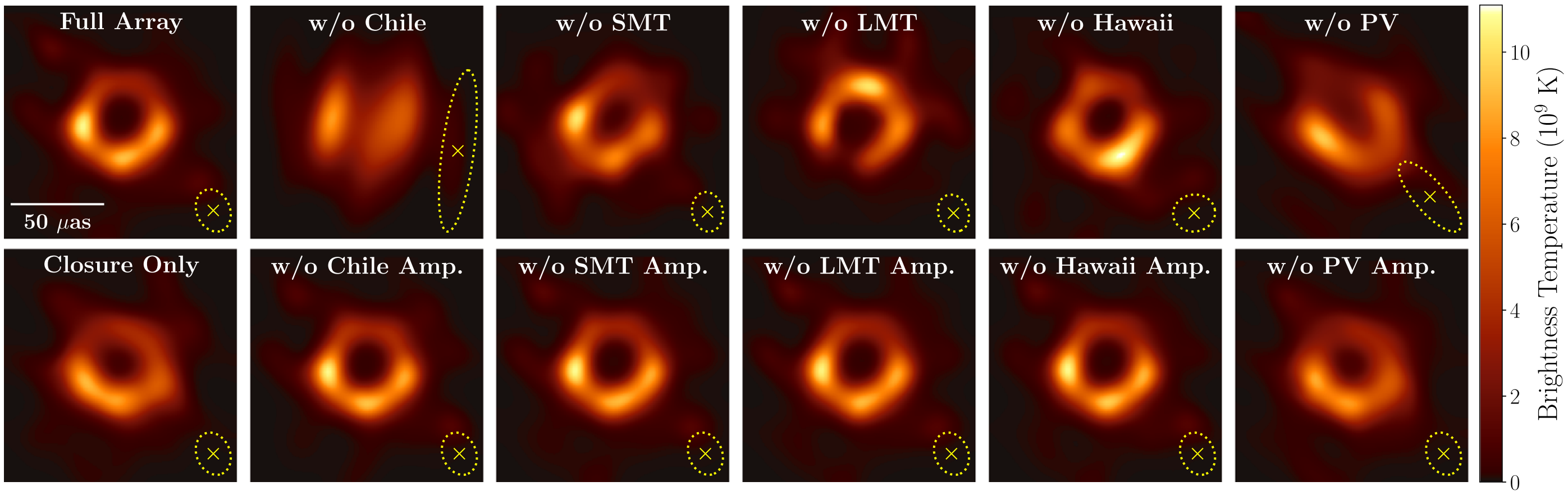

**Figure 22.** Example reconstructions of M87 on April 11 after omitting visibilities to each geographical site. The top row shows reconstructions that exclude all baselines to the indicated site (i.e., mimicking an observation without that site); the bottom row shows reconstructions that exclude all visibility amplitudes on baselines to the indicated site, while still including closure phases and closure amplitudes (i.e., exploring the limit of arbitrarily poor amplitude calibration). The lower-left panel shows an image reconstructed after excluding visibility amplitudes to all sites (i.e., using only closure quantities). These images were reconstructed using the `eht-imaging` pipeline with its fiducial parameters (see Section 6.3.3). The ellipse in each panel shows the corresponding synthesized beam with uniform weighting, but the image is not convolved with this beam. While the fiducial parameters used are not necessarily optimal for imaging these modified data sets, the reconstructed images of M87 are rather resilient to the loss of most sites, especially if the closure information is preserved, demonstrating that our results are insensitive to calibration assumptions. Figure 32 in Appendix D shows corresponding results for the `DIFMAP` and `SMILI` pipelines.

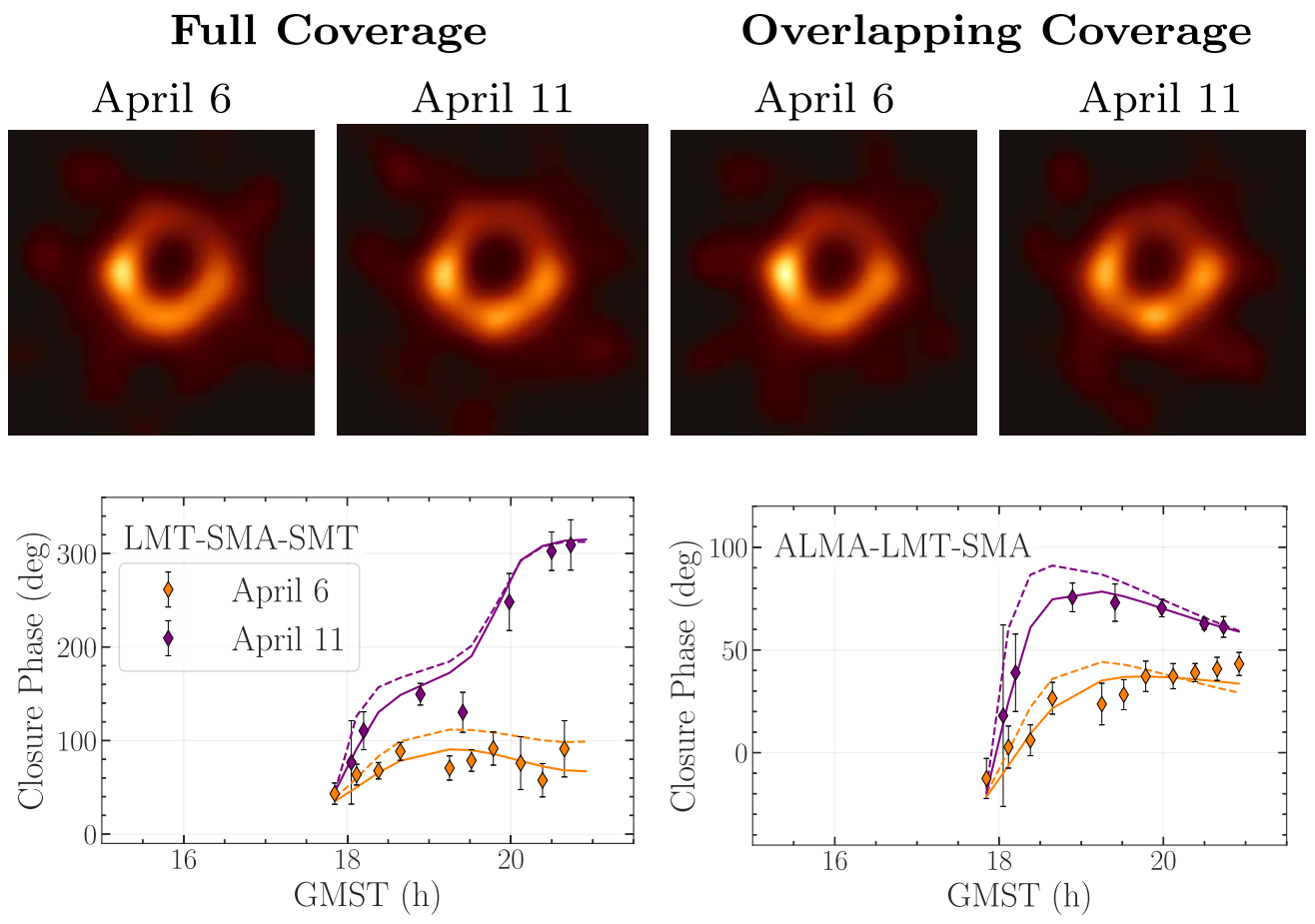

**Figure 23.** Validating the observed variation in the reconstructed images between April 6 and 11. Top-left panels: fiducial `eht-imaging` image reconstructions from observations on April 6 and 11 (Section 7.1). Top-right panels: reconstructions using the same fiducial script and parameters, but using data only on the ($u$, $v$) points which overlap on April 6 and 11. The reconstructions using only the overlapping ($u$, $v$) coverage show similar time evolution (a shift in brightness to the southwest; see Appendix E) to the reconstructions from the original data sets, indicating that the observed variation in images is intrinsic. Bottom panels: plots of two selected closure phases (LMT–SMA–SMT and ALMA–LMT–SMA) that clearly identify intrinsic structural evolution between April 6 and 11. The colored points identify the closure phases of the scan-averaged data, and the phases shown as solid and dotted colored lines are taken from the reconstructions of full and overlapping coverage data, respectively.

## 9. Image Analysis

To assess the consistency of our images with each other and determine which specific image features are most reliable, we define and measure several properties that characterize asymmetric rings: the diameter $d$, width $w$, orientation angle $\eta$, azimuthal brightness asymmetry $A$, and fractional central brightness $f_C$. This approach is also motivated by GRMHD simulations of M87, which predict an image dominated by a bright and nearly circular ring morphology across a large range of parameters (Paper V). In Section 9.1, we define these quantities, explain how they are measured, and give procedures to estimate their uncertainties. In Section 9.2, we evaluate biases and systematic errors in these measurements using tests with synthetic data. In Section 9.3, we apply our results to estimate properties of the fiducial images of M87 presented in Section 7. In Paper VI, we explore model fitting in more detail, both through visibility-domain fits of geometrical models and through image-domain analysis of the reconstructions from each imaging pipeline.

### 9.1. Ring Parameter Definitions

To identify a ring in a given reconstructed image, we search for the central point from which radial profiles are peaked at a similar distance. From a candidate ring center position ($x$, $y$), we sample the linearly interpolated image equally in azimuthal angle $\theta$ between 0° and 360° and in radius $r$ between 0 and 50 $\mu$as to obtain a transformed image: $I(r, \theta; x, y)$.[118] Then, for each radial profile at fixed $\theta$, we identify the distance $r_{pk}$ at which the angular profile assumes its peak brightness. We define the ring radius at a given position, $\bar{r}_{pk}(x, y)$ as the mean of these peak distances:

$$r_{pk}(\theta; x, y) = \mathrm{argmax}_r[I(r, \theta; x, y)],$$
$$\bar{r}_{pk}(x, y) = \langle r_{pk}(\theta; x, y)\rangle_{\theta \in [0, 2\pi]}. \quad (16)$$

To estimate an associated uncertainty, we use the standard deviation $\sigma_{\bar{r}}(x, y)$ of the $r_{pk}(\theta; x, y)$ values. For a similar

[118] We sample angles every 1° and radii every 0.5 $\mu$as between 0 and 50 $\mu$as.

method using Gaussian fits to the radial profiles, see Johannsen et al. (2016).

To find the ring center ($x_0$, $y_0$), we search over ($x$, $y$) and identify the position that minimizes the normalized radial peak dispersion:

$$(x_0, y_0) = \text{argmin}\left[\frac{\sigma_{\bar{r}}(x, y)}{\bar{r}_{\text{pk}}(x, y)}\right]_{(x,y)}. \quad (17)$$

The measured diameter $d$ is then twice $\bar{r}_{\text{pk}}$ measured from the identified center ($x_0$, $y_0$),

$$d = 2\bar{r}_{\text{pk}}(x_0, y_0), \quad (18)$$

and the associated uncertainty $\sigma_d$ in the diameter is

$$\sigma_d = 2\sigma_{\bar{r}}(x_0, y_0). \quad (19)$$

Although we have designed our analysis tools to specifically search for circular features, $\sigma_d/d$ can be interpreted as a measure of the circularity of the identified feature in an individual image. See Paper VI for further discussion of the circularity of these images and model fits to the visibility data.

To avoid spurious detections when searching for the center location (Equation (17)), we first blur the image with a 2 $\mu$as FWHM Gaussian (approximately the pixel size of the original reconstructions) and threshold the search image below 5% of the peak brightness. We also restrict the range of allowed diameters to 10–100 $\mu$as. We use the original (unconvolved and unthresholded) image for all subsequent analysis. Because the remaining features are all estimated using the fixed center coordinates estimated by this procedure, we suppress the center coordinates for the remainder of our discussion, and we denote the unwrapped, centered image as $I(r, \theta)$.

After identifying the ring center and diameter, we measure the remaining characteristic features. The ring width $w$ is determined by measuring the FWHM of each radial slice at constant $\theta$ and taking the mean. To avoid bias in the measurement from the resampled image $I(r, \theta)$ having a non-zero mean floor value outside the ring, we subtract the value $I_{\text{floor}} = \langle I(r_{\max} = 50\ \mu\text{as}, \theta)\rangle_\theta$ from each radial profile before computing the FWHM:

$$w = \langle \text{FWHM}[I(r, \theta) - I_{\text{floor}}]\rangle. \quad (20)$$

We estimate an uncertainty $\sigma_w$ using the standard deviation of the set of FWHMs. Note that the measured width is dependent upon both the intrinsic width of the source and the finite resolution of the array. Thus, $w$ should be viewed only as an upper limit on the intrinsic ring width, and it is upward biased by the application of a restoring beam (e.g., for DIFMAP reconstructions). We explore this bias further in Appendix G.

We define the ring orientation angle $\eta$ (measured east of north) by considering the angular profiles $I(r, \theta)$ at fixed $r$. At each radius, we find the argument of the first angular mode ($m = 1$) of the angular profile and then define the overall orientation angle $\eta$ as the circular mean of these angles over the ring width; from $r_{\text{in}} = (d - w)/2$ to $r_{\text{out}} = (d + w)/2$. That is,

$$\eta = \left\langle \text{Arg}\left[\int_0^{2\pi} I(\theta) e^{i\theta} d\theta\right]\right\rangle_{r\in[r_{\text{in}}, r_{\text{out}}]}. \quad (21)$$

Similarly, we define the associated uncertainty $\sigma_\eta$ as the circular standard deviation of the angle measurements for $r \in [r_{\text{in}}, r_{\text{out}}]$.

To characterize the degree of azimuthal asymmetry in a ring, we compute the normalized amplitude of the first angular mode for radii between $r_{\text{in}}$ and $r_{\text{out}}$ and again take the mean. That is,

$$A = \left\langle \frac{\left|\int_0^{2\pi} I(\theta) e^{i\theta} d\theta\right|}{\int_0^{2\pi} I(\theta) d\theta} \right\rangle_{r\in[r_{\text{in}}, r_{\text{out}}]}. \quad (22)$$

The associated uncertainty $\sigma_A$ is the standard deviation of the asymmetry at each $r \in [r_{\text{in}}, r_{\text{out}}]$. The asymmetry $A$ takes values in the range from 0 to 1, with 0 corresponding to perfect azimuthal symmetry and 1 corresponding to a delta function concentrating all of the flux density at a single orientation angle. The crescent models defined in Section 6.1 have $A = 0.23$, with angular brightness profiles $I(\theta) \propto 1 + 2A\cos(\theta - \eta)$.

Finally, we define a fractional central brightness (or inverse contrast ratio) $f_{\text{C}}$ as the ratio of the mean brightness interior to the ring to the mean brightness around the ring. To define the interior brightness, we average over a disk of radius 5 $\mu$as. That is,

$$f_{\text{C}} = \frac{\langle I(r, \theta)\rangle_{\theta, r\in[0,5\ \mu\text{as}]}}{\langle I(d/2, \theta)\rangle_{\theta\in[0,2\pi]}}. \quad (23)$$

We find that this statistic has an extremely large scatter across the Top Sets, primarily because the interior brightness can become arbitrarily low. Thus, our imaging methods only securely identify an upper limit on $f_{\text{C}}$.

### 9.2. Tests with Synthetic Data Reconstructions

We tested the analysis methods described in Section 9.1 on synthetic data from a crescent model with asymmetry parameter $A = 0.23$ oriented at a orientation angle $\eta = 150°$ (Appendix C), and on the GRMHD simulation image shown in Figure 10. For both sources, we measured the ring features defined in Section 9.1 for every image reconstructed with parameter combinations in the M87 Top Sets (see Section 6). While some of these images may not meet the $\chi^2 < 2$ constraint imposed for M87 images, using the M87 Top Set allows us to assess the expected variation resulting from imaging choices irrespective of image-data consistency.

The results of this analysis for the diameter ($d$), width ($w$), and orientation angle ($\eta$) are displayed in Figure 24. The points shown correspond to the measured quantities from the fiducial reconstructions. These measurements have two distinct sources of error: the intrinsic measurement uncertainty on each quantity from a single image, and the uncertainty in the quantity from varying imaging parameters across the Top Set. We combine these two sources of error in the error bars in Figure 24 by taking the quadrature sum of the measurement uncertainty from the fiducial image and the median absolute deviation of the parameter estimates across the Top Set.

In both simulated data tests, the ring diameter $d$ is the most accurately recovered quantity; however, the diameter measurement is correlated with the ring width and is biased downward by several microarcseconds when the image is blurred (see Appendix G). In the simulated crescent, the diameter of the unblurred model is 44 $\mu$as; the value measured using our approach on the ground truth image is pushed down to 43 $\mu$as because of the Gaussian convolution (Appendix G). Taking the median across all days, the diameters extracted from the fiducial DIFMAP crescent model reconstructions have a median value of $40 \pm 2\,\mu$as when restored with a $20\,\mu$as FWHM beam. When they are instead restored with a smaller $10\,\mu$as beam, the DIFMAP results become more accurate (compared to the model values), with a measured diameter of $43 \pm 2\,\mu$as. The RML crescent model reconstructions measure a median diameter of $42 \pm 1\,\mu$as.

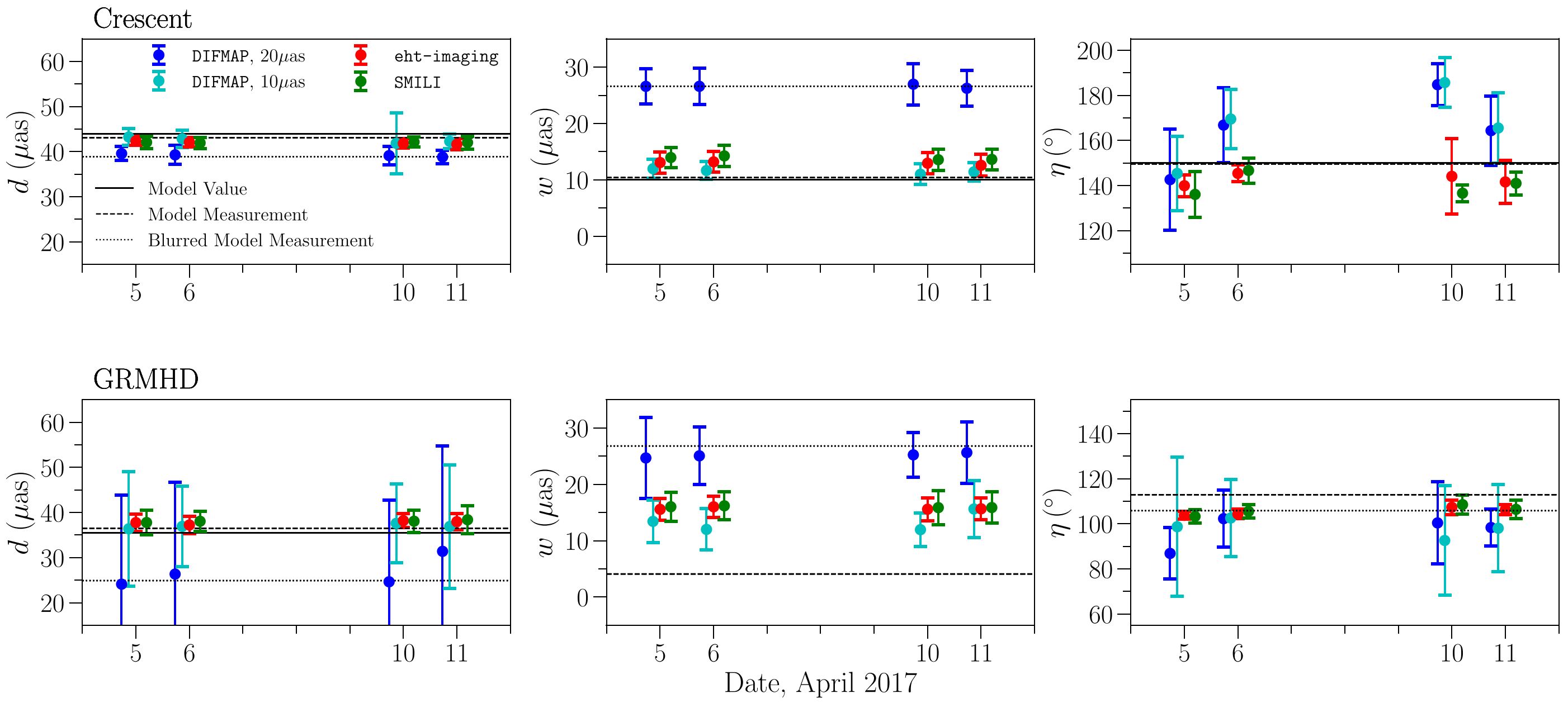

**Figure 24.** Measurements of ring features on fiducial reconstructions of a crescent model (top row) and a GRMHD simulation snapshot (bottom row) from images reconstructed with the fiducial parameters for three imaging pipelines. From left to right, panels show the measured ring diameter $d$, width $w$, and orientation angle $\eta$. The DIFMAP results are shown for images restored with both a 10 $\mu$as (cyan) and a 20 $\mu$as (blue) Gaussian beam. The eht-imaging (red) and SMILI (green) are shown for the unblurred images. Solid lines indicate the ground truth values of the three quantities in the crescent model, and the photon ring diameter of the GRMHD simulation. Dashed lines indicate the measured values from the ground truth images, and dotted lines indicate the measured values from the ground truth images convolved with a 20 $\mu$as FWHM circular Gaussian beam. The error bars are computed as the quadrature sum of the measurement uncertainty from the fiducial images (Section 9.1) and the median absolute deviation of the parameter estimates across the Top Set.

The GRMHD image that we use (Paper V) has a lensed photon ring with a diameter of $9.79M/D \approx 35.5\,\mu$as;[119] we recover a value of 36.5 $\mu$as from the ground truth image, indicating that even with a perfect image reconstruction, fine-scale substructure and extended flux in an image can bias the ring diameter measurement away from the photon ring value. The ring diameter measured from the eht-imaging and SMILI GRMHD fiducial reconstructions (Figure 24) has a median value of $38 \pm 2\,\mu$as across all four observed days. However, for this image, the DIFMAP results are strongly dependent upon the chosen restoring beam. The DIFMAP reconstructions restored with a $20\,\mu$as beam lack prominent rings, leading to poor recovery of the ring diameter ($26 \pm 20\,\mu$as). When blurred with a smaller 10 $\mu$as beam, the DIFMAP results align with the RML methods but have larger uncertainties due to larger scatter across the Top Set ($37 \pm 11\,\mu$as).

As shown in Figure 24, the measured orientation angles from the crescent model reconstructions using the RML imaging pipelines have a median value of $141° \pm 5°$, which is moderately discrepant from the true value of 150°. The values from the DIFMAP images are even more discrepant and are unstable among days. The GRMHD image has no defined a priori orientation angle; the measured orientation angles from all pipelines are more stable than for the simple crescent model, and are consistent with the value measured from the blurred simulation image (105°). Both of these cases indicate that our procedure may underestimate uncertainties on $\eta$.

Both the crescent and GRMHD models have intrinsic widths that are much narrower than the resolution of the EHT. As expected from the $20\,\mu$as restoring beam, the DIFMAP reconstructions lead to a larger measurement of ring width than the RML reconstructions. When a $10\,\mu$as FWHM beam is used to restore the DIFMAP images, the measured width aligns with the RML results. Nonetheless, all reconstructions have widths that are systematically biased upward from the true values, and the extracted ring widths from image-domain fitting can at best be viewed as upper limits.

Similarly, the brightness depression contrast ratio $f_C$ is highly sensitive to the image resolution and particular imaging choices. In particular, we find that the Top Sets from the RML imaging pipelines have an extremely large scatter in $f_C$. When blurred to the same $20\,\mu$as resolution, both RML and CLEAN methods give measurements of $f_C$ that are consistently in the range 0.2–0.5 (Appendix G). Consequently, we can only determine an upper limit for $f_C$.

For all the imaging pipelines, we find that the scatter in ring diameters across the Top Set (typically $\lesssim 1\,\mu$as) is subdominant to the intrinsic measurement uncertainty estimated from a single image (typically 1–2 $\mu$as). Thus, choices made in the imaging process do not significantly affect the measured ring diameter. In contrast, the other measured features have error budgets that are more evenly divided between intrinsic uncertainty in a single image and the scatter across the Top Set.

### 9.3. Results for M87

Table 7 lists the values of all ring parameters measured for the fiducial M87 reconstructions for each day from the three

[119] The simulation used an assumed black hole mass $M = 6.2 \times 10^9\,M_\odot$ and dimensionless spin $a_* = 0.9375$, observed at a distance $D = 16.9$ Mpc and 17° inclination (Paper V).

**Table 7**
Diameter $d$, Width $w$, Orientation Angle $\eta$, Asymmetry $A$, and Floor-to-ring Contrast Ratio $f_C$ Measured from the Fiducial M87 Images for Each Day from Each Method (Section 7.1)

| | $d$ ($\mu$as) | $w$ ($\mu$as) | $\eta$ (°) | $A$ | $f_C$ |
|---|---|---|---|---|---|
| DIFMAP | | | | | |
| Apr 5 | 37.2 ± 2.4 | 28.2 ± 2.9 | 163.8 ± 6.5 | 0.21 ± 0.03 | $5 \times 10^{-1}$ |
| April 6 | 40.1 ± 7.4 | 28.6 ± 3.0 | 162.1 ± 9.7 | 0.24 ± 0.08 | $4 \times 10^{-1}$ |
| April 10 | 40.2 ± 1.7 | 27.5 ± 3.1 | 175.8 ± 9.8 | 0.20 ± 0.04 | $4 \times 10^{-1}$ |
| April 11 | 40.7 ± 2.6 | 29.0 ± 3.0 | 173.3 ± 4.8 | 0.23 ± 0.04 | $5 \times 10^{-1}$ |
| eht-imaging | | | | | |
| April 5 | 39.3 ± 1.6 | 16.2 ± 2.0 | 148.3 ± 4.8 | 0.25 ± 0.02 | $8 \times 10^{-2}$ |
| April 6 | 39.6 ± 1.8 | 16.2 ± 1.7 | 151.1 ± 8.6 | 0.24 ± 0.02 | $6 \times 10^{-2}$ |
| April 10 | 40.7 ± 1.6 | 15.7 ± 2.0 | 171.2 ± 6.9 | 0.23 ± 0.03 | $4 \times 10^{-2}$ |
| April 11 | 41.0 ± 1.4 | 15.5 ± 1.8 | 168.0 ± 6.9 | 0.20 ± 0.02 | $4 \times 10^{-2}$ |
| SMILI | | | | | |
| April 5 | 40.5 ± 1.9 | 16.1 ± 2.1 | 154.2 ± 7.1 | 0.27 ± 0.03 | $7 \times 10^{-5}$ |
| April 6 | 40.9 ± 2.4 | 16.1 ± 2.1 | 151.7 ± 8.2 | 0.25 ± 0.02 | $2 \times 10^{-4}$ |
| April 10 | 42.0 ± 1.8 | 15.7 ± 2.4 | 170.6 ± 5.5 | 0.21 ± 0.03 | $4 \times 10^{-6}$ |
| April 11 | 42.3 ± 1.6 | 15.6 ± 2.2 | 167.6 ± 2.8 | 0.22 ± 0.03 | $6 \times 10^{-6}$ |

**Note.** The DIFMAP results are presented for a restoring beam of 20 $\mu$as. The estimated diameters and orientation angles are generally consistent across different imaging pipelines on all four days, but the measured width, contrast, and asymmetry differ between imaging pipelines.

parameter surveys. As in the previous section, for the fiducial images the uncertainties are computed by adding the scatter in the measured quantities across the Top Sets in quadrature to the intrinsic measurement uncertainty.

Figure 25 plots the measured diameter, width, and orientation angle from each method over all four days, and Figure 26 visualizes these parameters. Across all days, the DIFMAP reconstructions recover a median diameter of 40 ± 3 $\mu$as when restored with a 20 $\mu$as beam, and 44 ± 5 $\mu$as when restored with a 10 $\mu$as beam. The RML methods recover a median diameter of 41 ± 2 $\mu$as. For each imaging pipeline, there is slight upward trend in the diameter over time, increasing by ≈2 $\mu$as from the first to the last day of the observing campaign. However, this trend is well within the estimated uncertainty of the diameter measurements. The measured ring diameters are consistent among all imaging pipelines and largely unchanged from the first April 11 images produced from early-release engineering data (Section 5), for which we measure a median diameter of 41 ± 3 $\mu$as. The other parameters are less consistent from these first images to the final fiducial images selected by the parameter survey, indicating more sensitivity in these parameters to the data quality and the imaging method.

As in the synthetic data results presented in Figure 24, the beam-convolved DIFMAP reconstructions produce larger measured widths than the RML methods. The width measurements become consistent when the DIFMAP images are restored with a smaller 10 $\mu$as beam, but the ring width remains limited by the resolution of the EHT. Thus, from these imaging results, we can only firmly conclude that the ring width is ⩽20 $\mu$as.

In the reconstructions from all three imaging methods, there is a counterclockwise trend in the orientation angle from April 5 to 11, which is consistent with the apparent shift in brightness along the ring remarked on in Section 7.1. However, we emphasize that this ≈20° change could be the result of spurious azimuthal structure, and we have seen that our orientation angle uncertainties may be underestimated (see Figure 24). Even if this shift in angle is physical, it does *not* necessarily indicate continuous motion or a flow direction associated with the black hole in M87, and it is opposite to the inferred rotation direction for the large-scale jet (Walker et al. 2018).

Identifying a ring in the fiducial images allows us to "unwrap" the brightness distribution around the ring and display the image in $I(r, \theta)$ space. Figure 27 shows these unwrapped ring profiles of the fiducial images for all pipelines and days. In Appendix I, we present more diagnostic figures to characterize the radial and azimuthal structure of the recovered rings.

Even under the assumption that the ring in our images corresponds to the lensed photon ring of an SMBH (i.e., $d \approx (5.0 \pm 0.2)\ R_s/D$; Bardeen 1973; Johannsen & Psaltis 2010), translating the estimated ring diameter to a Schwarzschild radius—or, equivalently, to a black hole mass via $R_s = GM_{BH}/c^2$—requires evaluating biases and uncertainties related to the unresolved image structure, the black hole vector spin, and the estimated distance to M87. We perform this analysis separately, in Paper VI.

## 10. Summary and Conclusions

We have presented imaging analysis and results for EHT observations of M87 in 2017 April. These observations achieved an angular resolution (synthesized beam size) of ~20 $\mu$as, and they are the first EHT observations with sufficient sensitivity and baseline coverage to reconstruct images. Images can provide agnostic results, potentially identifying entirely unexpected features. However, a multitude of imaging techniques and parameters can lead to results that depend on specific imaging choices. Consequently, we adopted a staged imaging approach to control and evaluate potential biases.

We designed our first imaging stage to minimize shared bias among different groups of imaging experts and to test whether or not their independent inferences were consistent. In this stage, described in Section 5, we separately imaged M87 in four imaging teams. These imaging teams did not communicate about their methods or findings while analyzing the M87 data. Each team relied upon the judgment of its members to select strategies for data flagging, calibration, and imaging, using whatever software and techniques that they judged to be useful. The final images from the teams were blindly submitted for comparison. These first images of M87 all showed a prominent ring of ∼40 μas diameter with enhanced brightness in the south. The four teams made different assumptions about the total compact flux density of M87, varying from ∼0.4 to 0.9 Jy. The images also showed differences in the azimuthal ring structure.

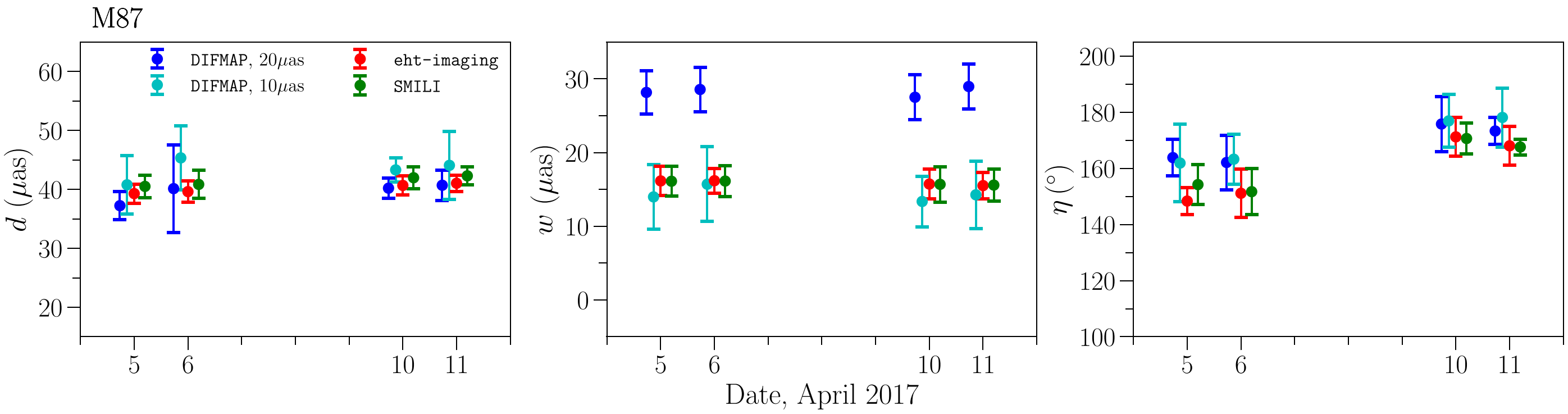

**Figure 25.** Measured ring properties on the fiducial images of M87 produced with all three imaging pipelines. From left to right, panels show the measured ring diameter $d$, width $w$, and orientation angle $\eta$. The DIFMAP results are shown for images restored with both a 10 $\mu$as (cyan) and a 20 $\mu$as (blue) Gaussian beam. The eht-imaging (red) and SMILI (green) are shown for the unblurred images. The three imaging pipelines produce consistent measurements of the ring diameter across all days. The measured orientation angles indicate a modest shift between April 5, 6 and 10, 11. The error bars are computed in the same way as in Figure 24.

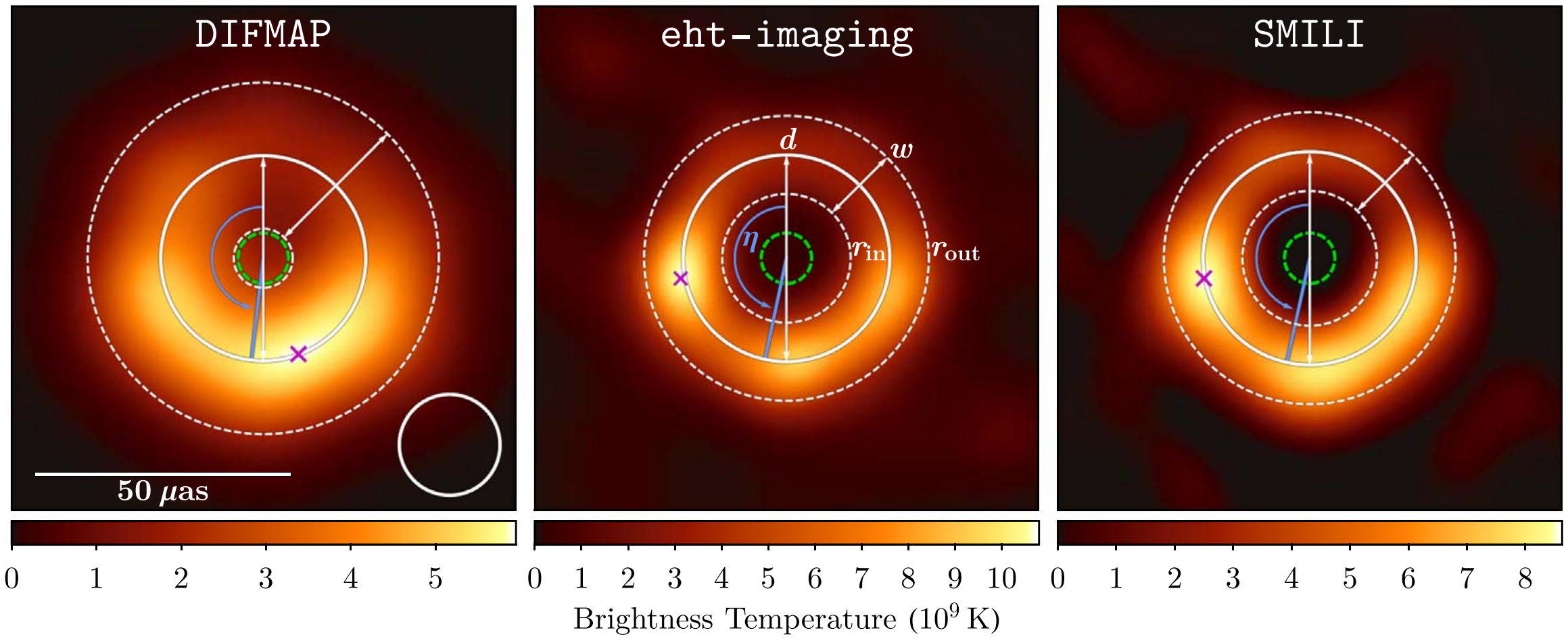

**Figure 26.** Summary of the estimated ring properties overlaid on the April 11 fiducial images from each imaging pipeline. Our procedure estimates the ring diameter $d$, width $w$, orientation angle $\eta$, and fractional central brightness $f_C$, as well as the asymmetry $A$ (not shown). In each panel, the magenta cross indicates the location of peak ring brightness, and the dashed green circle shows the region used to define interior brightness for $f_C$.

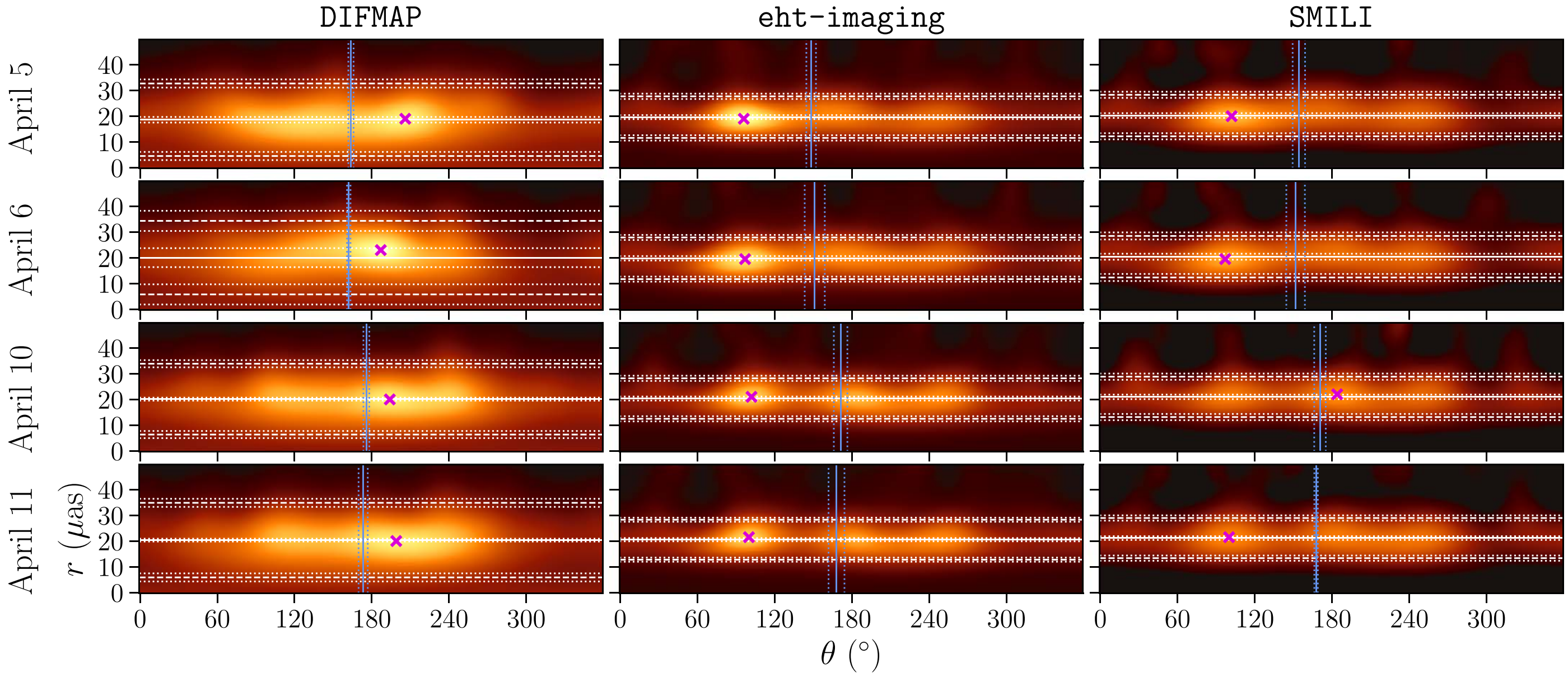

**Figure 27.** Unwrapped ring profiles of the fiducial images from April 5 to 11 (top to bottom) and for the three imaging pipelines (left to right). The columns are each scaled as in Figure 26. The estimated radius $d/2$ is shown with a horizontal line, with dotted lines denoting the associated uncertainty (Section 9.1). Horizontal dashed lines at $(d \pm w)/2$ show the measured ring width. Vertical blue lines give the orientation angle $\eta$ and its uncertainty. The magenta cross marks the peak brightness in each reconstruction.

We designed our second imaging stage to select imaging parameters objectively and to assess how parameter choices correspond to uncertainties in the final images. In this stage, described in Section 6, we developed scripted imaging pipelines for one established (`DIFMAP`) and two EHT-tailored imaging libraries (`eht-imaging` and `SMILI`). For each method, we conducted a parameter survey, exploring $\sim 10^3$–$10^4$ parameter combinations for each imaging pipeline. We then reconstructed images from synthetic data sets and compared them to the corresponding ground truth images. To determine which parameter combinations in the survey produced acceptable images, we imposed an image-data consistency requirement for reconstructions of M87, as well as an image fidelity requirement for reconstructions of synthetic data.

This procedure identified a "Top Set" of parameter choices as well as a single "fiducial" combination of parameters for each of the three surveys. When applied to M87 data, the Top Set reconstructions provide a measure of uncertainty corresponding to choices made in the imaging procedure. In addition, the fiducial image (Figure 11) acts as a single representative image from each imaging pipeline that has been identified based on objective performance, rather than expert judgment.

Our reconstructed images of M87 are consistently dominated by a ring of roughly 40 μas diameter, with enhanced brightness in the south. The ring is present in both imaging stages, across all imaging pipelines, and all observing days. The residual time-dependent station gains derived from our images are consistent between the interleaved EHT observations of M87 and the bright quasar 3C 279. Our reconstructed images of M87 show variation between the observations on April 5 and 11, as expected from the variations seen in the underlying data (Paper III). However, we cannot unambiguously associate the variability with evolution of a specific image feature.

These first images from the EHT achieve the highest angular resolution in the history of ground-based VLBI. The images reveal the central engine of M87 and are dominated by a ∼ 40 μas diameter ring with pronounced azimuthal asymmetry. Such an asymmetry is expected for relativistic rotation of emitting material near a black hole. For an assumed distance of 16.8 Mpc, the ring diameter is consistent with the expected "shadow" of a $\sim 6.5 \times 10^9\, M_\odot$ black hole.

The authors of this Letter thank the following organizations and programs: the Academy of Finland (projects 274477, 284495, 312496); the Advanced European Network of E-infrastructures for Astronomy with the SKA (AENEAS) project, supported by the European Commission Framework Programme Horizon 2020 Research and Innovation action under grant agreement 731016; the Alexander von Humboldt Stiftung; the Black Hole Initiative at Harvard University, through a grant (60477) from the John Templeton Foundation; the China Scholarship Council; Comisión Nacional de Investigación Científica y Tecnológica (CONICYT, Chile, via PIA ACT172033, Fondecyt 1171506, BASAL AFB-170002, ALMA-conicyt 31140007); Consejo Nacional de Ciencia y Tecnología (CONACYT, Mexico, projects 104497, 275201, 279006, 281692); the Delaney Family via the Delaney Family John A. Wheeler Chair at Perimeter Institute; Dirección General de Asuntos del Personal Académico-Universidad Nacional Autónoma de México (DGAPA-UNAM, project IN112417); the European Research Council Synergy Grant "BlackHoleCam: Imaging the Event Horizon of Black Holes" (grant 610058); the Generalitat Valenciana postdoctoral grant APOSTD/2018/177; the Gordon and Betty Moore Foundation (grants GBMF-3561, GBMF-5278); the Istituto Nazionale di Fisica Nucleare (INFN) sezione di Napoli, iniziative specifiche TEON-GRAV; the International Max Planck Research School for Astronomy and Astrophysics at the Universities of Bonn and Cologne; the Jansky Fellowship program of the National Radio Astronomy Observatory (NRAO); the Japanese Government (Monbukagakusho: MEXT) Scholarship; the Japan Society for the Promotion of Science (JSPS) Grant-in-Aid for JSPS Research Fellowship (JP17J08829); JSPS Overseas Research Fellowships; the Key Research Program of Frontier Sciences, Chinese Academy of Sciences (CAS, grants QYZDJ-SSW-SLH057, QYZDJ-SSW-SYS008); the Leverhulme Trust Early Career Research Fellowship; the Max-Planck-Gesellschaft (MPG); the Max Planck Partner Group of the MPG and the CAS; the MEXT/JSPS KAKENHI (grants 18KK0090, JP18K13594, JP18K03656, JP18H03721, 18K03709, 18H01245, 25120007); the MIT International Science and Technology Initiatives (MISTI) Funds; the Ministry of Science and Technology (MOST) of Taiwan (105-2112-M-001-025-MY3, 106-2112-M-001-011, 106-2119-M-001-027, 107-2119-M-001-017, 107-2119-M-001-020, and 107-2119-M-110-005); the National Aeronautics and Space Administration (NASA, Fermi Guest Investigator grant 80NSSC17K0649); the National Institute of Natural Sciences (NINS) of Japan; the National Key Research and Development Program of China (grant 2016YFA0400704, 2016YFA0400702); the National Science Foundation (NSF, grants AST-0096454, AST-0352953, AST-0521233, AST-0705062, AST-0905844, AST-0922984, AST-1126433, AST-1140030, DGE-1144085, AST-1207704, AST-1207730, AST-1207752, MRI-1228509, OPP-1248097, AST-1310896, AST-1312651, AST-1337663, AST-1440254, AST-1555365, AST-1715061, AST-1614868, AST-1615796, AST-1716327, OISE-1743747, AST-1816420); the Natural Science Foundation of China (grants 11573051, 11633006, 11650110427, 10625314, 11721303, 11725312, 11873028, 11873073, U1531245, 11473010); the Natural Sciences and Engineering Research Council of Canada (NSERC, including a Discovery Grant and the NSERC Alexander Graham Bell Canada Graduate Scholarships-Doctoral Program); the National Youth Thousand Talents Program of China; the National Research Foundation of Korea (grant 2015-R1D1A1A01056807, the Global PhD Fellowship Grant: NRF-2015H1A2A1033752, and the Korea Research Fellowship Program: NRF-2015H1D3A1066561); the Netherlands Organization for Scientific Research (NWO) VICI award (grant 639.043.513) and Spinoza Prize SPI 78-409; the New Scientific Frontiers with Precision Radio Interferometry Fellowship awarded by the South African Radio Astronomy Observatory (SARAO), which is a facility of the National Research Foundation (NRF), an agency of the Department of Science and Technology (DST) of South Africa; the Onsala Space Observatory (OSO) national infrastructure, for the provisioning of its facilities/observational support (OSO receives funding through the Swedish Research Council under grant 2017-00648)the Perimeter Institute for Theoretical Physics (research at Perimeter Institute is supported by the Government of Canada through the Department of Innovation, Science and Economic Development Canada and by the Province of Ontario through the Ministry of Economic Development, Job Creation and Trade); the Russian Science Foundation (grant 17-12-01029); the Spanish Ministerio de Economía y Competitividad (grants AYA2015-63939-C2-1-P, AYA2016-80889-P); the State Agency for Research of the Spanish MCIU through the "Center of Excellence Severo Ochoa" award for the Instituto de Astrofísica de Andalucía (SEV-2017-0709); the Toray Science Foundation; the US Department of Energy (USDOE) through the Los Alamos National Laboratory (operated by Triad National Security, LLC, for the National Nuclear Security Administration of the USDOE (Contract 89233218CNA000001)); the Italian Ministero dell'Istruzione Università e Ricerca through the grant Progetti Premiali 2012-iALMA (CUP C52I13000140001); ALMA North America Development Fund; Chandra TM6-17006X. This work used the Extreme Science and Engineering Discovery Environment (XSEDE), supported by NSF grant ACI-1548562, and CyVerse, supported by NSF grants DBI-0735191, DBI-1265383, and DBI-1743442. XSEDE Stampede2 resource at TACC was allocated through TG-AST170024 and TG-AST080026N. XSEDE JetStream resource at PTI and TACC was allocated through AST170028. The simulations were performed in part on the SuperMUC cluster at the LRZ in Garching, on the LOEWE cluster in CSC in Frankfurt, and on the HazelHen cluster at the HLRS in Stuttgart. This research was enabled in part by support provided by Compute Ontario (http://computeontario.ca), Calcul Quebec (http://www.calculquebec.ca) and Compute Canada (http://www.computecanada.ca). We thank the staff at the participating observatories, correlation centers, and institutions for their enthusiastic support. This Letter makes use of the following ALMA data: ADS/JAO.ALMA#2016.1.01154.V. ALMA is a partnership of the European Southern Observatory (ESO; Europe, representing its member states), NSF, and National Institutes of Natural Sciences of Japan, together with National Research Council (Canada), Ministry of Science and Technology (MOST; Taiwan), Academia Sinica Institute of Astronomy and Astrophysics (ASIAA; Taiwan), and Korea Astronomy and Space Science Institute (KASI; Republic of Korea), in cooperation with the Republic of Chile. The Joint ALMA Observatory is operated by ESO, Associated Universities, Inc. (AUI)/NRAO, and the National Astronomical Observatory of Japan (NAOJ). The NRAO is a facility of the NSF operated under cooperative agreement by AUI. APEX is a collaboration between the Max-Planck-Institut für Radioastronomie (Germany), ESO, and the Onsala Space Observatory (Sweden). The SMA is a joint project between the SAO and ASIAA and is funded by the Smithsonian Institution and the Academia Sinica. The JCMT is operated by the East Asian Observatory on behalf of the NAOJ, ASIAA, and KASI, as well as the Ministry of Finance of China, Chinese Academy of Sciences, and the National Key R&D Program (No. 2017YFA0402700) of China. Additional funding support for the JCMT is provided by the Science and Technologies Facility Council (UK) and participating universities in the UK and Canada. The LMT project is a joint effort of the Instituto Nacional de Astrófisica, Óptica, y Electrónica (Mexico) and the Universit The IRAM 30-m telescope on Pico Veleta, Spain is operated by IRAM and supported by CNRS (Centre National de la Recherche Scientifique, France), MPG (Max-Planck-Gesellschaft, Germany) and IGN (Instituto Geográfico Nacional, Spain). The SMT is operated by the Arizona Radio Observatory, a part of the Steward Observatory of the University of Arizona, with financial support of operations from the State of Arizona and financial support for instrumentation development from the NSF. Partial SPT support is provided by the NSF Physics Frontier Center award (PHY-0114422) to the Kavli Institute of Cosmological Physics at the University of Chicago (USA), the Kavli Foundation, and the GBMF (GBMF-947). The SPT hydrogen maser was provided on loan from the GLT, courtesy of ASIAA. The SPT is supported by the National Science Foundation through grant PLR-1248097. Partial support is also provided by the NSF Physics Frontier Center grant PHY-1125897 to the Kavli Institute of Cosmological Physics at the University of Chicago, the Kavli Foundation and the Gordon and Betty Moore Foundation grant GBMF 947. This work made use of Global Millimeter VLBI Array (GMVA) data (project MM07B; PI: A. Marscher). The GMVA consists of telescopes operated by the MPIfR, IRAM, Onsala, Metsahovi, Yebes, the Korean VLBI Network, the Green Bank Observatory and the VLBA. The VLBA is an instrument of the NRAO, which is a facility of the National Science Foundation operated by Associated Universities, Inc. This research has made use of data obtained with 12 radio telescopes from the East Asian VLBI Network (EAVN): 2 stations from the Chinese VLBI Network (the Tianma radio telescope operated by Shanghai Astronomical Observatory of Chinese Academy of Sciences (CAS) and the Nanshan radio telescope operated by Xinjiang Astronomical Observatory of CAS), 2 stations from the Japanese VLBI Network (the Hitachi station operated by Ibaraki University and the Kashima station operated by the National Institute of Information and Communications Technology), the Korean VLBI Network operated by the Korea Astronomy and Space Science Institute (KASI), the VLBI Exploration of Radio Astrometry (VERA) operated by the National Astronomical Observatory of Japan (NAOJ), and the Nobeyama 45-meter radio telescope operated by NAOJ. We are grateful to KVN staff who helped to operate the array and to correlate the data for quasi-simulataneous multi-wavelength observations of M87. The KVN is a facility operated by KASI. The KVN operations are supported by KREONET (Korea Research Environment Open NETwork), which is managed and operated by KISTI (Korea Institute of Science and Technology Information). The EHTC has received generous donations of FPGA chips from Xilinx Inc., under the Xilinx University Program. The EHTC has benefited from technology shared under open-source license by the Collaboration for Astronomy Signal Processing and Electronics Research (CASPER). The EHT project is grateful to T4Science and Microsemi for their assistance with Hydrogen Masers. This research has made use of NASA's Astrophysics Data System. We gratefully acknowledge the support provided by the extended staff of the ALMA, both from the inception of the ALMA Phasing Project through the observational campaigns of 2017 and 2018. We would like to thank A. Deller and W. Brisken for EHT-specific support with the use of DiFX. We acknowledge the significance that Maunakea, where the SMA and JCMT EHT stations are located, has for the indigenous Hawaiian people.

*Facility:* EHT.

*Software:* `eht-imaging`, `DIFMAP`, `SMILI`.

## Appendix A
## Regularizer Definitions

Here we give definitions for the regularization terms used in the RML methods that we employ in this Letter. In particular, these terms have normalization factors applied that differ among the imaging software libraries (`eht-imaging` and `SMILI`). The normalization factors are generally selected so that the associated regularization hyperparameter is of order unity, and to ensure that the regularization strength is independent of, e.g., the chosen pixel size and FOV of the reconstructed image.

*Total Flux Density Regularizer*: the total flux density regularizer favors reconstructed images that have total flux density near a specified value $F$:

$$S_{\rm tot\ flux} = -\frac{1}{\zeta}\left(\sum_i I_i - F\right)^2. \tag{24}$$

In `eht-imaging`, the normalization factor is

$$\zeta = F^2. \tag{25}$$

In `SMILI`, the normalization factor is

$$\zeta = (f_{\rm F} F)^2, \tag{26}$$

where $f_{\rm F}$ is the target fractional error and is fixed to be $10^{-2}$ for imaging in this Letter.

*Relative Entropy*: entropy (MEM) favors pixel-to-pixel similarity to a "prior image" with pixel values $P_i$ (in this work, the prior is always taken to be a circular Gaussian with a variable FWHM). The corresponding regularization term is

$$S_{\rm MEM} = -\frac{1}{\zeta}\sum_i I_i \log\left(\frac{I_i}{P_i}\right). \tag{27}$$

In `eht-imaging`, the normalization factor $\zeta$ for MEM is

$$\zeta = F, \tag{28}$$

where $F$ is the user-specified target total flux density. In this Letter, `SMILI` reconstructions do not use MEM regularization.

*$\ell_1$ Norm*: the $\ell_1$ norm favors image sparsity. In `eht-imaging`, it takes the form

$$S_{\ell_1} = -\frac{1}{\zeta}\sum_i |I_i|. \tag{29}$$

where the normalization factor $\zeta$ is simply

$$\zeta = F. \tag{30}$$

In `SMILI`, a generalized weighted $\ell_1$ norm ($\ell_1^{\rm w}$) is used, given by

$$S_{\ell_1^{\rm w}} = -\frac{1}{\zeta}\sum_i \frac{|I_i|}{|P_i| + \epsilon_s}, \tag{31}$$

where $P_i$ is a specified prior image, and $\epsilon_s$ is fixed to be $10^{-12}$ Jy/pixel. The total scaling factor $\zeta$ is given by

$$\zeta = \sum_i \frac{|P_i|}{|P_i| + \epsilon_s}. \tag{32}$$

For the absolute value operator $|...|$, `SMILI` adopts a smoothed-approximation operator, which is differentiable everywhere, given by

$$|x| \approx \sqrt{x^2 + \epsilon_a}, \tag{33}$$

where $\epsilon_a$ is fixed to be $10^{-12}$ Jy/pixel. We use a circular Gaussian with a specified FWFM for the prior image.

*Total Variation*: TV favors piecewise-smooth images with flat regions separated by sharp edges

$$S_{\rm TV} = -\frac{1}{\zeta}\sum_{\ell,m}[(I_{\ell+1,m} - I_{\ell,m})^2 + (I_{\ell,m+1} - I_{\ell,m})^2]^{1/2}. \tag{34}$$

In this equation, the two sums are taken over the two image dimensions and the image pixels $I_{\ell,m}$ are now indexed by their position $(l, m)$ in a 2D grid. This regularizer is not differentiable everywhere. In `eht-imaging`, the normalization factor $\zeta$ for TV is

$$\zeta = F\frac{\Delta\theta}{\Theta_{\rm beam}}, \tag{35}$$

where $\Delta\theta$ is the pixel size and $\Theta_{\rm beam}$ is the beam size.

On the other hand, `SMILI` adopts a smoothed TV that is differentiable everywhere:

$$S_{\rm TV} = -\frac{1}{\zeta}\sum_{\ell,m}[(I_{\ell+1,m} - I_{\ell,m})^2 + (I_{\ell,m+1} - I_{\ell,m})^2 + \epsilon]^{1/2}, \tag{36}$$

where $\epsilon$ is set to be $10^{-12}$ Jy/pixel for imaging in this Letter. The normalization factor for `SMILI` is

$$\zeta = F. \tag{37}$$

*Total Squared Variation*: TSV favors images with smooth edges:

$$S_{\rm TSV} = -\frac{1}{\zeta}\sum_{\ell,m}[(I_{\ell+1,m} - I_{\ell,m})^2 + (I_{\ell,m+1} - I_{\ell,m})^2]. \tag{38}$$

In `eht-imaging`, the normalization factor for TSV is

$$\zeta = F^2\left(\frac{\Delta\theta}{\Theta_{\rm beam}}\right)^4. \tag{39}$$

In `SMILI`, the normalization factor for TSV is

$$\zeta = F^2/N_{\rm pix}, \tag{40}$$

where $N_{\rm pix}$ is the total number of pixels in the reconstructed image.

## Appendix B
## Derivation of Constraints on the Total Compact Flux Density of M87

In this section, we use a combination of constraints directly accessible from EHT observations on long baselines (Section B.1) in combination with constraints from multi-wavelength observations (Section B.2) to constrain the compact ($\lesssim$100 $\mu$as) 230 GHz flux density of M87 during 2017 April.

### B.1. Constraints from EHT Observations

As discussed in Section 4.3, there are two relevant estimates of the total flux density of M87: the total flux density seen in the core on arcsecond scales, $F_{\rm tot} \approx 1.2$ Jy, and the total compact flux density, $F_{\rm cpct} \leqslant F_{\rm tot}$, which is the integrated flux density for a reconstructed image with an FOV limited to only $\sim$100 $\mu$as. Intra-site EHT baselines are sensitive to the former, while inter-site baselines are sensitive to the latter.

The shortest inter-site EHT baseline is SMT–LMT, which has a fringe spacing of (139–166) $\mu$as for M87. However, using this baseline to estimate $F_{\rm cpct}$ requires assumptions about the compact source structure and the amplitude calibration accuracy. In particular, while SMT gains are well constrained, the use of provisional equipment at the LMT led to severe uncertainty and variability in LMT gains, as is evident by anomalously large variations in amplitudes on LMT baselines (see Figure 2 and Paper III). Nevertheless, we can still obtain useful limits that are independent of source structure. To proceed, we will consider four separate constraints, each of which requires different assumptions about the data and image. Moreover, we will show that constraints on $F_{\rm cpct}$ are degenerate with constraints on the overall source size. For reference, the values of $F_{\rm tot}$ estimated using ALMA interferometric data are 1.13, 1.14, 1.17, and 1.21 Jy on our four observing days, April 5, 6, 10, and 11. Because of the short spacing combined with the relatively small difference in $F_{\rm tot}$ for adjacent days, we will analyze each pair of consecutive observations separately, allowing for the possibility of source evolution in the four-day observing gap.

In the following discussion, we will denote complex station gains by, e.g., $g_{\rm LMT}$, and measured complex visibilities by, e.g., $V_{\rm LMT-SMT}$. Apart from thermal noise, the measured visibilities are then related to the true visibilities $\mathcal{V}_{\rm LMT-SMT}$ via

$$V_{\rm LMT-SMT} = g_{\rm LMT} g^{*}_{\rm SMT} \mathcal{V}_{\rm LMT-SMT}. \tag{41}$$

In expressions using visibility amplitudes, we will assume that these are suitably debiased for the contribution of thermal noise (Equation (6)).

*Constraint 1.* Visibility amplitudes are maximum at zero baseline length. In addition, a priori calibration may overestimate station sensitivity (e.g., if there are unaccounted pointing errors), but it should not underestimate sensitivity; i.e., $|g_i| \leqslant 1$ for every station $i$, and every visibility amplitude should only underestimate the true amplitude (e.g., $|V_{\rm LMT-SMT}| \leqslant |\mathcal{V}_{\rm LMT-SMT}|$), apart from thermal noise. Combining these properties, we derive a lower limit to $F_{\rm cpct}$ by identifying the maximum LMT–SMT visibility amplitude measured for each of the four observing days: 0.34, 0.23, 0.43, and 0.43 Jy, respectively. Thus, $F_{\rm cpct} > 0.34$ Jy on the first two days and $F_{\rm cpct} > 0.43$ Jy on the latter two days.

These constraints can be strengthened under the additional assumption that the LMT–SMT baseline partially resolves the compact emission region. In this case, it is convenient to express the visibility amplitudes in terms of an equivalent Gaussian visibility function,

$$V_{\rm G}(\boldsymbol{u};\, I_0,\, \theta) = I_0 e^{-\frac{(\pi\theta|\boldsymbol{u}|)^2}{4\ln 2}}. \tag{42}$$

In this expression, $I_0$ is the total flux density of the Gaussian source, and $\theta$ is its FWHM in radians. Visibility amplitudes will match this generic Gaussian form until a source is at least moderately resolved, providing a characteristic image FWHM that does not make specific assumptions about image structure.[120] Because LMT–SMT amplitudes are up to ∼20%–40% of $F_{\rm tot}$ (and $F_{\rm cpct}$ may be considerably less than $F_{\rm tot}$), we expect LMT–SMT amplitudes to lie within this regime.

Combining these properties, we obtain a modified constraint on total compact flux density that is dependent upon the compact source size, $\theta_{\rm cpct}$. For any LMT–SMT visibility amplitude on a baseline $\boldsymbol{u}$,

$$F_{\rm cpct} > |V_{\rm LMT-SMT}| e^{\frac{(\pi|\boldsymbol{u}|\theta_{\rm cpct})^2}{4\ln 2}}. \tag{43}$$

*Constraint 2.* Our second constraint comes from the observation that the visibility amplitudes fall steeply with baseline length on short baselines. By comparing the relative visibility amplitudes of simultaneous ALMA–LMT and SMT–LMT measurements, we obtain an estimate of the relative source visibility amplitudes that only depends on the station gains for ALMA and the SMT:

$$\frac{|V_{\rm ALMA-LMT}|}{|V_{\rm SMT-LMT}|} = \frac{|g_{\rm ALMA}|}{|g_{\rm SMT}|} \frac{|\mathcal{V}_{\rm ALMA-LMT}|}{|\mathcal{V}_{\rm SMT-LMT}|}. \tag{44}$$

Both the SMT and ALMA are stable and have well-characterized performance. Moreover, ALMA visibility amplitudes have absolute calibration to within a few percent via "network calibration," using the short baseline to APEX. Consequently, we can make the approximation

$$\frac{|V_{\rm ALMA-LMT}|}{|V_{\rm SMT-LMT}|} \approx \frac{|\mathcal{V}_{\rm ALMA-LMT}|}{|\mathcal{V}_{\rm SMT-LMT}|}. \tag{45}$$

For M87, we find values for $|V_{\rm ALMA-LMT}|/|V_{\rm SMT-LMT}|$ ranging from 0.18 to 0.34 on the first two days, and from 0.17 to 0.28 on the last two days.

For each measurement, we can estimate a minimum image size by finding the equivalent Gaussian that achieves the same quotient on the corresponding baselines. This criterion is motivated by the property that long baselines will tend to measure more flux density than extrapolated by the equivalent Gaussian on short baselines, reflecting small-scale power from image substructure. This criterion also makes the simplifying assumption that the source has isotropic size; this assumption is both supported by our data (see Section 3.2) and is not especially problematic because the ALMA–LMT and SMT–LMT baselines sample similar orientation angles (see Figure 1).

We thereby estimate a minimum source size of $\theta \geqslant 38\ \mu$as for the first two days, and $\theta \geqslant 41\ \mu$as for the second two days. Note that these estimates are independent of the total (and total compact) flux density. These minimum size estimates can be applied to tighten the limits from Constraint 1, via Equation (43). Using these minimum source size estimates, we obtain $F_{\rm cpct} > 0.42$ Jy on the first two days and $F_{\rm cpct} > 0.56$ Jy on the last two days.

*Constraint 3.* For our third constraint, we again use the properties that LMT–SMT visibility amplitudes fall within the universal regime described via a Gaussian visibility amplitude function, and that LMT–SMT measurements will only under-predict true visibility amplitudes. Combining these properties, each measured amplitude $|V_{\rm LMT-SMT}|$ on an associated baseline $\boldsymbol{u}$ determines an upper limit on the image size $\theta_{\rm cpct}$. This upper limit occurs when the required total compact flux density must exceed $F_{\rm tot}$, giving

$$\theta_{\rm cpct} \leqslant \frac{2\sqrt{\ln 2}}{\pi|\boldsymbol{u}|} \sqrt{\ln \frac{F_{\rm tot}}{|V_{\rm LMT-SMT}|}}. \tag{46}$$

We thereby find $\theta \leqslant 82\ \mu$as for the first two days, and $\theta \leqslant 77\ \mu$as for the last two days.

*Constraint 4.* Our final constraint utilizes information from different baselines that sample nearly the same vector baseline $\boldsymbol{u}$, though not simultaneously. Specifically, the SMT–PV and LMT–PV tracks cross (see Figure 1), sampling a single baseline near $(u,\ v) = (5.9,\ 0.9)\,{\rm G}\lambda$ at times $t_1$ and $t_2$, respectively. Under the assumption of stable source structure, the true source visibility amplitudes at the crossing point must

[120] Formally, the corresponding FWHM is proportional to the second central moment of the image, projected along the baseline direction (see, e.g., Johnson et al. 2018).

**Table 8**
SMT–LMT Amplitudes before and after Applying a Correction Due to Crossing Baseline Tracks for Each Observing Day

| | Hour UT | Baseline | $u$ (G$\lambda$) | $v$ (G$\lambda$) | $\|(u, v)\|$ (G$\lambda$) | Amp (mJy) | $\sigma$ (mJy) |
|---|---|---|---|---|---|---|---|
| April 5 | 4.6 | SMT–PV | −6.02 | −0.77 | 6.07 | 103.7 | 8.7 |
| | 2.6 | LMT–PV | −6.24 | −1.31 | 6.37 | 18.4 | 3.4 |
| | 2.6 | SMT–LMT | −0.08 | 1.25 | 1.25 | 61.9 | 5.7 |
| | | [LMT SEFD Correction Factor: 31.7] | | | | **348.6** | **74.5** |
| April 6 | 5.1 | SMT–PV | −5.60 | −0.97 | 5.68 | 140.1 | 10.7 |
| | 3.6 | LMT–PV | −5.98 | −1.69 | 6.22 | 62.6 | 4.2 |
| | 3.6 | SMT–LMT | −0.38 | 1.23 | 1.29 | 174.0 | 6.9 |
| | | [LMT SEFD Correction Factor: 5.0] | | | | **389.6** | **50.1** |
| April 10 | 4.3 | SMT–PV | −6.03 | −0.76 | 6.08 | 118.7 | 9.5 |
| | 2.2 | LMT–PV | −6.23 | −1.26 | 6.36 | 16.1 | 6.8 |
| | 2.2 | SMT–LMT | −0.04 | 1.25 | 1.25 | 55.2 | 12.9 |
| | | [LMT SEFD Correction Factor: 54.6] | | | | **408.2** | **164.5** |
| April 11 | 4.3 | SMT–PV | −5.95 | −0.81 | 6.00 | 111.7 | 7.1 |
| | 2.4 | LMT–PV | −6.23 | −1.38 | 6.38 | 71.9 | 7.5 |
| | 2.4 | SMT–LMT | −0.14 | 1.25 | 1.26 | 247.1 | 10.6 |
| | | [LMT SEFD Correction Factor: 2.4] | | | | **383.8** | **57.9** |

**Note.** Non-simultaneous SMT–PV and LMT–PV visibilities at a crossing baseline location (Near 6 G$\lambda$; see Figure 1) determine the LMT SEFD at the time of the LMT–PV measurement, yielding correction factors on a priori estimates of LMT SEFDs ranging from 2.4 to 54.6. Substituting these measurements into Equation (48) gives an estimate of the LMT–SMT visibility amplitude, shown as bold values in the table. Stated uncertainties on original measurements include only thermal noise, but those on the corrected LMT–SMT amplitudes also account for uncertainties in the LMT and SMT SEFDs. On all four days, the corrected LMT–SMT visibility amplitude is approximately 400 mJy. Note that these corrections require a substantial adjustment to the LMT SEFD. In Section 8.1 we show that these large gain corrections are consistent with those recovered through imaging.

be equal, providing an estimate of the relative station gains at their corresponding times:

$$\frac{|V_{\mathrm{SMT-PV}}^{t_1}|}{|V_{\mathrm{LMT-PV}}^{t_2}|} = \frac{|\mathcal{V}_{\mathrm{SMT-PV}}^{t_1}|}{|\mathcal{V}_{\mathrm{LMT-PV}}^{t_2}|}\left|\frac{g_{\mathrm{SMT}}^{t_1} g_{\mathrm{PV}}^{t_1}}{g_{\mathrm{LMT}}^{t_2} g_{\mathrm{PV}}^{t_2}}\right| \approx \left|\frac{g_{\mathrm{SMT}}^{t_1}}{g_{\mathrm{LMT}}^{t_2}}\right|. \quad (47)$$

The approximation makes the assumptions of stable source structure between $t_1$ and $t_2$, negligible change in the visibility function on the two baselines sampled near the crossing point, and a stable gain amplitude between $t_1$ and $t_2$ at PV.

If we further assume that the SMT gain is near unity (the a priori error budget estimates $|g_{\mathrm{SMT}}| = 1.00 \pm 0.04$), then Equation (47) provides an estimate of the LMT gain amplitude at time $t_2$. We can then apply this estimate to a simultaneous measurement of the SMT–LMT amplitude at time $t_2$ to estimate $|\mathcal{V}_{\mathrm{SMT-LMT}}|$:

$$\begin{aligned} |\mathcal{V}_{\mathrm{LMT-SMT}}^{t_2}| &= |g_{\mathrm{SMT}}^{t_2}|^{-1}|g_{\mathrm{LMT}}^{t_2}|^{-1}|V_{\mathrm{LMT-SMT}}^{t_2}| \\ &\approx \frac{|V_{\mathrm{SMT-PV}}^{t_1}|}{|g_{\mathrm{SMT}}^{t_1}||g_{\mathrm{SMT}}^{t_2}||V_{\mathrm{LMT-PV}}^{t_2}|}|V_{\mathrm{LMT-SMT}}^{t_2}| \\ &\approx \frac{|V_{\mathrm{SMT-PV}}^{t_1}||V_{\mathrm{LMT-SMT}}^{t_2}|}{|V_{\mathrm{LMT-PV}}^{t_2}|}. \end{aligned} \quad (48)$$

However, because the gain of the LMT is substantially time variable, extrapolating the gain estimated in Equation (47) to other times (even nearby times) is not reliable. To summarize, this final approximation assumes the following: that the SMT–PV measurement at time $t_1$ samples a baseline close to the LMT–PV measurement at $t_2$, that the source visibility on that baseline is stable between $t_1$ and $t_2$, that the PV gain amplitude is stable between $t_1$ and $t_2$, and that the SMT gain amplitude is near unity at both times.

We derive an uncertainty on $|\mathcal{V}_{\mathrm{LMT-SMT}}^{t_2}|$ by summing errors in quadrature from all these effects in addition to the nominal thermal noise on each of the three visibility amplitudes. We assume a conservative uncertainty in the relative PV gain amplitude between $t_1$ and $t_2$ of 10% (the a priori error budget for absolute gain calibration of PV is 5%), and we assume 3.5% uncertainty in the SMT gain amplitude (added independently for $t_1$ and $t_2$). Because we require mutual visibility of SMT and LMT at $t_2$, the nearest suitable points are not necessarily near the crossing point.[121] Consequently, the nearest suitable SMT–PV and LMT–PV baselines are displaced by 0.5-0.8 G$\lambda$. Empirically, we estimate up to a 20 mJy difference from this displacement (which occurs at $|\boldsymbol{u}| \approx 6$ G$\lambda$); the resultant uncertainty on the estimate of $|\mathcal{V}_{\mathrm{LMT-SMT}}^{t_2}|$ depends on this amplitude as a fraction of the SMT–PV and LMT–PV amplitudes.

Table 8 shows a triplet of visibility amplitudes and corresponding estimate of $|\mathcal{V}_{\mathrm{LMT-SMT}}^{t_2}|$ for each of the four observing days. With no LMT gain correction, the measurements range from 55–247 mJy, while after the LMT gain

[121] The LMT–PV baseline crosses the SMT–PV track when the elevation of M87 is only 11°.5 for the SMT. However, for the SMT, M87 rises behind the Large Binocular Telescope, so the minimum observable elevation is ~25°. Consequently, the need for simultaneous SMT–LMT and LMT–PV visibilities necessitates using a measured LMT–PV visibility on a baseline that is slightly displaced from the crossing point.

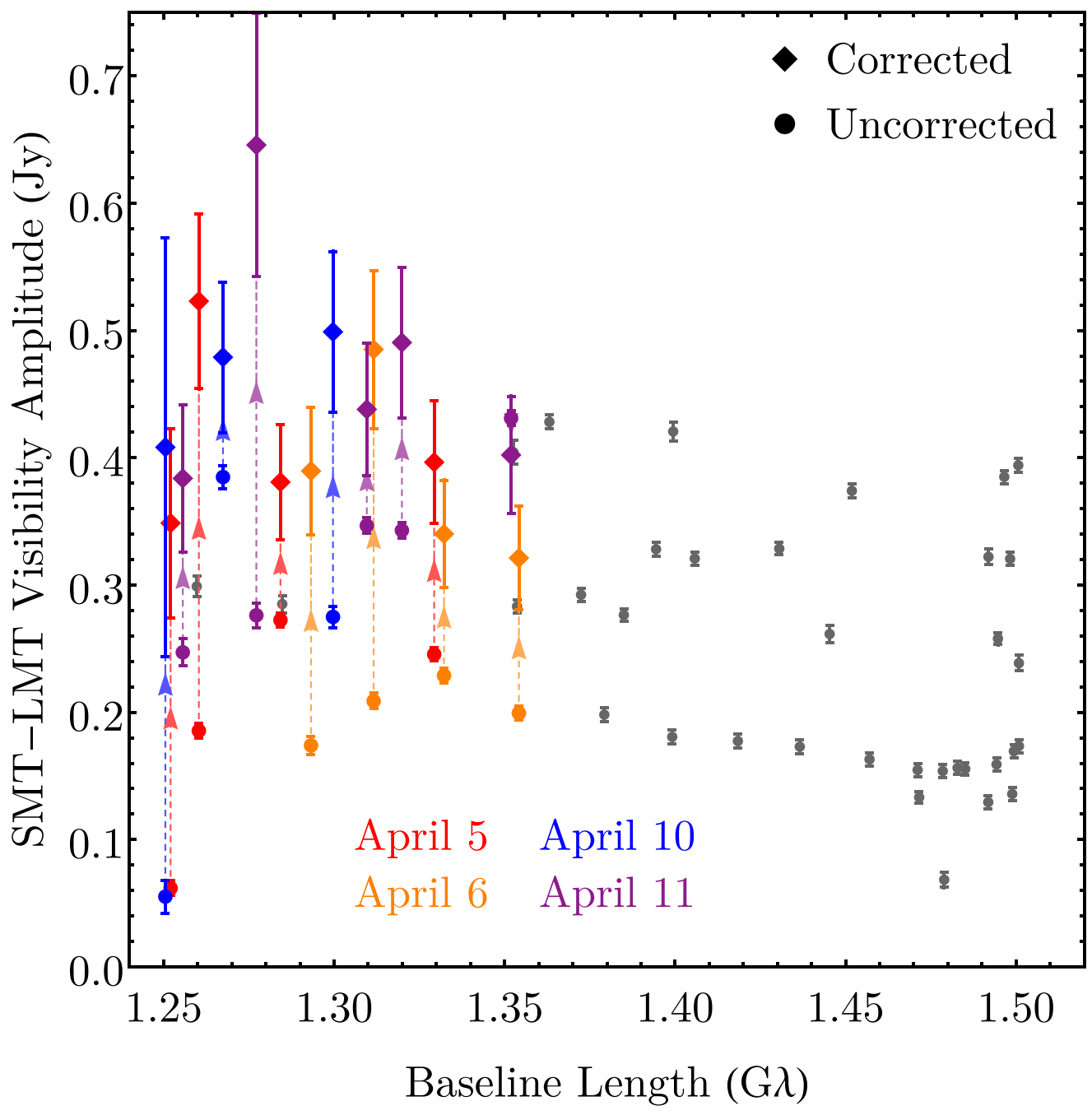

**Figure 28.** M87 visibility amplitudes measured on the SMT–LMT baseline, as a function of baseline length. Colored circles show measurements and their associated thermal noise ($\pm 1\sigma$) after a priori calibration; colored diamonds show the measurements after the SMT–PV/LMT–PV crossing track correction. Uncertainties in corrected amplitudes also account for propagated a priori calibration uncertainties in the derived correction factor. Correction factors are only applied if the nearest SMT–PV/LMT–PV baseline pair are located within 1 G$\lambda$. SMT–LMT amplitudes with no associated pair meeting this criterion are colored gray and have plotted uncertainties that only account for thermal noise. All uncorrected amplitudes (circles) have scatter far exceeding their indicated thermal noise, demonstrating that amplitude calibration uncertainties are the dominant source of error.

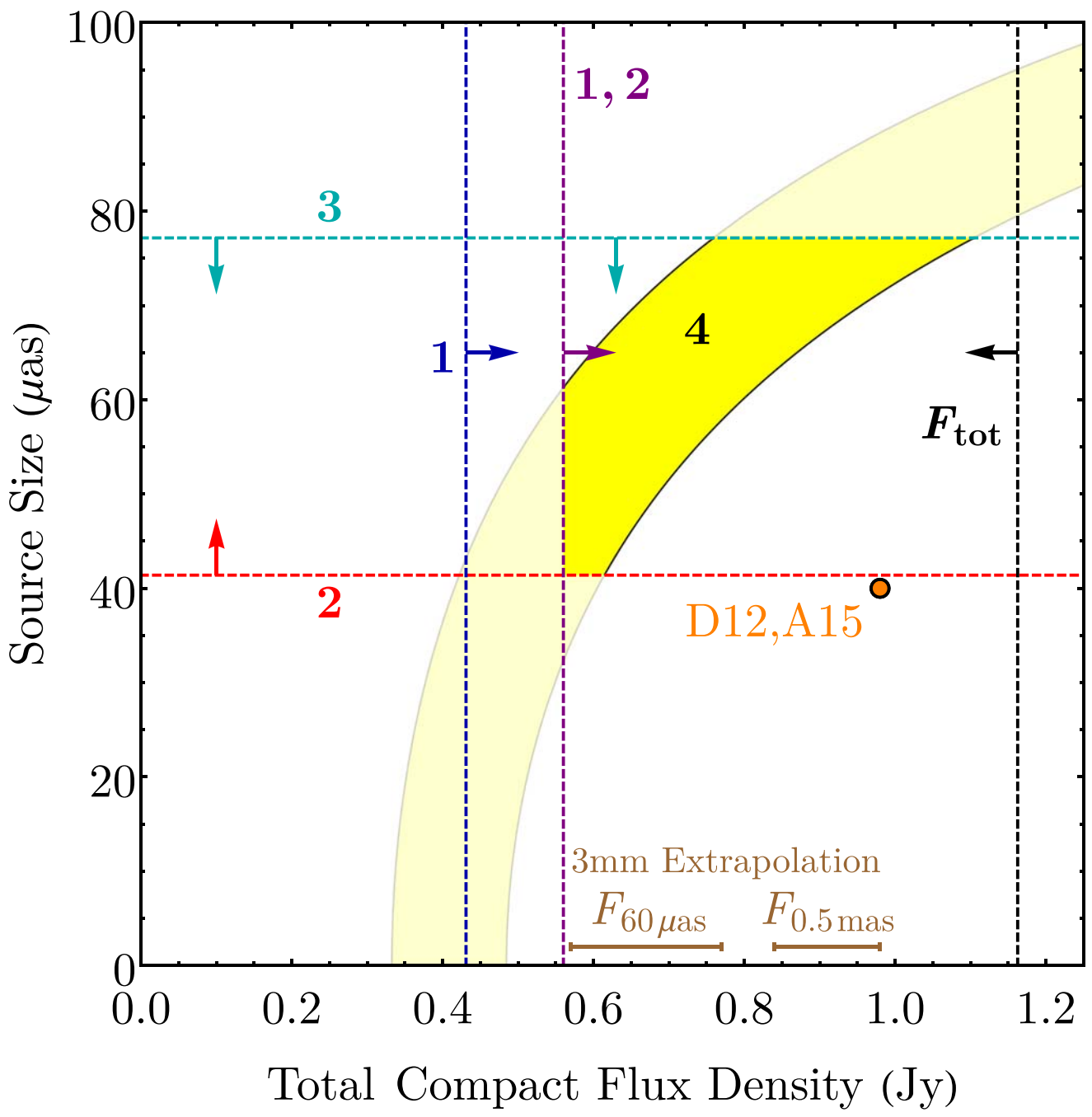

**Figure 29.** Joint constraints on the size and total compact flux density of 1.3 mm emission in M87. Numbered constraints (described in Appendix B.1) correspond to: (1) the maximum LMT–SMT amplitude, (2) the LMT–SMT/LMT–ALMA amplitude ratio, (3) LMT–SMT amplitudes coupled with the requirement $F_{\rm cpct} \leqslant F_{\rm tot}$, and (4) the SMT–LMT amplitude after LMT gain correction using the LMT–PV and SMT–PV crossing tracks in the $(u, v)$ plane. Estimates on $F_{\rm cpct}$ and $F_{\rm tot}$ (with $\pm 1\sigma$ uncertainties) from extrapolating VLBI observations at 3 mm are shown in brown (see Appendix B.2). The orange circle shows the estimates from earlier EHT observations by Doeleman et al. (2012) and Akiyama et al. (2015).

correction the measurements range from 349–408 mJy. The corrections also all raise the SMT–LMT amplitude, as is expected. Some of these points represent the most significant LMT amplitude losses in the entire experiment. However, Figure 28 shows the corrected SMT–LMT amplitudes for all points with a suitable crossing track correction, demonstrating that the corrected amplitudes are broadly consistent between points with large and small derived LMT correction factors.

In summary, constraints 1–3 lead to the conclusion that $0.42 \leqslant F_{\rm cpct} \leqslant 1.14$ Jy on April 5 and 6, and that $0.56 \leqslant F_{\rm cpct} \leqslant 1.21$ Jy on April 10 and 11. We also estimate that the compact source size has an equivalent Gaussian FWHM between 38 and 82 $\mu$as on the first two days and between 41 and 77 $\mu$as on the last two days. Constraint 4 provides an absolute estimate of the LMT–SMT amplitude at a single point for each day, which allows us to provide a joint constraint on source size and $F_{\rm cpct}$. Figure 29 summarizes these constraints. Combining these constraints and marginalizing over the unknown source size, we obtain $F_{\rm cpct} = 0.64^{+0.39}_{-0.08}$ Jy.

### *B.2. Constraints from Quasi-simultaneous Multi-wavelength Observations*

In addition to the non-imaging flux density constraints derived from EHT data (Section B.1), we can also utilize VLBI observations at longer wavelengths to estimate expected properties at 1.3 mm. Because both the compact and extended emission from M87 are variable, these observations must be quasi-simultaneous with the EHT observations to give precise constraints. Our approach assumes that the total compact emission of M87 at 230 GHz $F_{\rm tot}$ originates from the combination of (i) a compact emitter with unknown flux density $F_{\rm cpct}$ and spectral index at 230 GHz, and (ii) a larger and diffuse (milliarcsecond-scale) jet, which can be characterized by its total flux density $F_{\rm jet} = F_{\rm tot} - F_{\rm cpct}$ and a steep spectral index of $\alpha_{\rm jet} \sim -(0.7–1.0)$, as measured at cm- and mm-wavelengths (with typical errors of ~0.3; see Hovatta et al. 2014; Hada et al. 2016). Then, by measuring $F_{\rm jet}$ and $F_{\rm tot}$ at lower frequencies and extrapolating them to 230 GHz using a power-law model, we can estimate the total compact flux density $F_{\rm cpct}$ at 230 GHz: $F_{\rm cpct} = F_{\rm tot} - F_{\rm jet}$.

**Table 9**
Quasi-simultaneous Lower-frequency VLBI Flux Density Measurements of M87 within a 0.5 × 0.5 mas$^2$ Box around the Peak Flux Density

| Array | Freq. (GHz) | Flux (Jy) | Epoch |
|---|---|---|---|
| EAVN | 22 | 1.25 ± 0.13 | 2017 April 3 |
| | | 1.32 ± 0.13 | 2017 April 17 |
| EAVN | 43 | 1.21 ± 0.12 | 2017 April 4 |
| | | 1.10 ± 0.11 | 2017 April 9 |
| | | 1.18 ± 0.12 | 2017 April 14 |
| | | 1.12 ± 0.11 | 2017 April 18 |
| KVN | 86 | 1.12 ± 0.17 | 2017 April 19 |
| | 129 | 0.91 ± 0.27 | 2017 April 19 |

In parallel with the EHT campaign, we densely monitored M87 at 22 and 43 GHz with the East Asian VLBI Network (EAVN; see Wajima et al. 2016; Hada et al. 2017; and An et al. 2018 for more details), which included 12 stations from Japan, Korea, and China (Mizusawa, Iriki, Ogasawara, Ishigaki, Hitachi, Kashima, Nobeyama, Yonsei, Ulsan, Tamna, Tianma, and Nanshan). For each epoch, a continuous 5–7 hr track was observed. The angular resolution was as fine as 0.55 mas at 22 GHz and 0.63 mas at 43 GHz. At both frequencies, the total flux densities of the nuclear region (i.e., $F_{\rm tot} = F_{\rm jet} + F_{\rm cpct}$) were measured using a large window of 0.5 × 0.5 mas$^2$ around the peak of the image and summing up the flux densities of the CLEAN components within the window. From this, we obtain total flux densities of $F_{\rm tot,22} \approx 1.3$ Jy and $F_{\rm tot,43} \approx 1.1$–1.2 Jy at 22 and 43 GHz, respectively.

We also make use of quasi-simultaneously obtained data sets of M87 from Korean VLBI Network (KVN) observations at 22, 43, 86, and 129 GHz on 2017 April 19 (see Lee et al. 2016 and Kim et al. 2018b for the monitoring program, details of data analysis, typical angular resolutions, and discussions about the extended jet flux). At 22 and 43 GHz, the correlated flux densities were comparable to the results of the EAVN observations. At 86 and 129 GHz, we adopted the same window of 0.5 × 0.5 mas$^2$ and obtained $F_{\rm tot,86} \approx 1.1$ Jy and $F_{\rm tot,129} \approx 0.9$ Jy, respectively. These measurements and their uncertainties are summarized in Table 9.

We fit a power-law model to the milliarcsecond-scale flux densities (i.e., $F_{\rm tot} \propto \nu^{+\alpha_{\rm tot}}$) and find $\alpha_{\rm tot} = -(0.14 \pm 0.04)$. By extrapolating this model to 230 GHz, we estimate $F_{\rm tot,230} = 0.91 \pm 0.07$ Jy. While this value is slightly lower than the ALMA interferometric measurements during the EHT observations, it only includes the total flux density within the inner 0.5 × 0.5 mas$^2$, which is nearly three orders of magnitude below the resolution limit of ALMA.

Next, we estimate the compact flux density $F_{\rm cpct,86}$ at 86 GHz using previous GMVA observations of M87 (Kim et al. 2018a). To proceed, we compute the flux densities within a much smaller window of 60 × 60 $\mu$as$^2$ around the peak of the GMVA images from 2004 to 2015, but excluding the 2009 epoch when the core of the jet showed exceptionally high flux densities. The values of $F_{\rm cpct,86}$ are relatively stable from epoch to epoch, and therefore we adopt a value of $F_{\rm cpct,86} = 0.55 \pm 0.06$ Jy, where the mean and uncertainty correspond to the mean and standard deviation of the measured flux densities. We can then estimate $F_{\rm jet}$ at 86 GHz in 2017 April as $F_{\rm jet,86} = F_{\rm tot,86} - F_{\rm cpct,86}$. By propagating the errors in $F_{\rm tot,86}$ and $F_{\rm cpct,86}$, we obtain $F_{\rm jet,86} = 0.57 \pm 0.18$ Jy at 86 GHz.

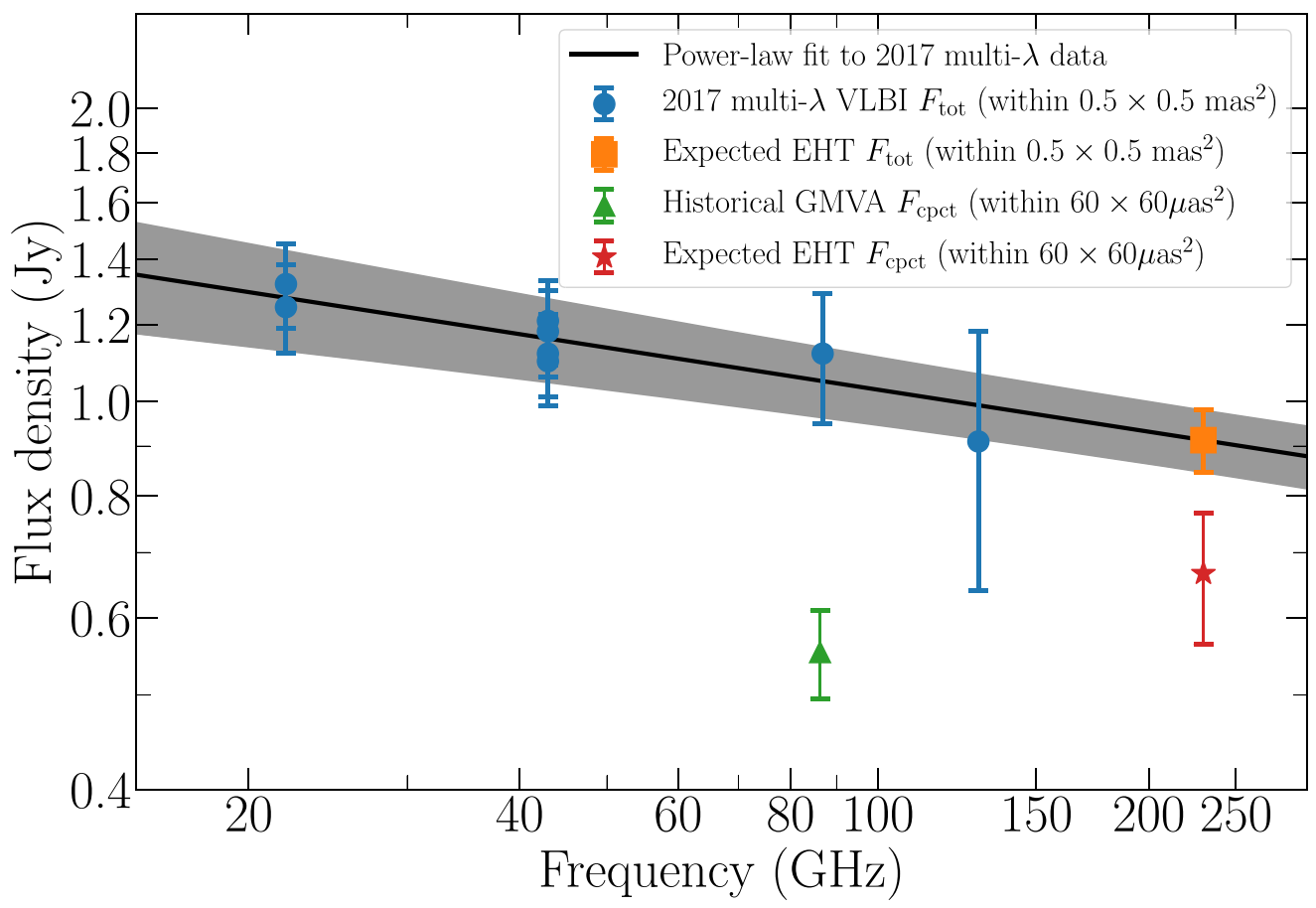

**Figure 30.** Multi-wavelength VLBI spectrum of M87 in 2017 and expected flux densities at 230 GHz (see Section B.2 for details). The black line and shaded region gives the model and its ±1$\sigma$ uncertainties for $F_{\rm tot}$.

Finally, we compute the expected $F_{\rm cpct,230}$ by estimating $F_{\rm jet,230}$ and subtracting it from the previously estimated $F_{\rm tot,230} = 0.91 \pm 0.07$ Jy. As explained above, we assume that $F_{\rm jet,230}$ follows a power law that can be extrapolated up to 230 GHz (i.e., $F_{\rm jet,230} \propto \nu^{+\alpha_{\rm jet}}$). For a range of $\alpha_{\rm jet} = -(0.7$–$1.0)$, we find $F_{\rm jet,230} = 0.25 \pm 0.08$ Jy. This value gives $F_{\rm cpct,230} = 0.67 \pm 0.10$ Jy within the central ∼60 × 60 $\mu$as$^2$. Figure 30 summarizes the spectrum of M87 at lower frequencies and the expected flux densities at 230 GHz. We emphasize that this estimate for $F_{\rm cpct,230}$ represents a characteristic value, and that the exact value may have varied slightly over the EHT observing window.

We note that this value for $F_{\rm cpct,230}$ is consistent with the constraints derived in Section B.1 using EHT measurements. Combining them, we obtain $F_{\rm cpct,230} = 0.66^{+0.16}_{-0.10}$ Jy.

## Appendix C
## Synthetic Data

To test and calibrate the parameters that we use in our final image reconstructions from imaging parameter surveys (Section 6), we generate synthetic data from different geometrical models and GRMHD simulations using two independent pipelines (as introduced in Section 6.1). We consider four geometric source models with parameters chosen such that their visibility data on EHT baselines match salient features of the observations, in particular the visibility nulls at ∼3.4 and ∼8.3 G$\lambda$.

### *C.1. Geometric Models*

We use four geometric models with different compact structure (displayed in Figure 5). The model parameters in each case were chosen such that their visibility data on EHT baselines match salient features of the observations (e.g., the visibility null at ∼3.4 G$\lambda$).

All of our models have 0.6 Jy of flux density in a compact component and 0.6 Jy in an extended jet that is heavily resolved by EHT inter-site baselines. The large-scale jet model is common to all images and consists of three elliptical Gaussian components aligned to images of the M87 jet at 3 mm (e.g., Kim et al. 2018a). The jet parameters are displayed in Table 10.

We used four different models for the compact flux density. These are as follows.

**Table 10**
Gaussian Parameters of the Large-scale Jet Model used in Synthetic Data Generation

| Flux Density (Jy) | $\theta_{\rm maj}$ ($\mu$as) | $\theta_{\rm min}$ ($\mu$as) | P.A. (°) | $\Delta_{\rm R.A.}$ ($\mu$as) | $\Delta_{\rm decl.}$ ($\mu$as) |
|---|---|---|---|---|---|
| 0.3 | 1000 | 600 | 140 | −688.77 | 80.86 |
| 0.2 | 400 | 200 | 93 | −295.44 | −52.09 |
| 0.1 | 400 | 200 | 115 | −215.80 | 208.40 |

1. A delta-function ring with radius $r_0 = 22\ \mu$as, convolved with a circular Gaussian with FWHM of 10 $\mu$as.
2. A crescent composed of a delta-function ring with radius $r_0 = 22\ \mu$as and a dipolar angular intensity profile

$$I(r, \theta) = I_0[1 + 2A\cos(\theta - \eta)]\frac{\delta(r - r_0)}{2\pi r_0}, \quad (49)$$

where $I_0 = 0.55$ Jy, the asymmetry parameter $A = 0.23$ (see Section 9.1), and $\eta$ is the orientation angle east of north. The crescent is then also convolved with a 10 $\mu$as Gaussian.
3. A uniform disk with radius $r = 35\ \mu$as, convolved with a circular Gaussian with FWHM of 10 $\mu$as.
4. Two circular Gaussian components, each with a FWHM of 20 $\mu$as. The first component is located at the origin and has a total flux density of 0.27 Jy, while the second is located at $\Delta_{\rm R.A.} = 30\ \mu$as and $\Delta_{\rm decl.} = -12\ \mu$as and has a total flux density of 0.33 Jy.

Both the jet and compact structure are polarized in order to produce non-closing systematic errors from polarimetric leakage. The jet components all have a constant fractional polarization of 20% and a random position angle that varies across the image with a coherence length of 400 $\mu$as. The compact components all have a constant fractional polarization of 40% and a random polarization position angle with a coherence length of 5 $\mu$as.

### C.2. Synthetic Data Generation with `eht-imaging`

The `eht-imaging` library produces synthetic visibilities from model images by computing their Fourier transform, adding random thermal noise (Equation (3)), and corrupting the data with systematic station-based effects using the Jones matrix formalism (Thompson et al. 2017). Off-diagonal terms in the Jones matrices mix the measured polarizations, which in the absence of noise and systematic error would correspond simply to the Fourier transform of the image polarizations by Equation (2).

We take circular polarizations as our primary basis, and consider a single baseline $ij$. The Fourier transform of the image gives the uncorrupted visibility of each polarization ($RR'_{ij}$, $LL'_{ij}$, $LR'_{ij}$, $RL'_{ij}$), which are assembled into a $2 \times 2$ correlation matrix:

$$V'_{ij} = \begin{pmatrix} RR'_{ij} & RL'_{ij} \\ LR'_{ij} & LL'_{ij} \end{pmatrix}. \quad (50)$$

The thermal noise standard deviation on each baseline and polarization is given by Equation (3) and assembled into a matrix $\boldsymbol{\sigma}_{ij}$ of the same form.

**Table 11**
Field Rotation Parameters for the EHT Stations

| Station | Receiver Mount | $f_{\rm par}$ | $f_{\rm el}$ | $\varphi_{\rm off}$ |
|---|---|---|---|---|
| ALMA | Cassegrain | 1 | 0 | 0 |
| APEX | Nasmyth-Right | 1 | 1 | 0 |
| JCMT | Cassegrain | 1 | 0 | 0 |
| LMT | Nasmyth-Left | 1 | −1 | 0 |
| PV | Nasmyth-Left | 1 | −1 | 0 |
| SMA | Nasmyth-Left | 1 | −1 | 45° |
| SMT | Nasmyth-Right | 1 | 1 | 0 |
| SPT | Cassegrain | 1 | 0 | 0 |

Including the effects of systematic and thermal noise, the simulated visibility matrix is $\boldsymbol{V}_{ij}$:

$$\boldsymbol{V}_{ij} = \boldsymbol{J}_i \boldsymbol{V}'_{ij} \boldsymbol{J}_j^{\dagger} + \mathcal{N}(\boldsymbol{\sigma}_{ij}), \quad (51)$$

where $\boldsymbol{J}_i$ and $\boldsymbol{J}_j$ are the station Jones matrices. Each Jones matrix has the form

$$\boldsymbol{J} = \begin{pmatrix} g_{\rm R} & e^{i\varphi}\, d_{\rm R}\, g_{\rm R} \\ e^{-i\varphi}\, d_{\rm L}\, g_{\rm L} & g_{\rm L} \end{pmatrix}, \quad (52)$$

where $g_{\rm R}$, $g_{\rm L}$ are the complex gain terms, $d_{\rm R}$ and $d_{\rm L}$ are the constant station $d$-terms, and $\varphi$ is a term from field rotation. We typically expect $g < 1$ due to losses in telescope sensitivity.

The complex gain terms include an absolute amplitude gain offset $|g|$, and a random atmospheric phase $\phi$. For the purposes of this Letter, we set $g_{\rm R} = g_{\rm L}$ because the atmosphere is not significantly birefringent at millimeter wavelengths and because differential gains between right and left polarizations are removed in the calibration process (Paper III):

$$g_{\rm R} = g_{\rm L} = |g(t)|\, e^{i\phi(t)}. \quad (53)$$

The phase error $\phi$ is drawn from a uniform distribution once per scan; it is kept stable within scans to mimic the ad hoc phase calibration procedure. The amplitude gain of a site at each time has a time-stable component $g_1$ and a time-varying component $g_2(t)$:

$$g(t) = |1 - |g_1| + g_2(t)|^{1/2}. \quad (54)$$

Both $g_1$ and $g_2$ are sampled from normal distributions with zero mean: $g_1 \sim \mathcal{N}(0, \sigma_{g_1})$ and $g_2(t) \sim \mathcal{N}(0, \sigma_{g_2})$. To simulate its higher median gain error and variability, for the LMT we set $\sigma_{g_1} = 0.6$ and $\sigma_{g_2} = 0.5$. For all other sites we set $\sigma_{g_1} = 0.15$ and $\sigma_{g_2} = 0.05$. $g_2$. Like the atmospheric phase, we fixed the gain errors so that $g_2(t)$ changes between scans but is kept stable across scans.

The complex $d$ terms, $d_{\rm R}$ and $d_{\rm L}$, are stationary in time and drawn from a complex normal distribution with zero mean and a standard deviation of 0.05 in each of the real and imaginary parts (motivated by the estimates in Johnson et al. 2015). The mean amplitude of the leakage terms is then 6.3%. The field rotation phase term $\varphi$ has three possible contributions depending upon the receiver mount:

$$\varphi = f_{\rm el}\,\theta_{\rm el} + f_{\rm par}\,\psi_{\rm par} + \varphi_{\rm off}, \quad (55)$$

where $\theta_{\rm el}$ is the elevation angle, $\psi_{\rm par}$ is the parallactic angle, and $\varphi_{\rm off}$ is a constant offset. Cassegrain mounts have $f_{\rm par} = 1$ and $f_{\rm el} = 0$. Nasmyth mounts have $f_{\rm par} = 1$ and $f_{\rm el} = \pm 1$, depending on the handedness. The EHT station field rotation parameters are listed in Table 11.

After being simulated with Equation (51), the synthetic data were network calibrated and converted to Stokes $I$, following the procedure used for the EHT data pipeline in Paper III.

### C.3. Synthetic Data Generation with MeqSilhouette and rPICARD

An mm-VLBI synthetic data pipeline built on the synthetic data generator `MeqSilhouette` (Blecher et al. 2017) and the EHT CASA calibration pipeline `rPICARD` (Janssen et al. 2019) has been developed (F. Roelofs et al. 2019, in preparation). Like `eht-imaging`, this pipeline is able to produce synthetic data sets with statistics that closely match real VLBI data, but it is done using an a priori physics-based approach instead of an a posteriori data-based approach.

`MeqSilhouette` (Blecher et al. 2017) is a synthetic VLBI data generator that predicts visibilities from both parametric and non-parametric sky models using `MeqTrees` (Noordam & Smirnov 2010) and `WSClean` (Offringa et al. 2014), respectively, and applies physically motivated propagation and instrumental effects. Uncorrupted frequency-resolved visibilities are generated from a full-Stokes sky model, antenna locations, and a real observing schedule. Elevation-dependent mean tropospheric delays, opacity, and sky temperatures are simulated from input weather parameters (ground pressure $P_g$, ground temperature $T_g$, and precipitable water vapor) using the `ATM` software (Pardo et al. 2001). In addition, turbulent phases are simulated based on input atmospheric coherence times $t_c$ and assuming a thin scattering screen model characterized by Kolmogorov turbulence. Station-dependent thermal noise drawn from a Gaussian distribution is included and contains contributions from both the receiver and the atmosphere. Scan-based antenna pointing errors are added using the primary beam sizes at 230 GHz and observatory-measured pointing uncertainties. Full polarization simulation of time-variable sources is also possible (as will be relevant to Sgr A$^*$ observations), with the option of introducing station-based, frequency-dependent polarization leakage and complex gain errors.

In the `MeqSilhouette+rPICARD` pipeline (F. Roelofs et al. 2019, in preparation), the full observation and calibration process is simulated by running the synthetic data generated by `MeqSilhouette` through the CASA VLBI calibration pipeline `rPICARD` (Janssen et al. 2019). In this calibration process, fringe fitting, amplitude calibration, and time and frequency averaging are done in the same way as for the real EHT data.

Figure 31 shows a comparison between synthetic data generated from the same model image using eht-imaging and

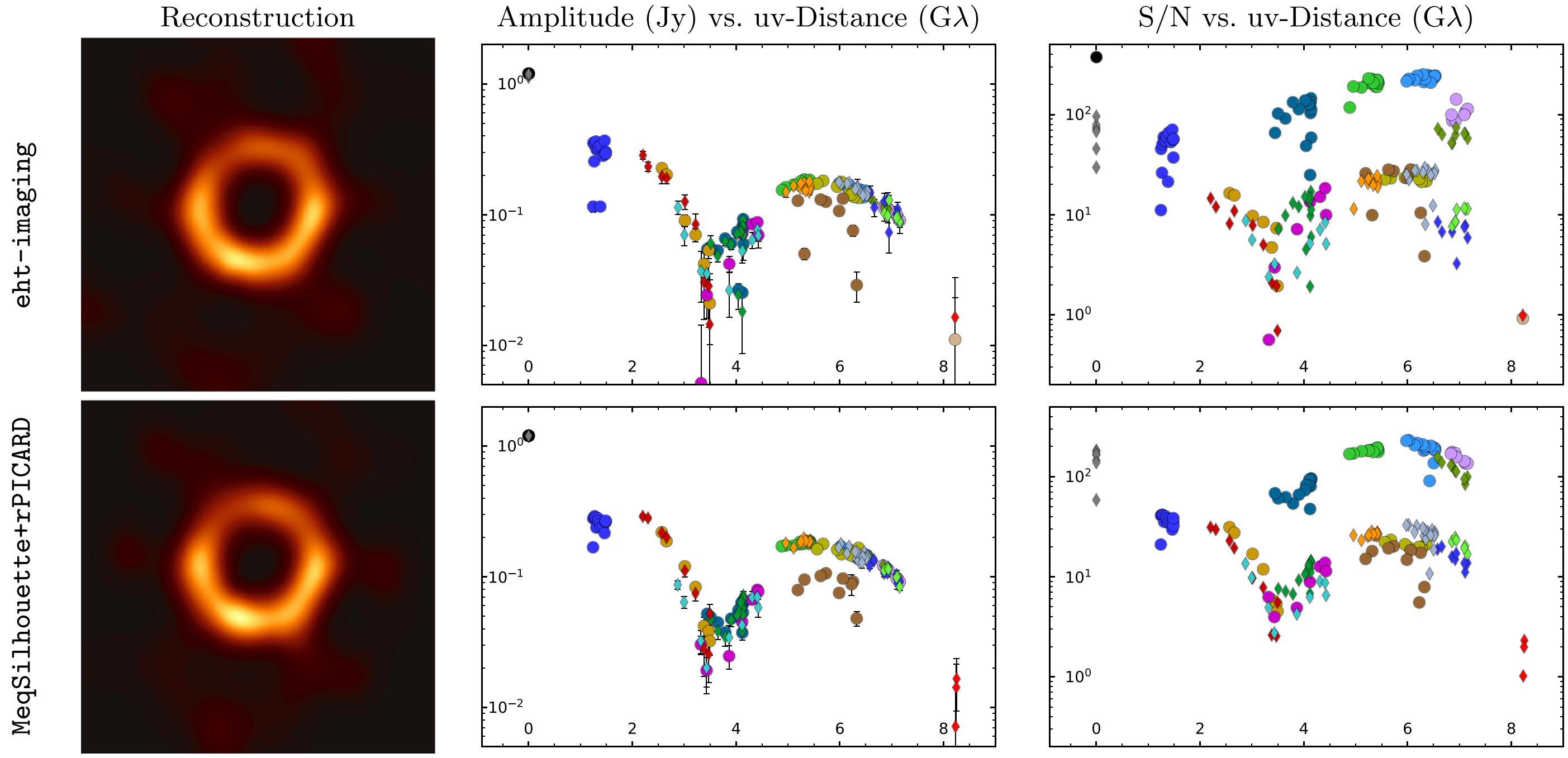

**Figure 31.** Comparison of the synthetic data from the `eht-imaging` and `MeqSilhouette` data generation libraries. Low-band data were generated from the same source image. The reconstructed images, visibility amplitudes, and S/Ns from the data generated with the `MeqSilhouette+rPICARD` pipeline are broadly consistent with those generated with `eht-imaging`, despite the different approaches of the software packages. Points are colored by baseline as in Figure 18. Images are reconstructed using the same script in `eht-imaging` using the fiducial parameters from Section 6.

**Table 12**
Weather Parameters used in the MeqSilhouette Simulations

| Antenna | PWV (mm) | $P_g$ (mb) | $T_g$ (K) | $t_c$ (s) |
|---|---|---|---|---|
| ALMA | 1.6 | 555 | 271 | 10 |
| APEX | 1.6 | 555 | 271 | 10 |
| JCMT | 1.1 | 626 | 278 | 5 |
| LMT | 4.7 | 604 | 275 | 6 |
| PV | 1.5 | 723 | 270 | 4 |
| SMA | 1.1 | 626 | 278 | 5 |
| SMT | 2.7 | 695 | 276 | 3 |

**Note.** The parameters PWV, $P_g$, and $T_g$ are median values for the EHT 2017 campaign as logged in VLBImonitor (Paper II).

the MeqSilhouette+rPICARD simulation software. For the MeqSilhouette simulations, input weather parameters (Table 12) were based on site measurements logged in VLBImonitor (Paper II) during the EHT campaign. Coherence times, SEFDs, and pointing offsets were based on measurements from individual EHT stations. A leakage of 5% was added to the visibilities with the field rotation parameters listed in Table 11, and an overall gain factor of 0.6 was set for the LMT.

The reconstructed images (generated with the M87 fiducial parameters of the eht-imaging imaging pipeline), visibility amplitudes, and S/Ns from the data generated with the MeqSilhouette and rPICARD pipeline are broadly consistent with those generated with eht-imaging, despite the different approaches used in the two software packages. Table 13 shows the normalized cross-correlation $\rho_{NX}$ and equivalent blurring $\alpha$ (given in $\mu$as), for images made from the eht-imaging and MeqSilhouette+rPICARD data simulation software. Each software package was used to generate three different realizations of visibilities from the same underlying crescent geometric model (see Figure 5). The $\rho_{NX}$ and $\alpha$ computed from reconstructions of this data show broad consistency. This consistency reinforces our confidence in both pipelines, particularly in corroborating that the image fidelity is primarily determined by the limited baselines coverage rather than data sensitivity or calibration limitations.

**Table 13**
Image Fidelity Metrics, Normalized Cross-correlation $\rho_{NX}$, and Equivalent Blurring $\alpha$ (given in $\mu$as), for Images Made from Data Simulated using eht-imaging and MeqSilhouette+rPICARD

| Data Simulation Software | Sample 1 | | Sample 2 | | Sample 3 | |
|---|---|---|---|---|---|---|
| | $\rho_{NX}$ | $\alpha$ | $\rho_{NX}$ | $\alpha$ | $\rho_{NX}$ | $\alpha$ |
| eht-imaging | 0.982 | 8.6 | 0.981 | 8.7 | 0.978 | 9.2 |
| MeqSilhouette +rPICARD | 0.980 | 8.9 | 0.980 | 9.0 | 0.983 | 8.4 |

**Note.** Images from three sets of visibilities of the crescent (+jet) model, with different realizations of noise (samples 1–3), were generated for each data simulation software package using the fiducial parameters from the eht-imaging imaging pipeline. The $\rho_{NX}$ and $\alpha$ fidelity metrics are insensitive to both the particular noise realization and the simulation software. Thus, our selected Top Sets and fiducial parameters are insensitive to the particular noise realization and the simulation software.

## Appendix D
## Image Dependence on Sites for SMILI and DIFMAP Pipelines

In Section 8.2, we use the eht-imaging pipeline with fiducial parameters to show that the ring structure in reconstructed images is resilient to the loss of most sites. Figure 32 shows the corresponding reconstructions from the other two pipelines on April 11. The ring structures produced by all three pipelines are similarly sensitive to dropping most individual sites, with Chile and PV being the most critical. The image degradation due to the loss of all data to a site tends to be larger for RML approaches, as these methods use weaker constraints on the allowed FOV for image flux.

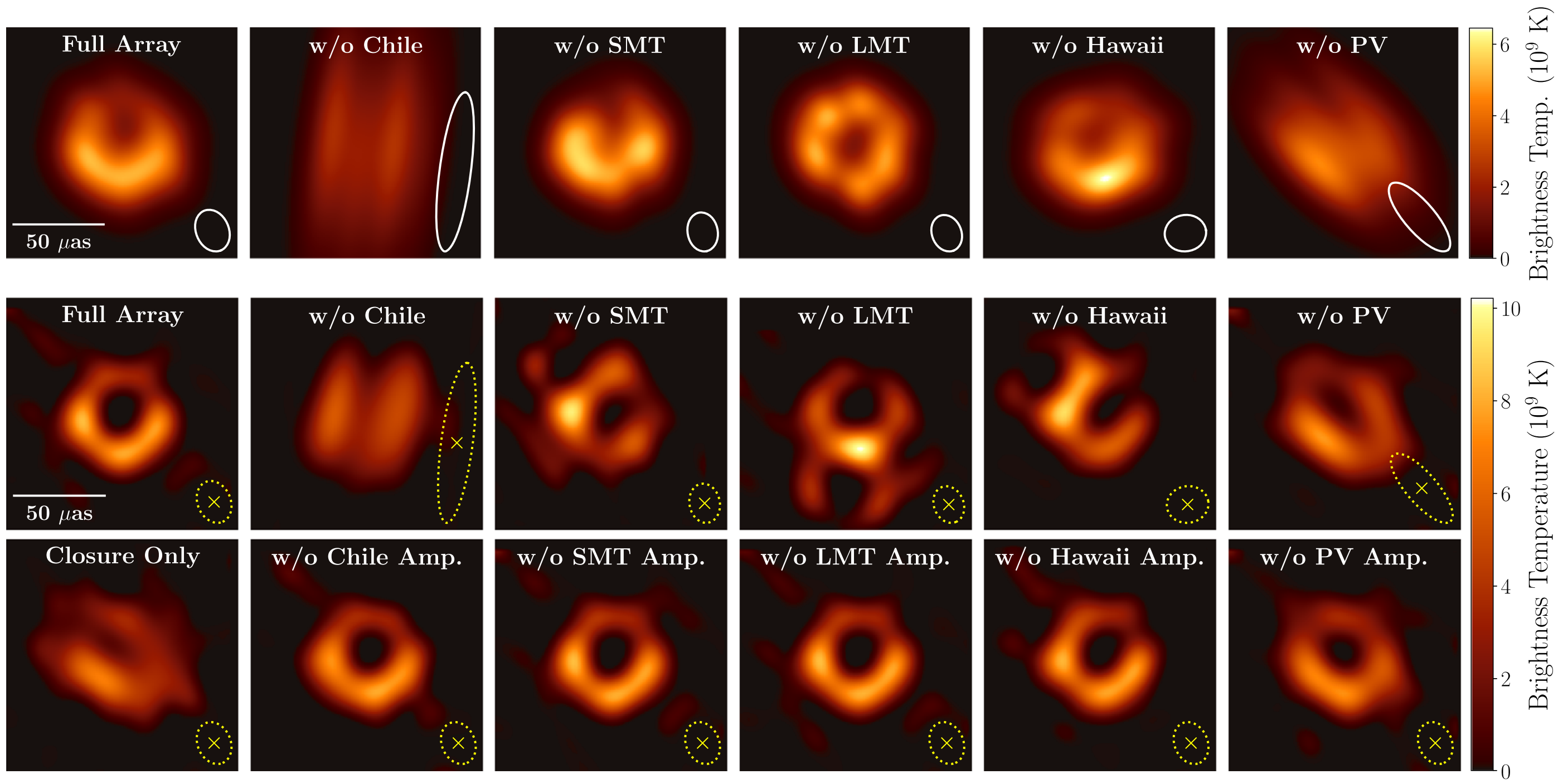

**Figure 32.** Example reconstructions of M87 on April 11 after omitting visibilities to each geographical site for the DIFMAP and SMILI pipelines. Following Figure 22, the top row of each pipeline shows reconstructions that exclude all baselines to the indicated site, while the bottom row shows reconstructions that exclude all visibility amplitudes on baselines to the indicated site but retain closure phases and closure amplitudes. The lower-left panel shows an image reconstructed after excluding visibility amplitudes to all sites (i.e., using only closure quantities). Because DIFMAP uses only visibilities, we only show the case of excluding baselines. The ellipse in each panel shows the corresponding synthesized beam with uniform weighting; only the DIFMAP images are restored with these beams.

## Appendix E
## Interday Variability in M87 Images

The closure phases of M87 show variation across the four observations (Paper III). Similarly, our fiducial images show corresponding variation in image structure between April 5, 6 and 10, 11 (Figure 11). These differences are also apparent in the averaged fiducial images restored to a common, conservative resolution (see Figure 15). In Section 8.3, we demonstrated that the image variability is not simply from changes in the baseline coverage. We now assess this image variation.

Figure 33 shows differences of reconstructed image pairs on different days. To suppress features that are specific to imaging parameter choices, we show the beam-convolved difference of two mean images, where the mean is taken over all corresponding Top Set images for the specified pipeline and day. For all pipelines, but especially for eht-imaging and SMILI, the differences are larger across the multi-day gap than across adjacent days, as expected if these differences reflect source evolution.

For both eht-imaging and SMILI, the difference images indicate a shift of the flux from the left to the right side of the ring (east to west) over the observations, which is consistent with the slight increase in $\eta$ over the days (Section 9). In contrast, DIFMAP difference images are dominated by larger-scale morphology changes. For all pipelines, the beam-convolved temporal differences are nearly an order of magnitude larger than the median standard deviation of differences seen across eht-imaging Top Set images ($1.6 \times 10^8$ K; see Figure 17). Nevertheless, in all cases, we cannot unambiguously associate the variability with smooth motion or evolution of a specific image feature.

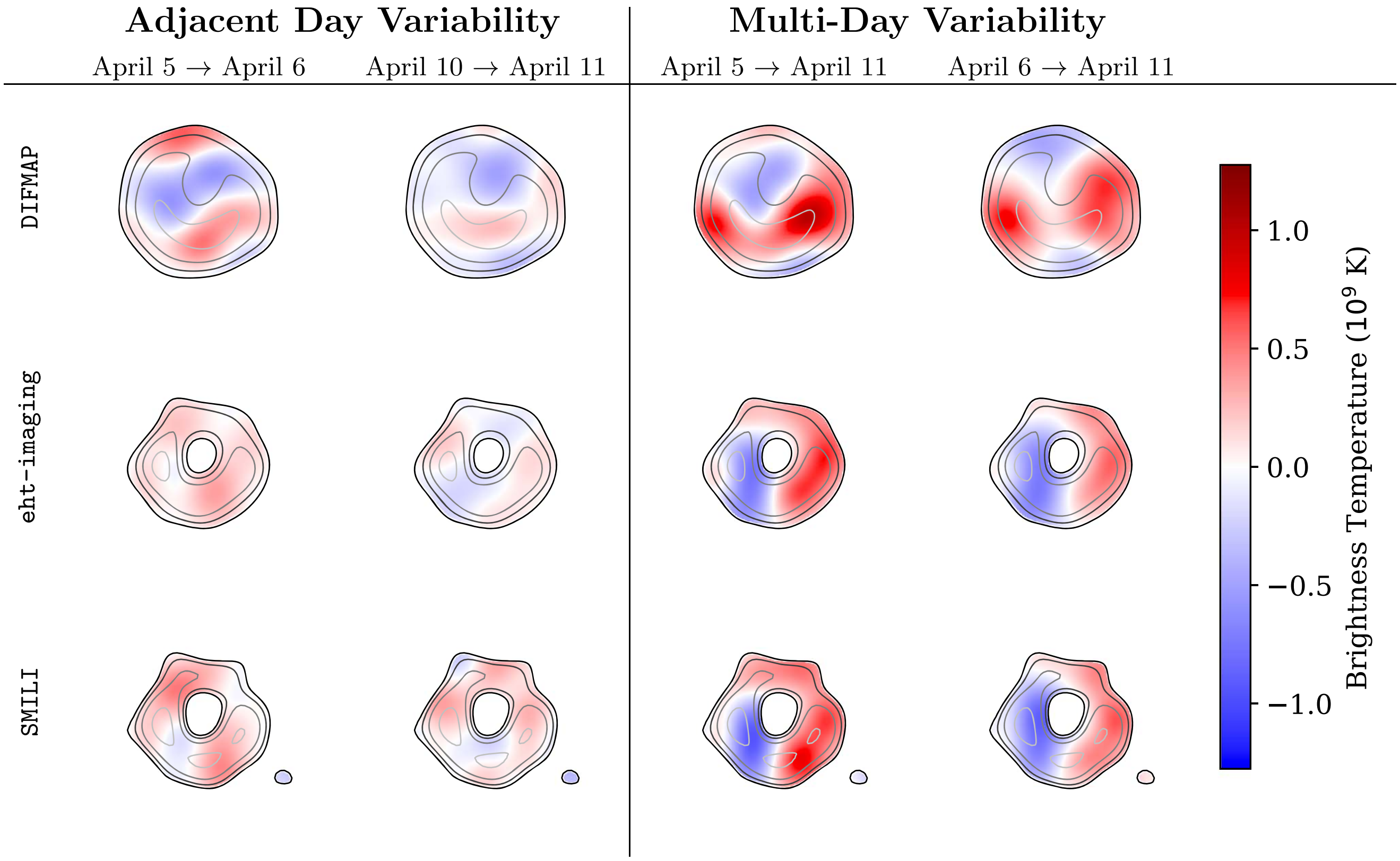

**Figure 33.** Variations seen in reconstructed images between days. Each row shows the beam-convolved difference of the aligned mean images from the Top Set of the corresponding parameter survey. All images are shown using the same color scale. Positive values (red) indicate an increase in brightness at the later date, and negative values (blue) indicate a decrease in brightness at the later date.

## Appendix F
## Multi-day Gain Comparisons with 3C 279

In Section 8, we compared the residual amplitude gains for the LMT and SMT on April 5 derived from separately imaging M87 and 3C 279. We now expand this comparison to all four days on which these sources were observed.

Figure 34 compares the interleaved multiplicative station gains for M87 and 3C 279 for each day; Table 14 presents the gain statistics for the two sources and compares them to the expected amplitude calibration error budget (Paper III). The trends of inferred residual gain amplitudes are consistent among the imaging pipelines, and those of M87 are similar to those of 3C 279. The inferred gains for M87 at the LMT have large excursions that are not seen in the 3C 279 gain trends; these excursions are likely due to poor pointing and tracking. These large LMT correction factors (up to $|g_{\mathrm{LMT}}|^{-1} \approx 6$) are also consistent with those estimated using the SMT–PV and LMT–PV crossing tracks (see Table 8 and Figure 28).

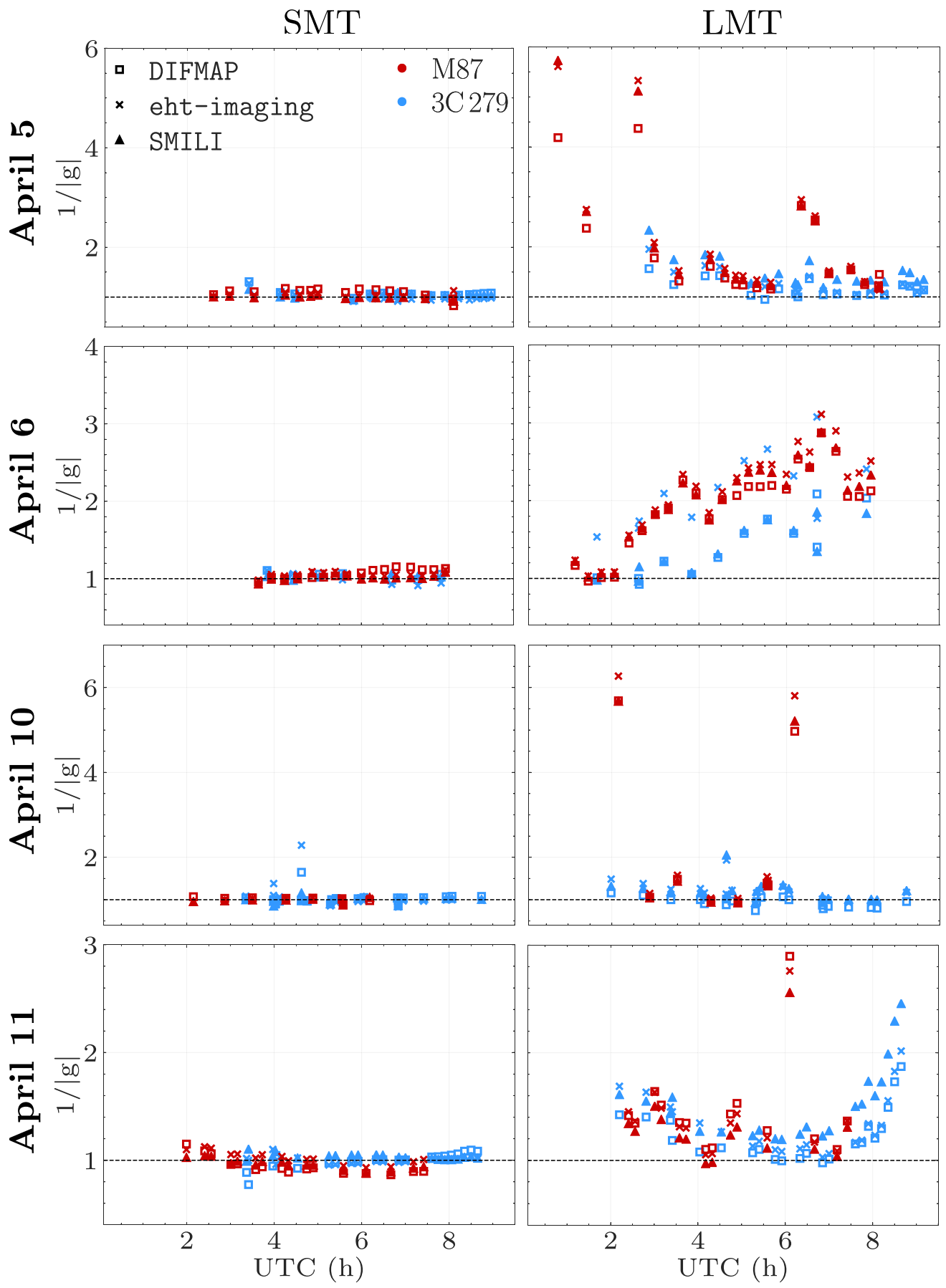

**Figure 34.** Derived residual gains for the SMT (left) and LMT (right) using self-calibration to images of M87 (red) and 3C 279 (blue). The M87 images are the fiducial images from each pipeline; the 3C 279 images were reconstructed separately, using adapted imaging scripts. The particularly large excursions on the LMT M87 gains are often due to poor pointing. For instance, excursions at ≈6 UTC are from difficulties tracking the source during transit (∼81° elevation). Note the different ranges shown for each observing day.

**Table 14**
Residual Gain Corrections Derived for M87 and 3C 279 for Observations on All Days (April 5, 6, 10, 11)

| Station | Fiducial M87 Median Gain | | | 3C 279 Median Gain | | | A Priori Budget (%) |
|---|---|---|---|---|---|---|---|
| April 5 | DIFMAP | eht-imaging | SMILI | DIFMAP | eht-imaging | SMILI | |
| ALMA | $0.97^{+0.02}_{-0.03}$ | $0.97^{+0.02}_{-0.01}$ | $0.98^{+0.01}_{-0.01}$ | $0.99^{+0.02}_{-0.09}$ | $1.11^{+0.04}_{-0.02}$ | $0.97^{+0.03}_{-0.01}$ | 5.0 |
| APEX | $1.05^{+0.04}_{-0.02}$ | $1.02^{+0.02}_{-0.02}$ | $1.01^{+0.01}_{-0.01}$ | $0.99^{+0.01}_{-0.00}$ | $0.90^{+0.01}_{-0.00}$ | $1.05^{+0.01}_{-0.02}$ | 5.5 |
| SMT | $1.13^{+0.01}_{-0.06}$ | $1.02^{+0.04}_{-0.01}$ | $0.99^{+0.02}_{-0.01}$ | $1.06^{+0.01}_{-0.02}$ | $0.97^{+0.01}_{-0.02}$ | $1.04^{+0.02}_{-0.01}$ | 3.5 |
| JCMT | $1.00^{+0.02}_{-0.00}$ | $1.00^{+0.00}_{-0.00}$ | $1.00^{+0.00}_{-0.00}$ | $1.00^{+0.01}_{-0.00}$ | $1.02^{+0.02}_{-0.01}$ | $1.00^{+0.00}_{-0.01}$ | 7.0 |
| LMT | $1.46^{+0.76}_{-0.21}$ | $1.55^{+0.93}_{-0.19}$ | $1.47^{+0.91}_{-0.19}$ | $1.08^{+0.16}_{-0.04}$ | $1.21^{+0.13}_{-0.11}$ | $1.35^{+0.27}_{-0.05}$ | 11.0 |
| SMA | $1.00^{+0.02}_{-0.01}$ | $1.00^{+0.01}_{-0.00}$ | $1.00^{+0.00}_{-0.00}$ | $1.01^{+0.01}_{-0.01}$ | $1.04^{+0.03}_{-0.00}$ | $0.99^{+0.00}_{-0.01}$ | 7.5 |
| PV | $1.14^{+0.04}_{-0.04}$ | $0.96^{+0.02}_{-0.07}$ | $0.98^{+0.02}_{-0.07}$ | $1.33^{+0.12}_{-0.04}$ | $1.14^{+0.04}_{-0.05}$ | $0.94^{+0.05}_{-0.02}$ | 5.0 |
| SPT | … | … | … | … | … | … | 7.5 |
| April 6 | | | | | | | |
| ALMA | $1.02^{+0.03}_{-0.02}$ | $0.98^{+0.02}_{-0.02}$ | $0.99^{+0.02}_{-0.02}$ | $1.00^{+0.02}_{-0.01}$ | $1.04^{+0.09}_{-0.04}$ | $0.99^{+0.01}_{-0.03}$ | 5.0 |
| APEX | $0.98^{+0.02}_{-0.03}$ | $1.01^{+0.03}_{-0.02}$ | $1.01^{+0.02}_{-0.02}$ | $1.00^{+0.01}_{-0.01}$ | $0.97^{+0.03}_{-0.09}$ | $1.01^{+0.03}_{-0.01}$ | 5.5 |
| SMT | $1.06^{+0.06}_{-0.04}$ | $1.05^{+0.02}_{-0.01}$ | $1.01^{+0.03}_{-0.02}$ | $1.06^{+0.00}_{-0.05}$ | $0.99^{+0.05}_{-0.05}$ | $1.02^{+0.01}_{-0.00}$ | 3.5 |
| JCMT | $1.00^{+0.00}_{-0.00}$ | $1.00^{+0.00}_{-0.00}$ | $1.00^{+0.00}_{-0.00}$ | $1.00^{+0.00}_{-0.00}$ | $1.00^{+0.01}_{-0.00}$ | $1.00^{+0.01}_{-0.00}$ | 7.0 |
| LMT | $2.06^{+0.12}_{-0.33}$ | $2.30^{+0.16}_{-0.49}$ | $2.16^{+0.20}_{-0.44}$ | $1.34^{+0.29}_{-0.30}$ | $2.13^{+0.30}_{-0.37}$ | $1.33^{+0.32}_{-0.20}$ | 11.0 |
| SMA | $1.01^{+0.01}_{-0.01}$ | $1.00^{+0.00}_{-0.01}$ | $1.00^{+0.00}_{-0.02}$ | $1.00^{+0.01}_{-0.00}$ | $1.04^{+0.00}_{-0.01}$ | $0.99^{+0.00}_{-0.00}$ | 7.5 |
| PV | $0.98^{+0.03}_{-0.04}$ | $0.97^{+0.01}_{-0.07}$ | $0.99^{+0.01}_{-0.06}$ | $0.81^{+0.12}_{-0.29}$ | $1.05^{+0.01}_{-0.02}$ | $0.91^{+0.06}_{-0.01}$ | 5.0 |
| SPT | … | … | … | … | … | … | 7.5 |
| April 10 | | | | | | | |
| ALMA | $1.03^{+0.02}_{-0.03}$ | $0.99^{+0.00}_{-0.01}$ | $0.99^{+0.01}_{-0.02}$ | $1.00^{+0.03}_{-0.00}$ | $1.08^{+0.06}_{-0.08}$ | $0.99^{+0.01}_{-0.02}$ | 5.0 |

**Table 14**
(Continued)

| Station | Fiducial M87 Median Gain | | | 3C 279 Median Gain | | | A Priori Budget (%) |
|---|---|---|---|---|---|---|---|
| APEX | $0.99^{+0.02}_{-0.03}$ | $1.02^{+0.01}_{-0.01}$ | $1.01^{+0.01}_{-0.02}$ | $1.00^{+0.00}_{-0.02}$ | $0.94^{+0.06}_{-0.06}$ | $1.00^{+0.02}_{-0.00}$ | 5.5 |
| SMT | $1.03^{+0.01}_{-0.02}$ | $1.03^{+0.00}_{-0.04}$ | $0.99^{+0.02}_{-0.02}$ | $1.03^{+0.02}_{-0.05}$ | $0.97^{+0.04}_{-0.02}$ | $1.00^{+0.03}_{-0.04}$ | 3.5 |
| JCMT | $1.00^{+0.01}_{-0.01}$ | $1.00^{+0.00}_{-0.00}$ | $1.00^{+0.00}_{-0.00}$ | $1.00^{+0.00}_{-0.04}$ | $0.98^{+0.01}_{-0.00}$ | $1.00^{+0.00}_{-0.07}$ | 7.0 |
| LMT | $1.37^{+0.98}_{-0.33}$ | $1.50^{+1.13}_{-0.38}$ | $1.35^{+1.02}_{-0.34}$ | $0.96^{+0.04}_{-0.12}$ | $1.17^{+0.09}_{-0.22}$ | $1.19^{+0.07}_{-0.18}$ | 11.0 |
| SMA | $1.01^{+0.04}_{-0.00}$ | $1.00^{+0.02}_{-0.00}$ | $1.00^{+0.02}_{-0.00}$ | $1.00^{+0.01}_{-0.01}$ | $1.04^{+0.02}_{-0.01}$ | $1.00^{+0.01}_{-0.01}$ | 7.5 |
| PV | $1.02^{+0.02}_{-0.00}$ | $0.94^{+0.03}_{-0.03}$ | $0.94^{+0.02}_{-0.01}$ | $0.97^{+0.06}_{-0.02}$ | $1.10^{+0.15}_{-0.04}$ | $0.96^{+0.04}_{-0.01}$ | 5.0 |
| SPT | … | … | … | $1.02^{+0.58}_{-0.09}$ | $1.02^{+0.28}_{-0.08}$ | $1.03^{+0.03}_{-0.10}$ | 7.5 |
| April 11 | | | | | | | |
| ALMA | $1.00^{+0.02}_{-0.01}$ | $1.00^{+0.02}_{-0.01}$ | $1.00^{+0.01}_{-0.01}$ | $1.01^{+0.01}_{-0.01}$ | $1.12^{+0.03}_{-0.09}$ | $0.99^{+0.01}_{-0.01}$ | 5.0 |
| APEX | $1.00^{+0.01}_{-0.02}$ | $1.00^{+0.01}_{-0.02}$ | $1.00^{+0.01}_{-0.01}$ | $1.00^{+0.01}_{-0.02}$ | $0.90^{+0.10}_{-0.03}$ | $1.01^{+0.01}_{-0.01}$ | 5.5 |
| SMT | $0.92^{+0.04}_{-0.03}$ | $1.02^{+0.03}_{-0.03}$ | $0.95^{+0.04}_{-0.01}$ | $1.01^{+0.03}_{-0.03}$ | $1.01^{+0.01}_{-0.03}$ | $1.03^{+0.02}_{-0.01}$ | 3.5 |
| JCMT | $1.00^{+0.01}_{-0.01}$ | $1.00^{+0.01}_{-0.00}$ | $1.01^{+0.01}_{-0.01}$ | $1.00^{+0.00}_{-0.00}$ | $1.00^{+0.01}_{-0.00}$ | $1.00^{+0.00}_{-0.00}$ | 7.0 |
| LMT | $1.35^{+0.16}_{-0.15}$ | $1.35^{+0.10}_{-0.18}$ | $1.23^{+0.11}_{-0.13}$ | $1.16^{+0.20}_{-0.09}$ | $1.25^{+0.25}_{-0.12}$ | $1.47^{+0.14}_{-0.21}$ | 11.0 |
| SMA | $1.00^{+0.01}_{-0.04}$ | $0.99^{+0.01}_{-0.01}$ | $1.00^{+0.02}_{-0.01}$ | $1.00^{+0.01}_{-0.00}$ | $1.04^{+0.00}_{-0.01}$ | $1.00^{+0.00}_{-0.01}$ | 7.5 |
| PV | $1.00^{+0.01}_{-0.05}$ | $0.92^{+0.02}_{-0.00}$ | $0.95^{+0.02}_{-0.03}$ | $0.91^{+0.04}_{-0.10}$ | $1.09^{+0.12}_{-0.04}$ | $1.00^{+0.02}_{-0.01}$ | 5.0 |
| SPT | … | … | … | $0.90^{+0.07}_{-0.03}$ | $1.00^{+0.00}_{-0.05}$ | $1.02^{+0.05}_{-0.06}$ | 7.5 |

**Note.** Numbers indicate the median of $1/|g|$, with stated uncertainties corresponding to the 25th and 75th percentiles. The multiplicative gains on the visibilities for each station are computed by self-calibrating the network-calibrated data sets to the reconstructions after rescaling the intra-site baseline amplitudes to match the total flux of the reconstructions. The percentage deviation from unity can be compared to the expectations from the station-based a priori error budget on the visibility amplitudes derived from Paper III. The SPT is omitted from gain comparisons on the first two days as it did not observe 3C 279 on April 5, and it observed only a single 3C 279 scan on April 6.

## Appendix G
## Finite Resolution Bias on Ring Parameters

In addition to uncertainties from thermal noise, systematic noise, and algorithmic imaging assumptions, estimated image properties are necessarily limited by the image resolution. Image structure at scales finer than the diffraction-limited resolution can bias properties such as the magnitude and location of maximum image brightness.

As a simple example, we will consider the finite resolution bias for a thin ring. Written in cylindrical coordinates $(r, \theta)$, the image of an infinitesimally thin ring of diameter $d$ takes the form

$$I_{\rm ring}(r, \theta; d) = \frac{1}{\pi d}\delta(r - d/2), \tag{56}$$

where the image is normalized to have a total flux density of unity. Next, suppose that this image is convolved with a circular Gaussian kernel having FWHM $\alpha$. The resulting blurred ring is given by

$$\begin{aligned} I_{\rm ring}(r, \theta; d, \alpha) &= \frac{4\ln 2}{\pi^2\alpha^2 d}\int r'\, dr' d\theta' \delta(r' - d/2) \\ &\quad\times \exp\left[-\frac{4\ln 2}{\alpha^2}(r^2 + r'^2 - 2rr'\cos\theta')\right] \\ &= \frac{4\ln 2}{\pi\alpha^2}\exp\left[-\frac{4\ln 2}{\alpha^2}(r^2 + d^2/4)\right] \\ &\quad\times I_0\left(4\ln 2\frac{rd}{\alpha^2}\right), \end{aligned}$$

where $I_0(x)$ in the final line is a modified Bessel function of the first kind. In the limit $\alpha \ll d$, we can simplify this expression using the identity $I_0(x) = \frac{e^x}{\sqrt{2\pi x}}\left[1 + \mathcal{O}\left(\frac{1}{x}\right)\right]$ (Abramowitz & Stegun 1972). Substituting $\Delta = r - d/2$ and only keeping terms to leading order in $\Delta/d$ and $\alpha/d$, we obtain

$$I_{\rm ring}(r; d, \alpha) \approx \sqrt{\frac{4\ln 2}{\pi^3}}\frac{1}{\alpha d}\left[1 - \frac{\Delta}{d}\right]e^{-4\ln 2(\Delta/\alpha)^2}. \tag{57}$$

Thus, when $\alpha \ll d$, the FWHM of the blurred ring is approximately equal to $\alpha$. More generally, convolving a ring that has a Gaussian profile of intrinsic FWHM $w_{\rm true}$ with a circular Gaussian kernel with FWHM $\alpha$ then gives an effective width

$$w_{\rm meas} \approx \sqrt{w_{\rm true}^2 + \alpha^2}. \tag{58}$$

Likewise, we can use Equation (57) to estimate the radius of peak brightness after blurring; to leading order in $\alpha/d$, $r_{\rm pk} \approx d/2 - \frac{1}{8\ln 2}\frac{\alpha^2}{d}$. Thus, letting $d_{\rm true}$ denote the angular diameter of the unconvolved ring and $d_{\rm meas}$ denote the angular diameter of peak brightness of the convolved ring, we obtain

$$d_{\rm true} \approx \frac{d_{\rm meas}}{1 - \frac{1}{4\ln 2}\left(\frac{\alpha}{d_{\rm meas}}\right)^2}, \tag{59}$$

where the approximation is accurate to leading order in $\alpha/d_{\rm meas}$.

As a concrete example, choosing parameters that are similar to our M87 measurements in Section 9.3, an infinitesimally thin ring with true diameter $d = 44\ \mu$as and width $\alpha = 15\ \mu$as will have its measured diameter biased downward by $\approx 2\ \mu$as.

While the simple Gaussian convolution assumed in this example only crudely approximates the effects of finite resolution on reconstructed images, it indicates that for images dominated by a thin ring, estimated diameters (widths) will be biased downward (upward) by the finite resolution of an image reconstruction. For the fiducial image measurements in Section 9, the diameter bias from this finite-resolution effect would be maximum in the case of an infinitesimal intrinsic ring, with $\alpha$ then approximately corresponding to the measured width. Thus, the finite-resolution diameter bias is at most a few $\mu$as.

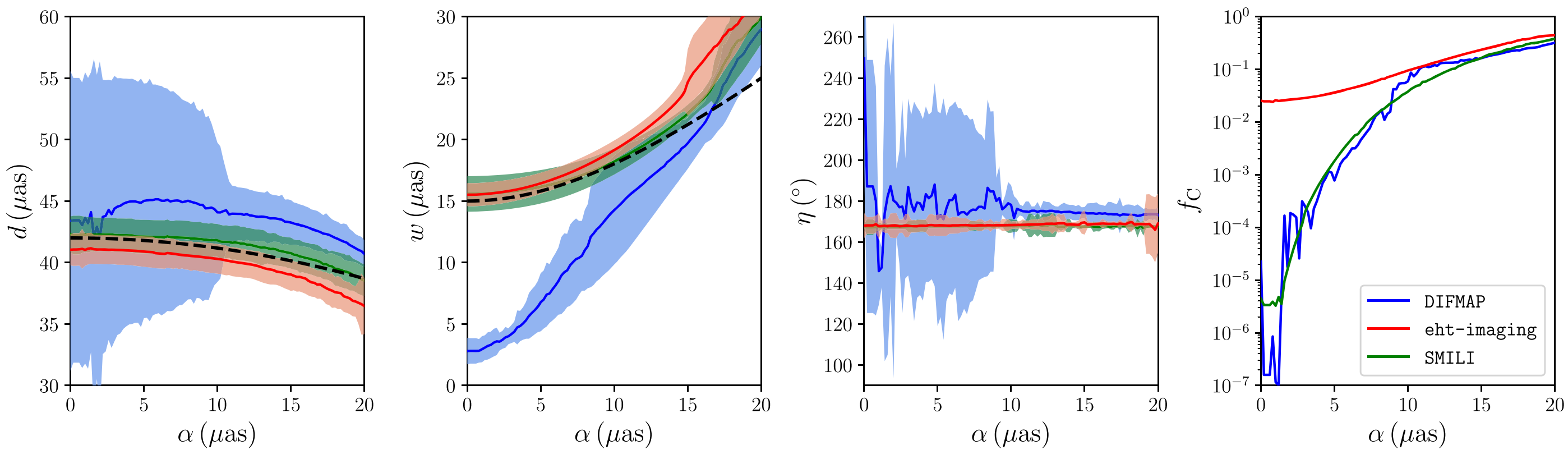

**Figure 35.** From left to right, measured diameter $d$, width $w$, orientation angle $\eta$, and central brightness ratio $f_C$ measured from the April 11 fiducial images blurred with circular Gaussian kernels of increasing FWHM $\alpha$. The solid lines indicate the measured value, and the shaded regions give the $1\sigma$ uncertainty as defined in Section 9.1 (these uncertainties do not include a contribution from scatter across the Top Set). The dashed line on the first panel shows the prediction of Equation (59) for the measured diameter $d_{\rm meas}$ as a function of the blur kernel assuming $d_{\rm true} = 42$ $\mu$as. The dashed line in the second panel shows the FWHM of a 15 $\mu$as Gaussian convolved with the kernel of FWHM $\alpha$ (Equation (58)). Because the DIFMAP images are fundamentally composed of point sources, the measurements from these images become highly uncertain when $\alpha \lesssim 10$ $\mu$as.

In Figure 35, we explore the dependence of estimated ring parameters on the size of an applied restoring beam. Using the fiducial M87 images on April 11 from all three imaging pipelines, we show parameters computed after convolving each image with a circular Gaussian kernel having FWHM $\alpha$ in the range 0–20 $\mu$as. The DIFMAP images are fundamentally composed of point sources ("CLEAN" components). When these point sources are restored with $\alpha \lesssim 10$ $\mu$as, the CLEAN components still appear in the convolved image as individual point sources. As a result, the feature extraction methods of Section 9, which assume a smooth ring structure, have large uncertainties when extracting parameters on these images. This uncertainty is particularly apparent in the orientation angle measurement. The RML images, by contrast, have a finite width and smooth structure even at $\alpha = 0$. As a result, their parameters vary smoothly and have similar uncertainties at all values of $\alpha$.

The leftmost panel of Figure 35 shows that the dependence of the measured diameter on $\alpha$ closely follows Equation (59). In addition, the fiducial images have a scatter of approximately 1 $\mu$as across the different imaging methods, but an uncertainty of $\approx$3–4 $\mu$as across the varying $\alpha$. Thus, when extracting physical parameters from the measured diameter, it is important to take into consideration the additional bias and uncertainty induced by this effect.

For small values of the restoring beam $\alpha$, Figure 35 shows that the measured width $w$ of the RML reconstructions follows the prediction of Equation (58). For larger $\alpha > 15$ $\mu$as, the kernel size approaches the ring radius, and higher-order effects become important (i.e., contributions from the opposite side of the ring in the convolved width). Because it is intrinsically built of point sources that are *not* confined to a $\delta$-function in radius, the DIFMAP image does not follow this simple prediction for the increase in width with blurring kernel size, and it instead increases more rapidly to converge with the RML result at $\alpha \approx 20$ $\mu$as. For the RML reconstructions (and for DIFMAP with $\alpha > 10$ $\mu$as), the measured orientation angle is relatively unaffected by the Gaussian convolution.

Of all the parameters defined in Section 9.1, the fractional central brightness $f_C$ between the average ring center brightness and the rim varies the most with resolution. In the absence of a restoring beam, both SMILI and DIFMAP produce rings with practically zero brightness in the ring center; as a result of this near-zero floor, $f_C$ is extremely small ($f_C < 10^{-5}$). As convolution with a finite Gaussian kernel fills in the center of the ring, $f_C$ increases rapidly with $\alpha$, by several orders of magnitude, for these images. By contrast, because they include inverse-tapering of the data and a final blurring with a 5 $\mu$as Gaussian, the eht-imaging fiducial reconstructions always have non-zero central brightness. All three imaging methods give $f_C \approx 0.3$ for $\alpha = 20$ $\mu$as. Because $f_C$ is sensitive to resolution and imaging methodology, we can at only securely conclude that $f_C \lesssim 0.3$ for M87.

## Appendix H
## Image Dependence on the Assumed Total Compact Flux Density

While all three parameter surveys explored different choices of total compact flux density from 0.4 to 0.8 Jy, they were tested against simulated images that all had $F_{\rm cpct} = 0.6$. For M87, the true value of $F_{\rm cpct}$ is unknown, adding additional uncertainty that is not explored systematically in the parameter surveys or Top Sets. In Figure 36, we explore how the estimated ring parameters of M87 from reconstructed images depend upon the assumed total compact flux density. In this analysis, we fix all other imaging parameters to their fiducial values.

Not all values of $F_{\rm cpct}$ produce acceptable images, with two primary problems. First, a reconstructed image may not be in acceptable agreement with the underlying data. Second, a reconstructed image may have corresponding station gains from self-calibration that are significantly different than a priori expectations. The SMT and LMT gains are most sensitive to $F_{\rm cpct}$ because the SMT–LMT baseline is the only inter-site EHT baseline that weakly resolves M87. To define acceptable images in this test, we require two conditions to be satisfied (similar to the requirements for the Top Sets; Section 6.3.1). First, we require that $\chi^2_{\rm CP}$, $\chi^2_{\rm log\,CA} < 2$ with no systematic uncertainty for eht-imaging and SMILI and 5% systematic uncertainty for DIFMAP. Second, we require that the median SMT gain amplitude satisfies $0.9 \leqslant \text{median}(1/|g_{SMT}|) \leqslant 1.1$ (we only assess the images based upon $|g_{SMT}|$ because the LMT a priori calibration is comparatively poor; see Appendix F). For eht-imaging and SMILI, the SMT gain constraint gives $0.40\ \text{Jy} \leqslant F_{cpct} \leqslant 0.81$ Jy and $0.40\ \text{Jy} \leqslant F_{cpct} \leqslant 0.85$ Jy, respectively; all reconstructions across these ranges satisfy the $\chi^2$ criterion. For DIFMAP, the SMT gain constraint gives $0.43\ \text{Jy} \leqslant F_{cpct} \leqslant 0.91$ Jy; however, only images with $0.18\ \text{Jy} \leqslant F_{cpct} \leqslant 0.56$ Jy satisfy the $\chi^2$ criterion. Thus, for DIFMAP, the range of images meeting both criteria is $0.43\ \text{Jy} \leqslant F_{cpct} \leqslant 0.56$ Jy. For all pipelines, the acceptable range of $F_{cpct}$ extends lower than that derived in Section 4 because we do not strictly enforce $1/|g_{SMT}| \geqslant 1$, instead enforcing the weaker constraint that the $\text{median}(1/|g_{SMT}|) \geqslant 0.9$.

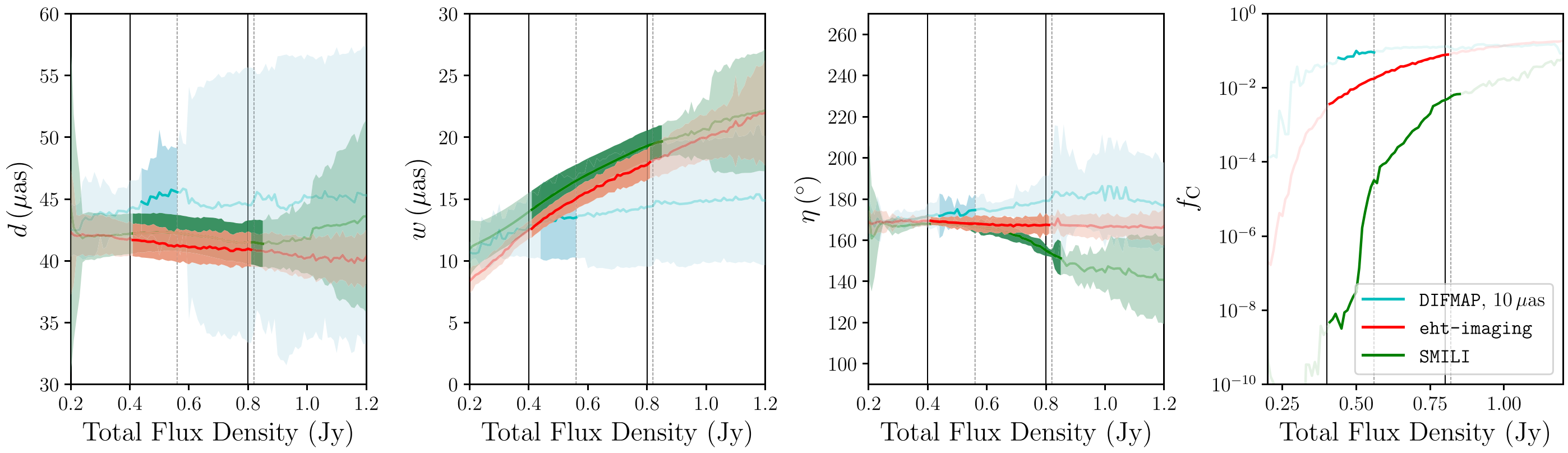

**Figure 36.** Measured diameter $d$, width $w$, orientation angle $\eta$, and central brightness ratio $f_C$ of M87, measured from image reconstructions with varying total compact flux density, $F_{cpct}$. All other imaging parameters were set to the fiducial parameters of the corresponding pipeline. DIFMAP values were measured after restoring with a 10 $\mu$as FWHM Gaussian beam. The solid lines indicate the measured value, and the shaded regions give the $\pm 1\sigma$ uncertainty as defined in Section 9.1 (these uncertainties do not include a contribution from scatter across the Top Set). The plotted values are dark in the range of $F_{cpct}$ that produces images that have (1) $\chi^2_{CP}$, $\chi^2_{\log CA} < 2$ (with 5% systematic uncertainty for DIFMAP), and (2) a corresponding self-calibration solution with $0.9 \leqslant \text{median}(1/|g_{SMT}|) \leqslant 1.1$. The vertical dashed lines indicate the 0.56–0.82 Jy constraint on the total compact flux density from pre-imaging considerations (Section 4); the vertical solid lines indicate the 0.4–0.8 Jy range explored in the parameter surveys (Section 6).

For DIFMAP, the ring diameter is relatively insensitive to $F_{cpct}$, although the azimuthal variations $\sigma_d$ increase significantly for $F_{cpct} \gtrsim 0.55$. For eht-imaging and SMILI, larger values of $F_{cpct}$ lead to broader rings with slightly lower diameters, as expected from the finite resolution bias discussed in Appendix G. Across the range of $F_{cpct}$ identified in Appendix B, the most significant variations are in the orientation angle $\eta$ estimated with the SMILI pipeline, and in the central brightness ratio $f_C$, which increases steeply with $F_{cpct}$ for all pipelines.

## Appendix I
## Image Radial and Azimuthal Profiles

In this section, we present additional radial and azimuthal profiles of the fiducial M87 images. These profiles use the ring centroids determined in Section 9.

Figure 37 shows radial profiles taken across the rings identified in the three April 11 fiducial images, with the DIFMAP image restored by the nominal 20 $\mu$as beam. For each image, we identified the orientation angle $\eta$, then divided the ring in two by the line perpendicular to the measured orientation angle. On each half-ring, we compute the median angular profile (solid lines in Figure 37), as well as the 25%–75% and 0%–100% percentile ranges of variation. Figure 37 shows that the peak-to-peak ring diameters measured from the three imaging methods are broadly consistent, as indicated by our measurements in Figure 25. The overall sense of ring asymmetry recovered from each imaging pipeline is also consistent; the ring is always brighter in the south.

However, the shape of the median radial profiles differ among the different imaging methods due to the different assumptions made in the imaging process. The DIFMAP reconstructions in Figure 37 are restored with a 20 $\mu$as beam, so they produce wider, shallower radial profiles. In contrast, the SMILI images tend to zero out low brightness regions, and the fiducial images have a near-zero brightness depression in the center of the ring. The eht-imaging results are of a higher resolution than the DIFMAP results, but they have a less dramatic brightness depression that the SMILI images as a result of the choice to include a 5 $\mu$as maximum resolution in the eht-imaging script and the use of entropy regularization.

Figure 38 shows the angular profiles along the ring from the three imaging pipelines on April 11 data. While Figure 37 showed variability in the radial profiles from the individual fiducial images, Figure 38 considers variability in the azimuthal structure across the entire Top Set of images. In each panel, the solid line represents the fiducial value of the angular profile, and the bands show variability across the entire Top Set.

The DIFMAP angular profile in Figure 38 is smooth and shallow, and it is distinct from those of the RML methods due to its 20 $\mu$as restoring beam. The dashed lines in the eht-imaging and SMILI panels show the angular profiles from the fiducial image blurred to match the DIFMAP resolution. When blurred, the angular profiles from the eht-imaging and SMILI images better match the broad DIFMAP profile, but they still differ in the measured orientation angle and the position of the brightest location on the ring.

The angular profiles for the unblurred eht-imaging and SMILI reconstructions are similar, with the 0.1 Jy difference in total flux density in these reconstructions manifesting in an overall lower profile for the SMILI reconstruction. The "knot" features in the eht-imaging and SMILI reconstructions show significant variation across the Top Sets. This suggests that these ring features are highly sensitive to imaging parameter choices, and that they are potentially artifacts of the limited $(u, v)$ coverage. In general, the azimuthal structure in the reconstructed images from all three imaging pipelines is more variable than the radial structure, making measurements of the orientation angle intrinsically more uncertain than measurements of the ring diameter.

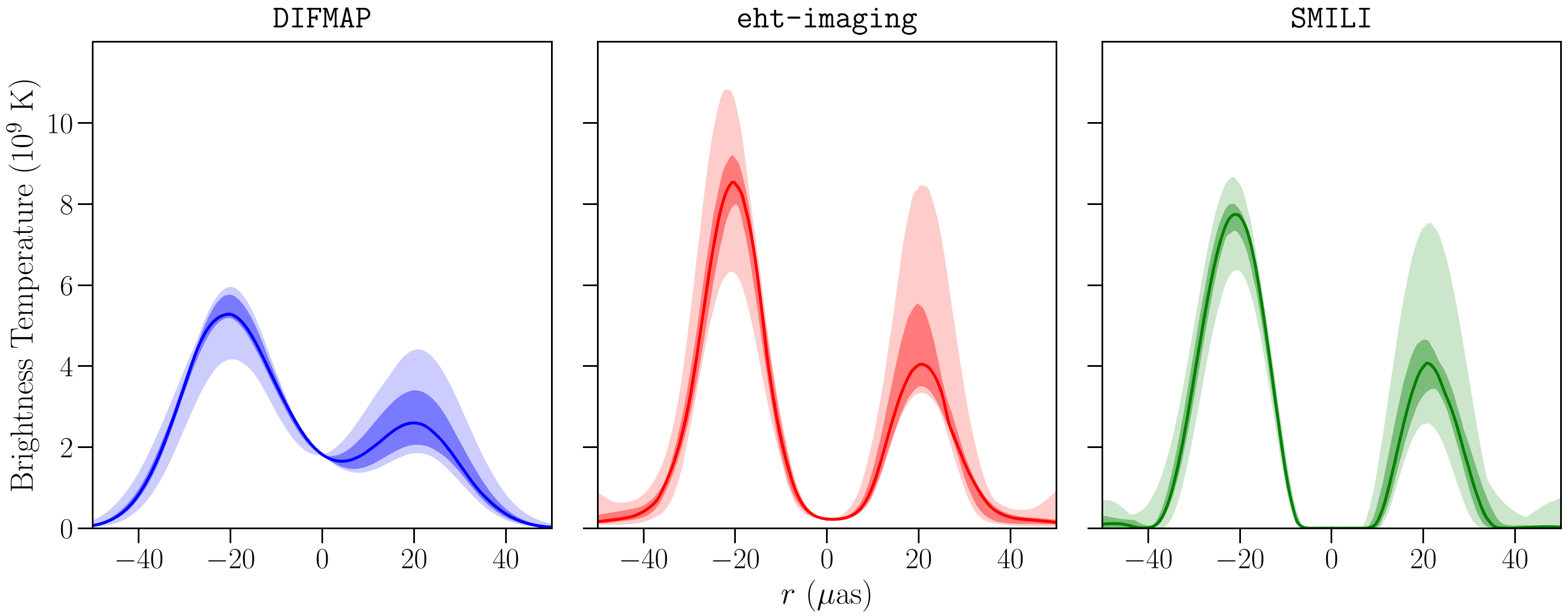

**Figure 37.** 1D radial brightness profiles of the three fiducial M87 images on April 11. For each image, radial profiles in the semi-circle centered on $\eta$ are plotted with negative values of $r$, and radial profiles through the opposing semi-circle centered on $\eta + 180°$ are plotted with positive $r$. The solid curves show the median profile over the corresponding semi-circle, the darker band shows the 25th to 75th percentile range, and the lighter band shows the full range of profiles in the fiducial images.

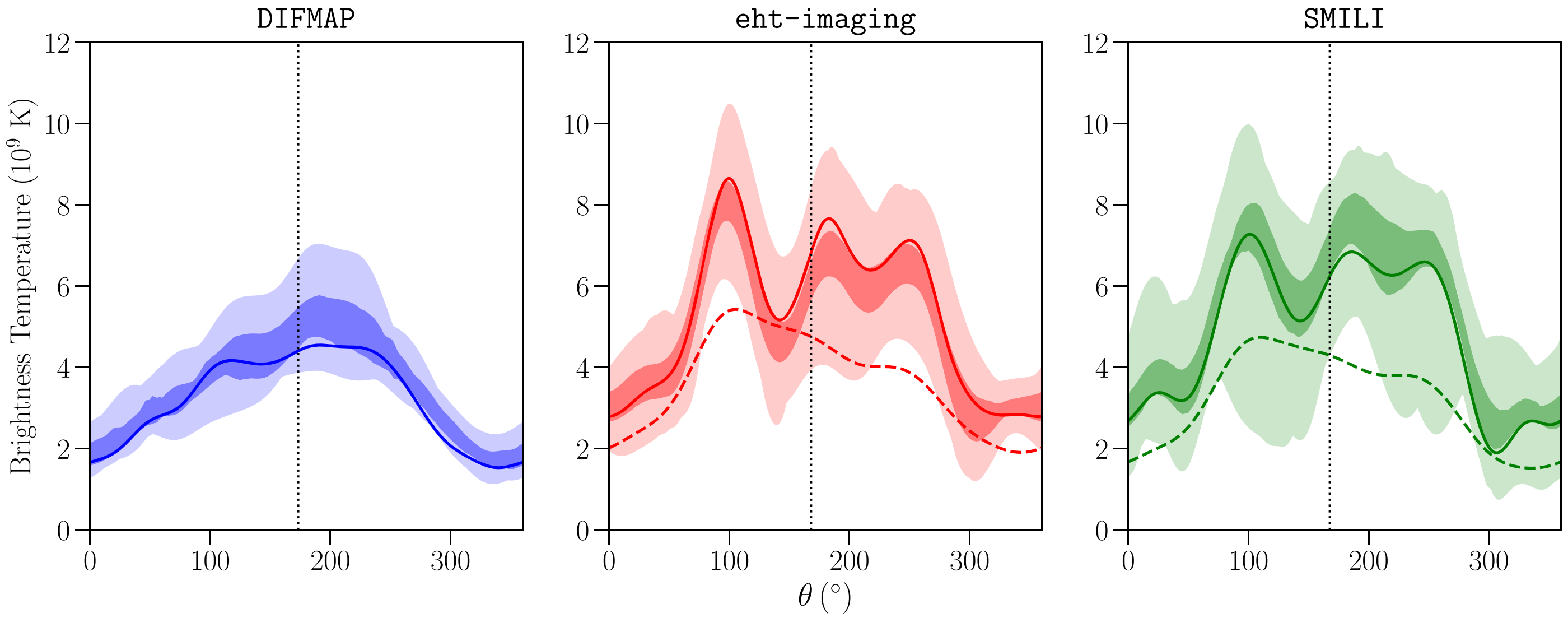

**Figure 38.** 1D angular profiles of M87 on April 11 from the three imaging methods. For each method, the solid line shows the angular profile obtained from the fiducial image, the darker band shows the 25th to 75th percentile range across the Top Set, and the lighter band shows the full Top Set range. For the RML methods, the dashed line shows the angular profile from the fiducial image blurred to the resolution of the `DIFMAP` image. The "knot" features in the unblurred `eht-imaging` and `SMILI` reconstructions show significant variation across the Top Set, suggesting that these features are sensitive to imaging parameter choices.

## ORCID iDs

Kazunori Akiyama https://orcid.org/0000-0002-9475-4254
Antxon Alberdi https://orcid.org/0000-0002-9371-1033
Rebecca Azulay https://orcid.org/0000-0002-2200-5393
Anne-Kathrin Baczko https://orcid.org/0000-0003-3090-3975
Mislav Baloković https://orcid.org/0000-0003-0476-6647
John Barrett https://orcid.org/0000-0002-9290-0764
Lindy Blackburn https://orcid.org/0000-0002-9030-642X
Katherine L. Bouman https://orcid.org/0000-0003-0077-4367
Geoffrey C. Bower https://orcid.org/0000-0003-4056-9982
Christiaan D. Brinkerink https://orcid.org/0000-0002-2322-0749
Roger Brissenden https://orcid.org/0000-0002-2556-0894
Silke Britzen https://orcid.org/0000-0001-9240-6734
Avery E. Broderick https://orcid.org/0000-0002-3351-760X
Do-Young Byun https://orcid.org/0000-0003-1157-4109
Andrew Chael https://orcid.org/0000-0003-2966-6220
Chi-kwan Chan https://orcid.org/0000-0001-6337-6126
Shami Chatterjee https://orcid.org/0000-0002-2878-1502

Ilje Cho https://orcid.org/0000-0001-6083-7521
Pierre Christian https://orcid.org/0000-0001-6820-9941
John E. Conway https://orcid.org/0000-0003-2448-9181
Geoffrey B. Crew https://orcid.org/0000-0002-2079-3189
Yuzhu Cui https://orcid.org/0000-0001-6311-4345
Jordy Davelaar https://orcid.org/0000-0002-2685-2434
Mariafelicia De Laurentis https://orcid.org/0000-0002-9945-682X
Roger Deane https://orcid.org/0000-0003-1027-5043
Jessica Dempsey https://orcid.org/0000-0003-1269-9667
Gregory Desvignes https://orcid.org/0000-0003-3922-4055
Jason Dexter https://orcid.org/0000-0003-3903-0373
Sheperd S. Doeleman https://orcid.org/0000-0002-9031-0904
Ralph P. Eatough https://orcid.org/0000-0001-6196-4135
Heino Falcke https://orcid.org/0000-0002-2526-6724
Vincent L. Fish https://orcid.org/0000-0002-7128-9345
Raquel Fraga-Encinas https://orcid.org/0000-0002-5222-1361
José L. Gómez https://orcid.org/0000-0003-4190-7613
Peter Galison https://orcid.org/0000-0002-6429-3872
Charles F. Gammie https://orcid.org/0000-0001-7451-8935
Boris Georgiev https://orcid.org/0000-0002-3586-6424
Roman Gold https://orcid.org/0000-0003-2492-1966
Minfeng Gu (顾敏峰) https://orcid.org/0000-0002-4455-6946
Mark Gurwell https://orcid.org/0000-0003-0685-3621
Kazuhiro Hada https://orcid.org/0000-0001-6906-772X
Ronald Hesper https://orcid.org/0000-0003-1918-6098
Luis C. Ho (何子山) https://orcid.org/0000-0001-6947-5846
Mareki Honma https://orcid.org/0000-0003-4058-9000
Chih-Wei L. Huang https://orcid.org/0000-0001-5641-3953
Shiro Ikeda https://orcid.org/0000-0002-2462-1448
Sara Issaoun https://orcid.org/0000-0002-5297-921X
David J. James https://orcid.org/0000-0001-5160-4486
Michael Janssen https://orcid.org/0000-0001-8685-6544
Britton Jeter https://orcid.org/0000-0003-2847-1712
Wu Jiang (江悟) https://orcid.org/0000-0001-7369-3539
Michael D. Johnson https://orcid.org/0000-0002-4120-3029
Svetlana Jorstad https://orcid.org/0000-0001-6158-1708
Taehyun Jung https://orcid.org/0000-0001-7003-8643
Mansour Karami https://orcid.org/0000-0001-7387-9333
Ramesh Karuppusamy https://orcid.org/0000-0002-5307-2919
Tomohisa Kawashima https://orcid.org/0000-0001-8527-0496
Garrett K. Keating https://orcid.org/0000-0002-3490-146X
Mark Kettenis https://orcid.org/0000-0002-6156-5617
Jae-Young Kim https://orcid.org/0000-0001-8229-7183
Junhan Kim https://orcid.org/0000-0002-4274-9373
Motoki Kino https://orcid.org/0000-0002-2709-7338
Jun Yi Koay https://orcid.org/0000-0002-7029-6658
Patrick M. Koch https://orcid.org/0000-0003-2777-5861
Shoko Koyama https://orcid.org/0000-0002-3723-3372
Michael Kramer https://orcid.org/0000-0002-4175-2271
Carsten Kramer https://orcid.org/0000-0002-4908-4925
Thomas P. Krichbaum https://orcid.org/0000-0002-4892-9586
Tod R. Lauer https://orcid.org/0000-0003-3234-7247
Sang-Sung Lee https://orcid.org/0000-0002-6269-594X
Yan-Rong Li (李彦荣) https://orcid.org/0000-0001-5841-9179
Zhiyuan Li (李志远) https://orcid.org/0000-0003-0355-6437
Michael Lindqvist https://orcid.org/0000-0002-3669-0715
Kuo Liu https://orcid.org/0000-0002-2953-7376
Elisabetta Liuzzo https://orcid.org/0000-0003-0995-5201
Laurent Loinard https://orcid.org/0000-0002-5635-3345
Ru-Sen Lu (路如森) https://orcid.org/0000-0002-7692-7967
Nicholas R. MacDonald https://orcid.org/0000-0002-6684-8691
Jirong Mao (毛基荣) https://orcid.org/0000-0002-7077-7195
Sera Markoff https://orcid.org/0000-0001-9564-0876
Daniel P. Marrone https://orcid.org/0000-0002-2367-1080
Alan P. Marscher https://orcid.org/0000-0001-7396-3332
Iván Martí-Vidal https://orcid.org/0000-0003-3708-9611
Lynn D. Matthews https://orcid.org/0000-0002-3728-8082
Lia Medeiros https://orcid.org/0000-0003-2342-6728
Karl M. Menten https://orcid.org/0000-0001-6459-0669
Yosuke Mizuno https://orcid.org/0000-0002-8131-6730
Izumi Mizuno https://orcid.org/0000-0002-7210-6264
James M. Moran https://orcid.org/0000-0002-3882-4414
Kotaro Moriyama https://orcid.org/0000-0003-1364-3761
Monika Moscibrodzka https://orcid.org/0000-0002-4661-6332
Cornelia Müller https://orcid.org/0000-0002-2739-2994
Hiroshi Nagai https://orcid.org/0000-0003-0292-3645
Neil M. Nagar https://orcid.org/0000-0001-6920-662X
Masanori Nakamura https://orcid.org/0000-0001-6081-2420
Ramesh Narayan https://orcid.org/0000-0002-1919-2730
Iniyan Natarajan https://orcid.org/0000-0001-8242-4373
Chunchong Ni https://orcid.org/0000-0003-1361-5699
Aristeidis Noutsos https://orcid.org/0000-0002-4151-3860
Héctor Olivares https://orcid.org/0000-0001-6833-7580
Daniel C. M. Palumbo https://orcid.org/0000-0002-7179-3816
Ue-Li Pen https://orcid.org/0000-0003-2155-9578
Dominic W. Pesce https://orcid.org/0000-0002-5278-9221
Oliver Porth https://orcid.org/0000-0002-4584-2557
Ben Prather https://orcid.org/0000-0002-0393-7734
Jorge A. Preciado-López https://orcid.org/0000-0002-4146-0113
Hung-Yi Pu https://orcid.org/0000-0001-9270-8812
Venkatessh Ramakrishnan https://orcid.org/0000-0002-9248-086X
Ramprasad Rao https://orcid.org/0000-0002-1407-7944
Alexander W. Raymond https://orcid.org/0000-0002-5779-4767
Bart Ripperda https://orcid.org/0000-0002-7301-3908
Freek Roelofs https://orcid.org/0000-0001-5461-3687
Eduardo Ros https://orcid.org/0000-0001-9503-4892
Mel Rose https://orcid.org/0000-0002-2016-8746
Alan L. Roy https://orcid.org/0000-0002-1931-0135
Chet Ruszczyk https://orcid.org/0000-0001-7278-9707
Benjamin R. Ryan https://orcid.org/0000-0001-8939-4461
Kazi L. J. Rygl https://orcid.org/0000-0003-4146-9043
David Sánchez-Arguelles https://orcid.org/0000-0002-7344-9920
Mahito Sasada https://orcid.org/0000-0001-5946-9960
Tuomas Savolainen https://orcid.org/0000-0001-6214-1085
Lijing Shao https://orcid.org/0000-0002-1334-8853
Zhiqiang Shen (沈志强) https://orcid.org/0000-0003-3540-8746
Des Small https://orcid.org/0000-0003-3723-5404
Bong Won Sohn https://orcid.org/0000-0002-4148-8378
Jason SooHoo https://orcid.org/0000-0003-1938-0720

Fumie Tazaki https://orcid.org/0000-0003-0236-0600
Paul Tiede https://orcid.org/0000-0003-3826-5648
Remo P. J. Tilanus https://orcid.org/0000-0002-6514-553X
Michael Titus https://orcid.org/0000-0002-3423-4505
Kenji Toma https://orcid.org/0000-0002-7114-6010
Pablo Torne https://orcid.org/0000-0001-8700-6058
Sascha Trippe https://orcid.org/0000-0003-0465-1559
Ilse van Bemmel https://orcid.org/0000-0001-5473-2950
Huib Jan van Langevelde https://orcid.org/0000-0002-0230-5946
Daniel R. van Rossum https://orcid.org/0000-0001-7772-6131
John Wardle https://orcid.org/0000-0002-8960-2942
Jonathan Weintroub https://orcid.org/0000-0002-4603-5204
Norbert Wex https://orcid.org/0000-0003-4058-2837
Robert Wharton https://orcid.org/0000-0002-7416-5209
Maciek Wielgus https://orcid.org/0000-0002-8635-4242
George N. Wong https://orcid.org/0000-0001-6952-2147
Qingwen Wu (吴庆文) https://orcid.org/0000-0003-4773-4987
André Young https://orcid.org/0000-0003-0000-2682
Ken Young https://orcid.org/0000-0002-3666-4920
Ziri Younsi https://orcid.org/0000-0001-9283-1191
Feng Yuan (袁峰) https://orcid.org/0000-0003-3564-6437
J. Anton Zensus https://orcid.org/0000-0001-7470-3321
Guangyao Zhao https://orcid.org/0000-0002-4417-1659
Shan-Shan Zhao https://orcid.org/0000-0002-9774-3606
Joseph R. Farah https://orcid.org/0000-0003-4914-5625
Daniel Michalik https://orcid.org/0000-0002-7618-6556
Andrew Nadolski https://orcid.org/0000-0001-9479-9957
Rurik A. Primiani https://orcid.org/0000-0003-3910-7529
Paul Yamaguchi https://orcid.org/0000-0002-6017-8199

---

The Event Horizon Telescope Collaboration

Kazunori Akiyama[1,2,3,4], Antxon Alberdi[5], Walter Alef[6], Keiichi Asada[7], Rebecca Azulay[8,9,6], Anne-Kathrin Baczko[6], David Ball[10], Mislav Baloković[4,11], John Barrett[2], Dan Bintley[12], Lindy Blackburn[4,11], Wilfred Boland[13], Katherine L. Bouman[4,11,14], Geoffrey C. Bower[15], Michael Bremer[16], Christiaan D. Brinkerink[17], Roger Brissenden[4,11], Silke Britzen[6], Avery E. Broderick[18,19,20], Dominique Broguiere[16], Thomas Bronzwaer[17], Do-Young Byun[21,22], John E. Carlstrom[23,24,25,26], Andrew Chael[4,11], Chi-kwan Chan[10,27], Shami Chatterjee[28], Koushik Chatterjee[29], Ming-Tang Chen[15], Yongjun Chen (陈永军)[30,31], Ilje Cho[21,22], Pierre Christian[10,11], John E. Conway[32], James M. Cordes[28], Geoffrey B. Crew[2], Yuzhu Cui[33,34], Jordy Davelaar[17], Mariafelicia De Laurentis[35,36,37], Roger Deane[38,39], Jessica Dempsey[12], Gregory Desvignes[6], Jason Dexter[40], Sheperd S. Doeleman[4,11], Ralph P. Eatough[6], Heino Falcke[17], Vincent L. Fish[2], Ed Fomalont[1], Raquel Fraga-Encinas[17], William T. Freeman[41,42], Per Friberg[12], Christian M. Fromm[36], José L. Gómez[5], Peter Galison[4,43,44], Charles F. Gammie[45,46], Roberto García[16], Olivier Gentaz[16], Boris Georgiev[19,20], Ciriaco Goddi[17,47], Roman Gold[36], Minfeng Gu (顾敏峰)[30,48], Mark Gurwell[11], Kazuhiro Hada[33,34], Michael H. Hecht[2], Ronald Hesper[49], Luis C. Ho (何子山)[50,51], Paul Ho[7], Mareki Honma[33,34], Chih-Wei L. Huang[7], Lei Huang (黄磊)[30,48], David H. Hughes[52], Shiro Ikeda[3,53,54,55], Makoto Inoue[7], Sara Issaoun[17], David J. James[4,11], Buell T. Jannuzi[10], Michael Janssen[17], Britton Jeter[19,20], Wu Jiang (江悟)[30], Michael D. Johnson[4,11], Svetlana Jorstad[56,57], Taehyun Jung[21,22], Mansour Karami[18,19], Ramesh Karuppusamy[6], Tomohisa Kawashima[3], Garrett K. Keating[11], Mark Kettenis[58], Jae-Young Kim[6], Junhan Kim[10], Jongsoo Kim[21], Motoki Kino[3,59], Jun Yi Koay[7], Patrick M. Koch[7], Shoko Koyama[7], Michael Kramer[6], Carsten Kramer[16], Thomas P. Krichbaum[6], Cheng-Yu Kuo[60], Tod R. Lauer[61], Sang-Sung Lee[21], Yan-Rong Li (李彦荣)[62], Zhiyuan Li (李志远)[63,64], Michael Lindqvist[32], Kuo Liu[6], Elisabetta Liuzzo[65], Wen-Ping Lo[7,66], Andrei P. Lobanov[6], Laurent Loinard[67,68], Colin Lonsdale[2], Ru-Sen Lu (路如森)[30,6], Nicholas R. MacDonald[6], Jirong Mao (毛基荣)[69,70,71], Sera Markoff[29,72], Daniel P. Marrone[10], Alan P. Marscher[56], Iván Martí-Vidal[32,73], Satoki Matsushita[7], Lynn D. Matthews[2], Lia Medeiros[10,74], Karl M. Menten[6], Yosuke Mizuno[36], Izumi Mizuno[12], James M. Moran[4,11], Kotaro Moriyama[33,2], Monika Moscibrodzka[17], Cornelia Müller[6,17], Hiroshi Nagai[3,34], Neil M. Nagar[75], Masanori Nakamura[7], Ramesh Narayan[4,11], Gopal Narayanan[76], Iniyan Natarajan[39], Roberto Neri[16], Chunchong Ni[19,20], Aristeidis Noutsos[6], Hiroki Okino[33,77], Héctor Olivares[36], Tomoaki Oyama[33], Feryal Özel[10], Daniel C. M. Palumbo[4,11], Nimesh Patel[11], Ue-Li Pen[18,78,79,80], Dominic W. Pesce[4,11], Vincent Piétu[16], Richard Plambeck[81], Aleksandar PopStefanija[76], Oliver Porth[36,29], Ben Prather[45], Jorge A. Preciado-López[18], Dimitrios Psaltis[10], Hung-Yi Pu[18], Venkatessh Ramakrishnan[75], Ramprasad Rao[15], Mark G. Rawlings[12], Alexander W. Raymond[4,11], Luciano Rezzolla[36], Bart Ripperda[36], Freek Roelofs[17], Alan Rogers[2], Eduardo Ros[6], Mel Rose[10], Arash Roshanineshat[10], Helge Rottmann[6], Alan L. Roy[6], Chet Ruszczyk[2], Benjamin R. Ryan[82,83], Kazi L. J. Rygl[65], Salvador Sánchez[84], David Sánchez-Arguelles[52,85], Mahito Sasada[33,86], Tuomas Savolainen[6,87,88], F. Peter Schloerb[76], Karl-Friedrich Schuster[16], Lijing Shao[6,51], Zhiqiang Shen (沈志强)[29,30], Des Small[58], Bong Won Sohn[21,22,89], Jason SooHoo[2], Fumie Tazaki[33], Paul Tiede[18,19,20], Remo P. J. Tilanus[17,47,90], Michael Titus[2], Kenji Toma[91,92], Pablo Torne[6,84], Tyler Trent[10], Sascha Trippe[93], Shuichiro Tsuda[33], Ilse van Bemmel[58], Huib Jan van Langevelde[58,94], Daniel R. van Rossum[17], Jan Wagner[6], John Wardle[95], Jonathan Weintroub[4,11], Norbert Wex[6], Robert Wharton[6], Maciek Wielgus[4,11], George N. Wong[45], Qingwen Wu (吴庆文)[96], André Young[17], Ken Young[11], Ziri Younsi[97,36], Feng Yuan (袁峰)[30,48,98], Ye-Fei Yuan (袁业飞)[99], J. Anton Zensus[6], Guangyao Zhao[21], Shan-Shan Zhao[17,63], Ziyan Zhu[44], Joseph R. Farah[11,100,4], Zheng Meyer-Zhao[7,101], Daniel Michalik[102,103], Andrew Nadolski[46], Hiroaki Nishioka[7], Nicolas Pradel[7], Rurik A. Primiani[104], Kamal Souccar[76], Laura Vertatschitsch[11,104], and Paul Yamaguchi[11]

[1] National Radio Astronomy Observatory, 520 Edgemont Rd, Charlottesville, VA 22903, USA
[2] Massachusetts Institute of Technology Haystack Observatory, 99 Millstone Road, Westford, MA 01886, USA
[3] National Astronomical Observatory of Japan, 2-21-1 Osawa, Mitaka, Tokyo 181-8588, Japan
[4] Black Hole Initiative at Harvard University, 20 Garden Street, Cambridge, MA 02138, USA
[5] Instituto de Astrofísica de Andalucía–CSIC, Glorieta de la Astronomía s/n, E-18008 Granada, Spain
[6] Max-Planck-Institut für Radioastronomie, Auf dem Hügel 69, D-53121 Bonn, Germany
[7] Institute of Astronomy and Astrophysics, Academia Sinica, 11F of Astronomy-Mathematics Building, AS/NTU No. 1, Sec. 4, Roosevelt Rd, Taipei 10617, Taiwan, R.O.C.
[8] Departament d'Astronomia i Astrofísica, Universitat de València, C. Dr. Moliner 50, E-46100 Burjassot, València, Spain
[9] Observatori Astronòmic, Universitat de València, C. Catedrático José Beltrán 2, E-46980 Paterna, València, Spain
[10] Steward Observatory and Department of Astronomy, University of Arizona, 933 N. Cherry Ave., Tucson, AZ 85721, USA
[11] Center for Astrophysics | Harvard & Smithsonian, 60 Garden Street, Cambridge, MA 02138, USA
[12] East Asian Observatory, 660 N. A'ohoku Pl., Hilo, HI 96720, USA
[13] Nederlandse Onderzoekschool voor Astronomie (NOVA), PO Box 9513, 2300 RA Leiden, The Netherlands
[14] California Institute of Technology, 1200 East California Boulevard, Pasadena, CA 91125, USA
[15] Institute of Astronomy and Astrophysics, Academia Sinica, 645 N. A'ohoku Place, Hilo, HI 96720, USA
[16] Institut de Radioastronomie Millimétrique, 300 rue de la Piscine, F-38406 Saint Martin d'Hères, France
[17] Department of Astrophysics, Institute for Mathematics, Astrophysics and Particle Physics (IMAPP), Radboud University, P.O. Box 9010, 6500 GL Nijmegen, The Netherlands
[18] Perimeter Institute for Theoretical Physics, 31 Caroline Street North, Waterloo, ON, N2L 2Y5, Canada
[19] Department of Physics and Astronomy, University of Waterloo, 200 University Avenue West, Waterloo, ON, N2L 3G1, Canada
[20] Waterloo Centre for Astrophysics, University of Waterloo, Waterloo, ON N2L 3G1 Canada
[21] Korea Astronomy and Space Science Institute, Daedeok-daero 776, Yuseong-gu, Daejeon 34055, Republic of Korea
[22] University of Science and Technology, Gajeong-ro 217, Yuseong-gu, Daejeon 34113, Republic of Korea
[23] Kavli Institute for Cosmological Physics, University of Chicago, 5640 South Ellis Avenue, Chicago, IL 60637, USA
[24] Department of Astronomy and Astrophysics, University of Chicago, 5640 South Ellis Avenue, Chicago, IL 60637, USA
[25] Department of Physics, University of Chicago, 5720 South Ellis Avenue, Chicago, IL 60637, USA
[26] Enrico Fermi Institute, University of Chicago, 5640 South Ellis Avenue, Chicago, IL 60637, USA
[27] Data Science Institute, University of Arizona, 1230 N. Cherry Ave., Tucson, AZ 85721, USA
[28] Cornell Center for Astrophysics and Planetary Science, Cornell University, Ithaca, NY 14853, USA
[29] Anton Pannekoek Institute for Astronomy, University of Amsterdam, Science Park 904, 1098 XH, Amsterdam, The Netherlands
[30] Shanghai Astronomical Observatory, Chinese Academy of Sciences, 80 Nandan Road, Shanghai 200030, People's Republic of China
[31] Key Laboratory of Radio Astronomy, Chinese Academy of Sciences, Nanjing 210008, People's Republic of China
[32] Department of Space, Earth and Environment, Chalmers University of Technology, Onsala Space Observatory, SE-439 92 Onsala, Sweden
[33] Mizusawa VLBI Observatory, National Astronomical Observatory of Japan, 2-12 Hoshigaoka, Mizusawa, Oshu, Iwate 023-0861, Japan
[34] Department of Astronomical Science, The Graduate University for Advanced Studies (SOKENDAI), 2-21-1 Osawa, Mitaka, Tokyo 181-8588, Japan
[35] Dipartimento di Fisica "E. Pancini", Universitá di Napoli "Federico II", Compl. Univ. di Monte S. Angelo, Edificio G, Via Cinthia, I-80126, Napoli, Italy
[36] Institut für Theoretische Physik, Goethe-Universität Frankfurt, Max-von-Laue-Straße 1, D-60438 Frankfurt am Main, Germany
[37] INFN Sez. di Napoli, Compl. Univ. di Monte S. Angelo, Edificio G, Via Cinthia, I-80126, Napoli, Italy
[38] Department of Physics, University of Pretoria, Lynnwood Road, Hatfield, Pretoria 0083, South Africa
[39] Centre for Radio Astronomy Techniques and Technologies, Department of Physics and Electronics, Rhodes University, Grahamstown 6140, South Africa
[40] Max-Planck-Institut für Extraterrestrische Physik, Giessenbachstr. 1, D-85748 Garching, Germany
[41] Department of Electrical Engineering and Computer Science, Massachusetts Institute of Technology, 32-D476, 77 Massachussetts Ave., Cambridge, MA 02142, USA
[42] Google Research, 355 Main St., Cambridge, MA 02142, USA
[43] Department of History of Science, Harvard University, Cambridge, MA 02138, USA
[44] Department of Physics, Harvard University, Cambridge, MA 02138, USA
[45] Department of Physics, University of Illinois, 1110 West Green St, Urbana, IL 61801, USA
[46] Department of Astronomy, University of Illinois at Urbana-Champaign, 1002 West Green Street, Urbana, IL 61801, USA
[47] Leiden Observatory—Allegro, Leiden University, P.O. Box 9513, 2300 RA Leiden, The Netherlands
[48] Key Laboratory for Research in Galaxies and Cosmology, Chinese Academy of Sciences, Shanghai 200030, People's Republic of China
[49] NOVA Sub-mm Instrumentation Group, Kapteyn Astronomical Institute, University of Groningen, Landleven 12, 9747 AD Groningen, The Netherlands
[50] Department of Astronomy, School of Physics, Peking University, Beijing 100871, People's Republic of China
[51] Kavli Institute for Astronomy and Astrophysics, Peking University, Beijing 100871, People's Republic of China
[52] Instituto Nacional de Astrofísica, Óptica y Electrónica. Apartado Postal 51 y 216, 72000. Puebla Pue., México
[53] The Institute of Statistical Mathematics, 10-3 Midori-cho, Tachikawa, Tokyo, 190-8562, Japan
[54] Department of Statistical Science, The Graduate University for Advanced Studies (SOKENDAI), 10-3 Midori-cho, Tachikawa, Tokyo 190-8562, Japan
[55] Kavli Institute for the Physics and Mathematics of the Universe, The University of Tokyo, 5-1-5 Kashiwanoha, Kashiwa, 277-8583, Japan
[56] Institute for Astrophysical Research, Boston University, 725 Commonwealth Ave., Boston, MA 02215, USA
[57] Astronomical Institute, St. Petersburg University, Universitetskij pr., 28, Petrodvorets,198504 St. Petersburg, Russia
[58] Joint Institute for VLBI ERIC (JIVE), Oude Hoogeveensedijk 4, 7991 PD Dwingeloo, The Netherlands
[59] Kogakuin University of Technology & Engineering, Academic Support Center, 2665-1 Nakano, Hachioji, Tokyo 192-0015, Japan
[60] Physics Department, National Sun Yat-Sen University, No. 70, Lien-Hai Rd, Kaosiung City 80424, Taiwan, R.O.C
[61] National Optical Astronomy Observatory, 950 North Cherry Ave., Tucson, AZ 85719, USA
[62] Key Laboratory for Particle Astrophysics, Institute of High Energy Physics, Chinese Academy of Sciences, 19B Yuquan Road, Shijingshan District, Beijing, People's Republic of China
[63] School of Astronomy and Space Science, Nanjing University, Nanjing 210023, People's Republic of China
[64] Key Laboratory of Modern Astronomy and Astrophysics, Nanjing University, Nanjing 210023, People's Republic of China
[65] Italian ALMA Regional Centre, INAF-Istituto di Radioastronomia, Via P. Gobetti 101, 40129 Bologna, Italy
[66] Department of Physics, National Taiwan University, No.1, Sect.4, Roosevelt Rd., Taipei 10617, Taiwan, R.O.C
[67] Instituto de Radioastronomía y Astrofísica, Universidad Nacional Autónoma de México, Morelia 58089, México
[68] Instituto de Astronomía, Universidad Nacional Autónoma de México, CdMx 04510, México
[69] Yunnan Observatories, Chinese Academy of Sciences, 650011 Kunming, Yunnan Province, People's Republic of China
[70] Center for Astronomical Mega-Science, Chinese Academy of Sciences, 20A Datun Road, Chaoyang District, Beijing, 100012, People's Republic of China
[71] Key Laboratory for the Structure and Evolution of Celestial Objects, Chinese Academy of Sciences, 650011 Kunming, People's Republic of China

[72] Gravitation Astroparticle Physics Amsterdam (GRAPPA) Institute, University of Amsterdam, Science Park 904, 1098 XH Amsterdam, The Netherlands
[73] Centro Astronómico de Yebes (IGN), Apartado 148, E-19180 Yebes, Spain
[74] Department of Physics, Broida Hall, University of California Santa Barbara, Santa Barbara, CA 93106, USA
[75] Astronomy Department, Universidad de Concepción, Casilla 160-C, Concepción, Chile
[76] Department of Astronomy, University of Massachusetts, 01003, Amherst, MA, USA
[77] Department of Astronomy, Graduate School of Science, The University of Tokyo, 7-3-1 Hongo, Bunkyo-ku, Tokyo 113-0033, Japan
[78] Canadian Institute for Theoretical Astrophysics, University of Toronto, 60 St. George Street, Toronto, ON M5S 3H8, Canada
[79] Dunlap Institute for Astronomy and Astrophysics, University of Toronto, 50 St. George Street, Toronto, ON M5S 3H4, Canada
[80] Canadian Institute for Advanced Research, 180 Dundas St West, Toronto, ON M5G 1Z8, Canada
[81] Radio Astronomy Laboratory, University of California, Berkeley, CA 94720, USA
[82] CCS-2, Los Alamos National Laboratory, P.O. Box 1663, Los Alamos, NM 87545, USA
[83] Center for Theoretical Astrophysics, Los Alamos National Laboratory, Los Alamos, NM, 87545, USA
[84] Instituto de Radioastronomía Milimétrica, IRAM, Avenida Divina Pastora 7, Local 20, E-18012, Granada, Spain
[85] Consejo Nacional de Ciencia y Tecnología, Av. Insurgentes Sur 1582, 03940, Ciudad de México, México
[86] Hiroshima Astrophysical Science Center, Hiroshima University, 1-3-1 Kagamiyama, Higashi-Hiroshima, Hiroshima 739-8526, Japan
[87] Aalto University Department of Electronics and Nanoengineering, PL 15500, FI-00076 Aalto, Finland
[88] Aalto University Metsähovi Radio Observatory, Metsähovintie 114, FI-02540 Kylmälä, Finland
[89] Department of Astronomy, Yonsei University, Yonsei-ro 50, Seodaemun-gu, 03722 Seoul, Republic of Korea
[90] Netherlands Organisation for Scientific Research (NWO), Postbus 93138, 2509 AC Den Haag, The Netherlands
[91] Frontier Research Institute for Interdisciplinary Sciences, Tohoku University, Sendai 980-8578, Japan
[92] Astronomical Institute, Tohoku University, Sendai 980-8578, Japan
[93] Department of Physics and Astronomy, Seoul National University, Gwanak-gu, Seoul 08826, Republic of Korea
[94] Leiden Observatory, Leiden University, Postbus 2300, 9513 RA Leiden, The Netherlands
[95] Physics Department, Brandeis University, 415 South Street, Waltham, MA 02453, USA
[96] School of Physics, Huazhong University of Science and Technology, Wuhan, Hubei, 430074, People's Republic of China
[97] Mullard Space Science Laboratory, University College London, Holmbury St. Mary, Dorking, Surrey, RH5 6NT, UK
[98] School of Astronomy and Space Sciences, University of Chinese Academy of Sciences, No. 19A Yuquan Road, Beijing 100049, People's Republic of China
[99] Astronomy Department, University of Science and Technology of China, Hefei 230026, People's Republic of China
[100] University of Massachusetts Boston, 100 William T, Morrissey Blvd, Boston, MA 02125, USA
[101] ASTRON, Oude Hoogeveensedijk 4, 7991 PD Dwingeloo, The Netherlands
[102] Science Support Office, Directorate of Science, European Space Research and Technology Centre (ESA/ESTEC), Keplerlaan 1, 2201 AZ Noordwijk, The Netherlands
[103] University of Chicago, 5640 South Ellis Avenue, Chicago, IL 60637, USA
[104] Systems and Technology Research, 600 West Cummings Park, Woburn, MA 01801, USA